\documentclass{article}
\usepackage{graphicx}
\usepackage{xcolor}
\usepackage{amsfonts}
\usepackage{amssymb}
\def\encadre#1#2{%
\setbox100=\hbox{\kern#1{#2}\kern#1}
\dimen100=\ht100 \advance \dimen100 by #1
\dimen101=\dp100 \advance \dimen101 by #1
\setbox100=\hbox{\vrule height \dimen100 depth \dimen101\box100\vrule}
\setbox100=\vbox{\hrule\box100\hrule}
\advance \dimen100 by .4pt \ht100=\dimen100
\advance \dimen101 by .4pt \dp100=\dimen101
\box100
\relax
}
\setbox221=\hbox{\includegraphics{cercle_L20.eps}}

\def\encercle#1#2{\hbox{\raise-5pt\copy221\hskip#2#1}}

\newtheorem{thm}{\textbf{Theorem}}
\newtheorem{lem}{\textbf{Lemma}}

\newtheorem{prop}{\textbf{Proposition}}
\newtheorem{defn}{\textbf{Definition}}

\newtheorem{fig}{\textbf{Figure}}

\newtheorem{const}{\textbf{Construction}}
\def\leurre{\noindent\leftskip0pt\rmix\mathix\baselineskip 10pt\parindent0pt}
\def\lanote #1 {\noindent{\leftskip0pt\rmix\mathix\baselineskip 10pt
\parindent0pt
{\sc Note:} #1\par}}

\makeatletter
\newcount\lapage\lapage=413
\newcount\pagei\pagei=\lapage\advance\lapage by 1
\newcount\pageii\pageii=\lapage\advance\lapage by 1
\newcount\pageiii\pageiii=\lapage\advance\lapage by 1
\newcount\pageiv\pageiv=\lapage\advance\lapage by 1
\newcount\pagev\pagev=\lapage\advance\lapage by 1
\newcount\pagevi\pagevi=\lapage\advance\lapage by 1
\newcount\pagevii\pagevii=\lapage\advance\lapage by 1
\newcount\pageviii\pageviii=\lapage\advance\lapage by 1
\newcount\pageix\pageix=\lapage\advance\lapage by 1
\newcount\pagex\pagex=\lapage\advance\lapage by 1
\makeatother
 

\makeindex
\begin{document}
\parindent=12pt
\def\ligne#1{\hbox to \hsize{#1}}
\def\PlacerEn#1 #2 #3 {\rlap{\kern#1\raise#2\hbox{#3}}}
\font\bfxiv=cmbx12 at 12pt
\font\bfxii=cmbx12
\font\bfix=cmbx9
\font\bfvii=cmbx7
\font\bfvi=cmbx6
\font\bfviii=cmbx8
\font\pc=cmcsc10
\font\itix=cmti9
\font\itviii=cmti8
\font\rmix=cmr9 \font\mmix=cmmi9 \font\symix=cmsy9
\def\mathix{\textfont0=\rmix \textfont1=\mmix \textfont2=\symix}
\font\rmviii=cmr8 \font\mmviii=cmmi8 \font\symviii=cmsy8
\def\mathviii{\textfont0=\rmviii \textfont1=\mmviii \textfont2=\symviii}
\font\rmvii=cmr7 \font\mmvii=cmmi7 \font\symvii=cmsy7
\def\mathvii{\textfont0=\rmvii \textfont1=\mmvii \textfont2=\symvii}
\font\rmvi=cmr6 \font\mmvi=cmmi6 \font\symvi=cmsy6
\def\mathvi{\textfont0=\rmvi \textfont1=\mmvi \textfont2=\symvi}
\font\rmv=cmr5 \font\mmv=cmmi5 \font\symv=cmsy5
\def\mathv{\textfont0=\rmv \textfont1=\mmv \textfont2=\symv}
\font\rmvii=cmr7
\font\rmv=cmr5
\ligne{\hfill}
\pagenumbering{arabic}
\begin{center}\bf\Large
The domino problem of the hyperbolic plane is undecidable, new proof
\end{center}
\vskip 20pt
\begin{center}\bf\large
Maurice Margenstern
\end{center}

\vskip10pt
\begin{center}\small
Laboratoire de G\'enie Informatique, de Production et de Maintenance, EA 3096,\\
Universit\'e de Lorraine, B\^atiment A de l'UFR MIM,\\
3 rue Augustin Fresnel,\\ 
BP 45112,\\ 
57073 Metz Cedex 03, France,\\
{\it e-mail} : maurice.margenstern@univ-lorraine.fr
\end{center}

\def\cqfd{\hbox{\kern 2pt\vrule height 6pt depth 2pt width 8pt\kern 1pt}}
\def\HH{\hbox{$I\!\!H$}}
\def\Hii{\hbox{$I\!\!H^2$}}
\def\Hiii{\hbox{$I\!\!H^3$}}
\def\Hiv{\hbox{$I\!\!H^4$}}
\vskip 10pt
\noindent

\begin{abstract}
\it The present paper is a new version of the arXiv paper revisiting the proof given in 
a paper of the author published in 2008 proving that the general tiling problem of the 
hyperbolic plane is undecidable by proving a slightly stronger version using only a 
regular polygon as the basic shape of the tiles. The problem was raised by a paper of
Raphael Robinson in 1971, in his famous simplified proof that the general tiling
problem is undecidable for the Euclidean plane, initially proved by Robert Berger
in~1966. The present construction simplifies that of the recent arXiv paper. It again
strongly reduces the number of prototiles.
\end{abstract}

\def\blank{\hbox{$\,$\_$\!$\_$\,$}}

\section{Introduction}

Whether it is possible to tile the plane with copies of a fixed
set of tiles was a question raised by Hao {\sc Wang}, \cite{wang} in the
late 50's of the previous century. {\sc Wang} solved the {\it origin-constrained}
problem, which consists in fixing an initial tile in the above finite set of
tiles. Indeed, fixing one tile is enough to entail the undecidability of the
problem.  Also called the {\bf general tiling problem} further in this paper,
the general case, free of any condition, in particular with no fixed initial tile,
was proved undecidable by Robert {\sc Berger} in 1966, \cite{berger}.
Both {\sc Wang}'s and {\sc Berger}'s proofs deal with the problem in the Euclidean
plane. In 1971, Raphael {\sc Robinson} found an alternative, simpler proof of the
undecidability of the general problem in the Euclidean plane, see \cite{robinson1}.
In that 1971 paper, Robinson raised the question of the general problem for the
hyperbolic plane. Seven years later, in 1978, he proved that in the hyperbolic
plane, the origin-constrained problem is undecidable, see \cite{robinson2}.
Since then, the problem had remained open.

   In 2007, I proved the undecidability of the tiling problem in the hyperbolic plane,
published in 2008, see \cite{mmundecTCS}. In a recent arXiv paper~\cite{mmarXiv22}, I 
presented a new proof of what was established in \cite{mmundecTCS}. I follow the same 
general idea but the tiling itself is changed. It is that of~\cite{mmarXiv22} but several
details of the presentation are changed in the present version whose result is a new 
significant reduction of the number of prototiles.

   In order the paper to be self-contained, I repeat the frame of the paper as well as the
strategy used to address the tiling problem. 

   In the present introduction, we remind the reader the general strategy to attack the
tiling problem, as already established in the famous proofs dealing with the Euclidean 
case. We assume that the reader is familiar with the tiling $\{7,3\}$ of the hyperbolic 
plane. That tiling is the frame which our solution of the problem lies in. The reader 
familiar with hyperbolic geometry can skip that part of the paper. We also refer the 
reader to~\cite{mmbook1} and to~\cite{mmDMTCS} for a more detailed introduction and for 
other bibliographical references. Also, in order that the paper can be self-contained, 
we sketch the notion of a space-time diagram of a Turing machine.

   With respect to paper~\cite{mmarXiv22}, I append a new section devoted to the 
construction of an aperiodic tiling. Hao Wang mentioned in~\cite{wang} that if 
any tiling of the hyperbolic plane were necessarily periodic then, the tiling problem 
would be decidable. Accordingly, the unsolvability of the problem entails the existence
of an aperiodic tiling of the hyperbolic plane. Section~\ref{s_aperiod} deals with that
point.

   In Section~\ref{s_proof}, I reuse the construction of Section~\ref{s_aperiod} to
establish the properties of the particular tiling which we consider within the 
tiling $\{7,3\}$ and which are later used for the proof of Theorem~\ref{undec}.
 
In Section~\ref{s_proof}, we proceed to the proof itself, leaning on
the definition of the needed tiles. In Subsection~\ref{ss_tiles} we proceed to the
counting of the needed prototiles. That allows us to prove: 

\begin{thm}\label{undec}
\it The domino problem of the hyperbolic plane is undecidable.
\end{thm}

   Reproducing the similar section of \cite{mmarXiv22} for self-containedness, Section~\ref{s_corol} gives
several corollaries of Theorem~\ref{undec}. We conclude about the difference between the present paper,
paper~\cite{mmarXiv22} and paper~\cite{mmundecTCS}.

   From Theorem~\ref{undec}, we immediately conclude that the general
tiling problem is undecidable in the hyperbolic plane.

\section{An aperiodic tiling of the hyperbolic plane}\label{s_aperiod}

    Subsection~\ref{ss_heptatil} briefly mentions the frame of our constructions.
Then, in Subsection~\ref{ss_aperiod}, we proceed to the construction of an aperiodic
tiling of the hyperbolic plane.

\subsection{The tiling $\{7,3\}$}\label{ss_heptatil}

   We assume the reader to be familiar with hyperbolic geometry. We can refer him/her to
\cite{mmbook1} for an introduction.


   {\bf Regular tessellations} are a particular case of tilings. They
are generated from a regular polygon by reflection in its sides
and, recursively, of the images in their sides. In the Euclidean
case, there are, up to isomorphism and up to similarities, three
tessellations, respectively based on the square, the equilateral
triangle and on the regular hexagon. Later on we say {\bf tessellation},
for short.

   In the hyperbolic plane, there are infinitely many tessellations.
They are based on the regular polygons with $p$ sides and with
$\displaystyle{{2\pi}\over q}$ as vertex angle and they are
denoted by $\{p,q\}$. This is a consequence of a famous theorem
by Poincar\'e which characterises the triangles starting from
which a tiling can be generated by the recursive reflection
process which we already mentioned. Any triangle tiles the
hyperbolic plane if its vertex angles are of the form
$\displaystyle{\pi\over p}$, $\displaystyle{\pi\over q}$ and
$\displaystyle{\pi\over r}$ with the condition that
$\displaystyle{1\over p}+\displaystyle{1\over q}+
\displaystyle{1\over r}< 1$.

   Among these tilings, we choose the tiling $\{7,3\}$ which
we called the {\bf ternary heptagrid} in \cite{ibkmACRI}. It is
below illustrated by Figure~\ref{splittil_73}. From now on we call it
the {\bf heptagrid}.

\vskip 0pt
\vtop{
\ligne{\hfill
\includegraphics[scale=0.5]{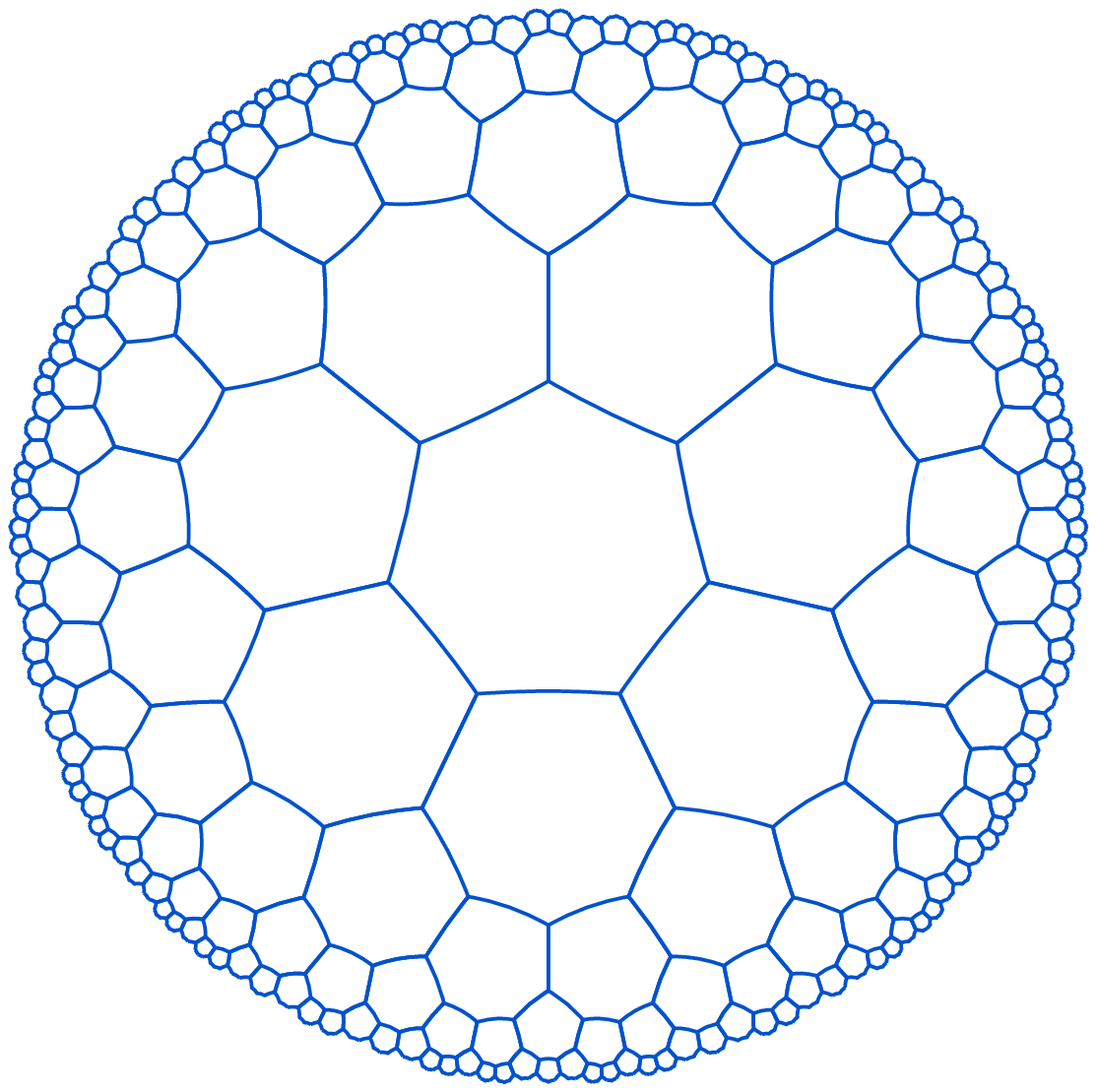}
\hfill}
\begin{fig}\label{splittil_73}
\leurre
The heptagrid in the Poincar\'e's disc
model.
\end{fig}
}

   In \cite{ibkmACRI,mmbook1}, many properties of the
heptagrid are described. An important tool to establish them is
the splitting method, prefigured in \cite{mmJUCSii} and for
which we refer to~\cite{mmbook1}. Here, we just suggest the
use of this method which allows us to exhibit a tree, spanning the
tiling: the {\bf Fibonacci tree}. Below, the left-hand side
of Figure~\ref{til_73} illustrates the splitting of \Hii\ into
a central tile~$T$ and seven sectors dispatched around~$T$.
Each sector is spanned by a Fibonacci tree. The right-hand side
of Figure~\ref{til_73} illustrates how the sector can be split
into sub-regions. Now, we notice that two of these regions are copies
of the same sector and that the third region~$S$ can be split into
a tile and then a copy of a sector and a copy of~$S$. Such
a process gives rise to a tree whose nodes are in bijection with the tiles
of the sector. The tree structure will be used in the sequel and
other illustrations will allow the reader to better understand
the process.

\ligne{\hfill}
\vtop{
\vspace{-195pt}
\ligne{\hfill
\includegraphics[scale=0.5]{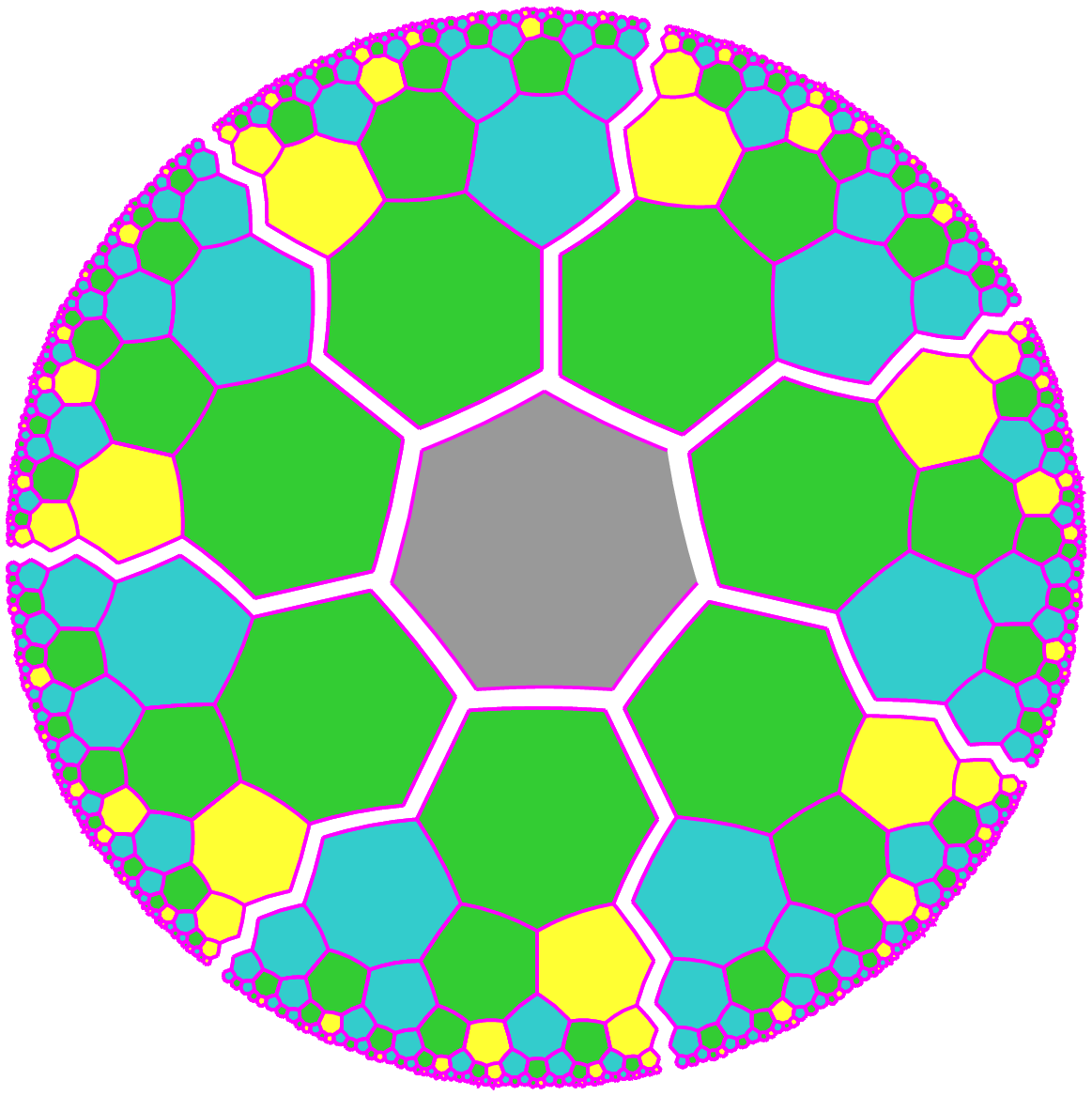}
\hskip-40pt
\includegraphics[scale=1.1]{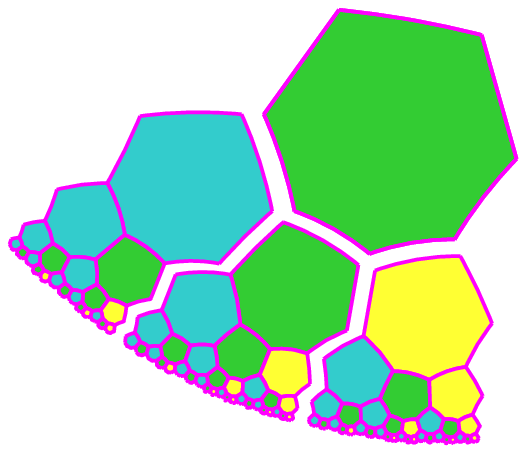}
\hfill}
\vskip 5pt
\begin{fig}\label{til_73}
\leurre
Left-hand side: the standard Fibonacci trees which span the 
heptagrid. Right-hand side: the 
splitting of a sector, spanned by a Fibonacci tree.
\end{fig}
}

\vtop{
\ligne{\hfill
\includegraphics[scale=0.55]{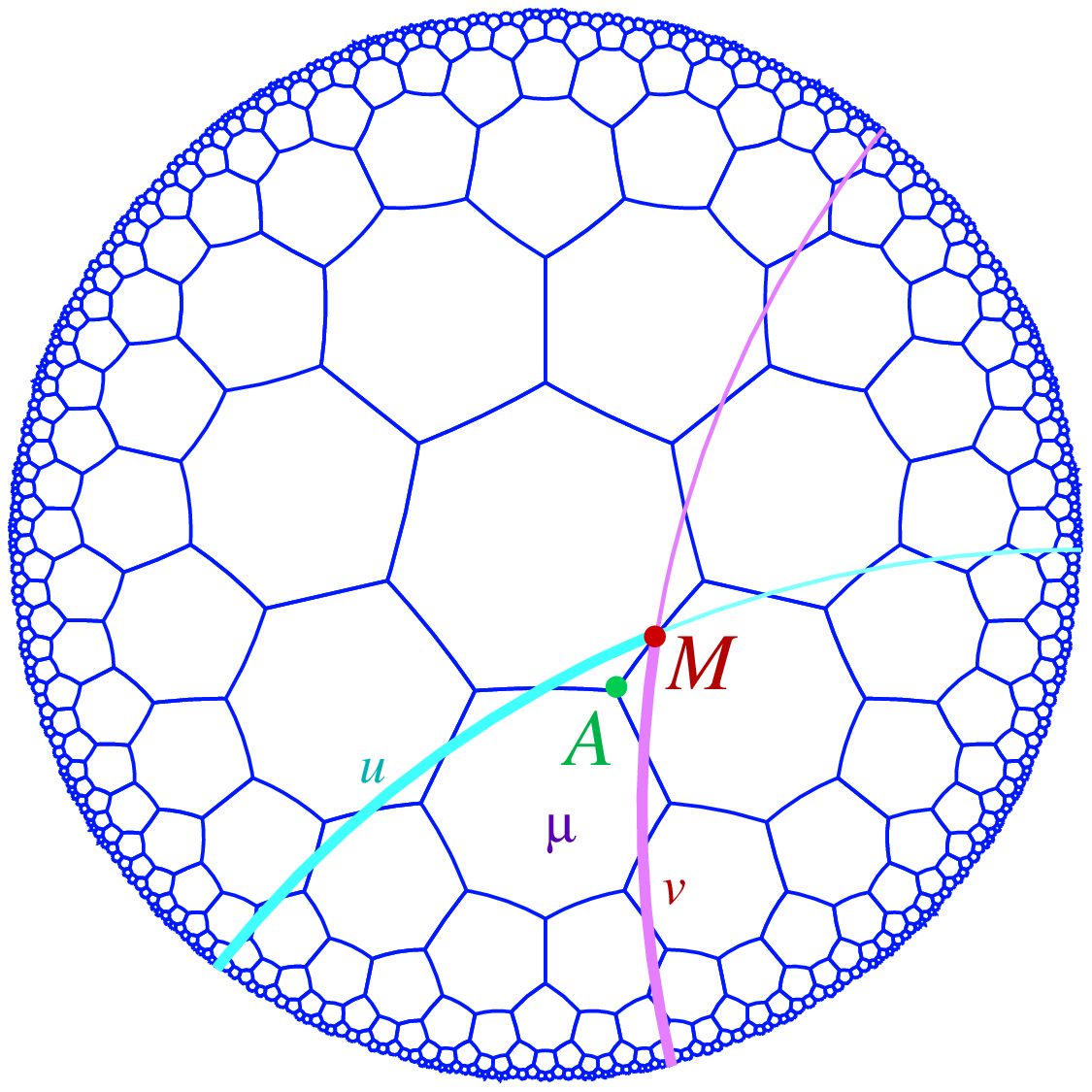}
\hfill}
\begin{fig}\label{the_mids}
\leurre
The mid-point lines delimiting a sector of the heptagrid. The rays $u$ and $v$ are supported by mid-point
lines.
\end{fig}
}
\vskip 10pt
Another important tool to study the tiling $\{7,3\}$ is given by
the {\bf mid-point} lines, which are illustrated by
Figure~\ref{the_mids}. The lines have that name as far as they
join the mid-points of contiguous edges of tiles. Let $s_0$ be a side of a tile.
Let $M$ be the mid-point of~$s_0$ and let $A$ be one of the vertices defined by~$s_0$. Two sides~$s_1$ 
and~$s_2$ of tiles of the heptagrid also share $A$. They define a tile~$\mu$. Let~$u$, $v$ be the rays 
issued from~$A$ which crosses the mid-point of~$s_1$, $s_2$ respectively, see Figure~\ref{the_mids}.
There, we can see how such rays us allow to delimit a sector, a property which is proved 
in~\cite{ibkmACRI,mmbook1}. Later on, such a sector will be called a {\bf sector of the heptagrid}. 
We say that $\mu$ is its {\bf root} or that the sector is {\bf rooted at}~$\mu$.

\subsection{Generating the heptagrid with tiles}
\label{ss_pre_tiles}

\def\GG{{\bf G}}
\def\YY{{\bf Y}}
\def\BB{{\bf B}}
\def\OO{{\bf O}}
\def\MM{{\bf M}}
\def\RR{{\bf R}}
   Now, we show that the tiling which we have described in general
terms in the previous section can effectively be generated from a small
finite set of tiles which we call the {\bf prototiles}: we simply need 4~of them. 
The basic colours we consider are {\bf green}, {\bf yellow}, {\bf blue} and 
{\bf orange}:
we denote them by \GG, \YY, \BB{} and \OO{} respectively.

\subsubsection{Trees of the heptagrid}
\label{sss_trees}

Using the tiles defined previously, we define a tiling by applying the rules~$(R_0)$
also illustrated by Figure~\ref{f_proto_0}. The tiles we use are copies of the prototiles.
\vskip 5pt
\ligne{\hfill
$\vcenter{\vtop{\leftskip 0pt\parindent 0pt\hsize 250pt
\ligne{\hfill
\GG{} $\rightarrow$ \YY\BB\GG,\hskip 20pt
\BB{} $\rightarrow$ \BB\OO,\hskip 20pt
\YY{} $\rightarrow$ \YY\BB\GG,\hskip 20pt
\OO{} $\rightarrow$ \YY\BB\OO
\hfill}
}}$
\hfill$(R_0)$\hskip 10pt}
\vskip 10pt
\ligne{\hfill
\vtop{\hsize=300pt
\ligne{\hfill
\includegraphics[scale=0.3]{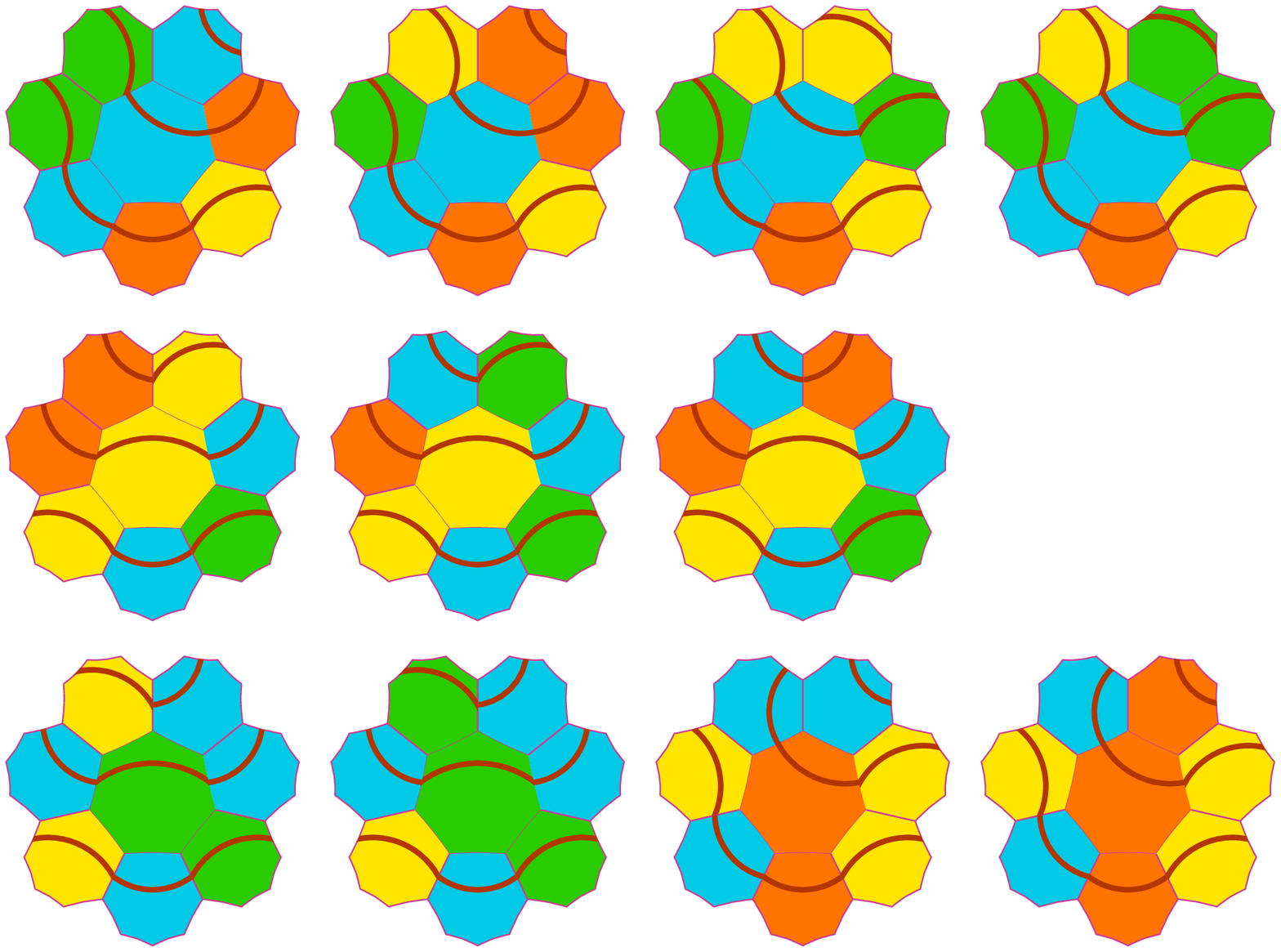}
\hfill}
\vspace{-10pt}
\begin{fig}\label{f_proto_0}
\leurre
The prototiles generating the tiling: all cases for the neighbourhood of
a tile are considered, whatever the tile: \BB, \YY, \OO{} or \GG. The neighbourhoods 
around a tile of the same colour correspond to the different occurrences of that colour 
in the right-hand side part of the rules~$(R_0)$. 
\end{fig}
}
\hfill}

   Infinitely many tilings of the heptagrid can be constructed by applying the 
rules~$(R_0)$ with the help of Figure~\ref{f_proto_0}. That figure illustrates all 
possible neighbourhoods
of tiles which are copies of the proto-tiles \GG, \YY, \OO{} and \BB. We later 
on say that such a symbol constitutes the {\bf status} of the tile. The first line
of the figure deals with the four possible neighbourhoods for a \BB-tile: the central
tile is a \BB-tile and the father is, successively, a \BB-, an \OO-, a \YY- and a \GG-tile.
As far as \YY{} occurs three times in the right side of the rules of~$(R)$, the second 
line of Figure~\ref{f_proto_0} illustrates
the  possible neighbourhoods for a \YY-tile. Next, both \GG{} and \OO{} appear twice only
in the right-hand side part of a rule of~$(R_0)$. It is the reason why in the third line
of the figure, the first two neighbourhoods concern a \GG-tile while the next two ones
deal with an \OO-tile.

\def\nng{\hbox{\bf\small g} } 
\def\nny{\hbox{\bf\small y} }
\def\nnb{\hbox{\bf\small b} }
\def\nno{\hbox{\bf\small o} }
   In order to get some order in the tiling, we introduce a numbering of the sides of
each tile of the heptagrid. Assume that the side which is given number~1 is fixed.
We then number the other sides increasingly while counterclockwise turning around the 
tile. In a tile, side~1 is the side shared with its father. As far as the central tile has
no father, its side~1 is arbitrarily fixed. The neighbours are numbered after the side
it shares with the tile we consider. A side receives different numbers in the tiles which
share it. 
Table~$(N)$ indicates the correspondence between those numbers depending on the status
of a tile~$\tau$ and a neighbour~$\nu$. The index \nng, \nny, \nnb{} or \nno{} refers
to the status of~$\nu$. When the number in $\nu$ is~1 it means that $\nu$
is a son of~$\tau$. 
\newdimen\ntlsizea\ntlsizea=45pt
\newdimen\ntlsizeb\ntlsizeb=35pt
\def\numtiline #1 #2 #3 #4 #5 #6 #7 #8 {%
\ligne{%
\hbox to \ntlsizea{#1\hfill}
\hbox to \ntlsizea{#2\hfill}
\hbox to \ntlsizeb{#3\hfill}
\hbox to \ntlsizeb{#4\hfill}
\hbox to \ntlsizeb{#5\hfill}
\hbox to \ntlsizeb{#6\hfill}
\hbox to \ntlsizeb{#7\hfill}
\hbox to \ntlsizeb{#8\hfill}
\hfill}
}
\vskip 5pt
\ligne{\hfill
$\vcenter{\vtop{\leftskip 0pt\parindent 0pt\hsize 320pt
\numtiline {}           {1}                     {2}          {3}           {4}                      {5}        {6}         {7} 
\numtiline {\GG-tile}   {6$_{\nny}$,7$_{\nng}$} {7$_{\nnb}$} {1$_{\nny}$}  {1$_{\nnb}$}             {1$_{\nng}$} {2$_{\nnb}$}  {3$_{\nnb}$}  
\numtiline {\YY-tile}   {4$_{\nny}$,3$_{\nng,\nno}$} {6$_{\nno,\nnb}$} {7$_{\nno}$}         {1$_{\nny}$}   {1$_{\nnb}$} {1$_{\nng}$}  {2$_{\nnb}$}  
\numtiline {\BB-tile}   {5$_{\nny}$,4$_{\nng,\nnb,\nno}$}  {7$_{\nny}$,6$_{\nng}$}           {7$_{\nng}$}  {1$_{\nnb}$} {1$_{\nno}$} {2$_{\nny}$} {2$_{\nng,\nno}$}  
\numtiline {\OO-tile}   {5$_{\nnb,\nno}$}  {7$_{\nnb}$}   {1$_{\nny}$}  {1$_{\nnb}$}        {1$_{\nno}$} {2$_{\nny}$}  {3$_{\nny}$}  
}}$
\hfill$(N)$\hskip 15pt}
\vskip 10pt
   Let us start with a few definitions needed to explain the properties which the
construction needs for proving the result of the present section.

   Let $\tau_0$ and $\tau_1$ be two tiles of the heptagrid. A {\bf path} between $\tau_0$
and $\tau_1$ is a finite sequence $\{\mu_i\}_{i\in\{0..n\}}$ such that $\mu_i$ and 
$\mu_{i+1}$ when \hbox{$0\leq i<n$}, share a side, say in that case the tiles are
{\bf adjacent}, and such that \hbox{$\mu_0=\tau_0$} and \hbox{$\mu_n=\tau_1$}. In that
case, we say that $n$+1 is the {\bf length} of the path and we also say that the path
{\bf joins} $\tau_0$ to~$\tau_1$. The {\bf distance} between two tiles $\tau_0$ and $\tau_1$ is the 
smallest $m$ for which there is a path joining~$\tau_0$ to~$\tau_1$ whose length is~$m$.

Considering a sector~$\mathcal S$ as above defined, rooted at a tile~$\mu$, see
Figure~\ref{the_mids}. We know that the set of tiles whose centre is contained in 
$\mathcal S$ is spanned by a tree rooted at~$\mu$. In~\cite{mmJUCSii,mmbook1}, it is 
proved that such a tree is spanned by the rules of~$(R_0)$. We call such a tree a 
{\bf tree of the heptagrid} when its root is not a \BB-tile. We indifferently say that 
$A$, see the figure, is the origin of the tree or of the sector and that $A$ points at 
$\mu$. We say that the origin of the tree points at the root of the tree. We denote that 
tree by $T(\mu)$. Note, that a tree of the heptagrid is a set of tiles, it is not the set 
of points contained in those tiles. We call left-, right-hand side {\bf border} of 
$T(\mu)$ the set of tiles of the tree which are crossed  by the left-, right hand side 
ray respectively which delimit the sector defining the tree.

In a tree of the heptagrid $\mathcal T$, the {\bf level~$m$} is the 
set of tiles of~$\mathcal T$ which are at the distance~$m$ from the root. By induction, it is easy to
prove that the sons of a tile on the level~$m$ belong to the level~$m$+1. 

   A tiling of the heptagrid can be defined by the following process:

\begin{const}\label{cons_til}
\phantom{construction}\\
	$-$ Time~$0$: fix a tile~$\tau$ as a root of a tree~$T(\tau_0)$ of the heptagrid;
that root is the level~$0$ of $T(\tau_0)$ and choose its status among \YY, \GG{} or \OO;

	$-$ time~$m$+1, $m\in \mathbb N$: construct $\tau_{m+1}$ as a father 
of~$\tau_m$, taking as $\tau_{m+1}$ with a status which is compatible with that 
of~$\tau_m$; construct the level~$1$ of $T(\tau_{m+1})$ and for $i$ in $\{0..m\}$, if 
$h_i$ is the level of $T(\tau_i)$ constructed at time~$m$, construct the 
level~$h_i$$+$$1$.
\end{const}

   It is not difficult to establish the following property:
\def\dprop #1 {\begin{prop}\label{#1}}
\def\fprop{\end{prop}}

\dprop {psubtree_levels}
Let $T_0$, $T_1$ be two trees of the heptagrid with $T_1\subset T_0$. Then a level of 
$T_1$ is contained in a level of $T_0$. More precisely, let $\rho_1$ be the root of $T_1$.
Let $h$ be the level of~$\rho_1$ in $T_0$. Let $\tau$ be a tile of~$T_1$ and let be $k$
its level in $T_1$. Then, the level of $\tau$ in~$T_0$ is $h$+$k$.
\fprop

\noindent
Proof. Let $\rho_0$, $\rho_1$ be the roots of~$T_0$, $T_1$ respectively. If $\rho_1$
belongs to the level~1 of $T_0$, the sons of~$\rho_1$, which belong to the level~1 
of~$T_1$, belong to the level~2 of~$T_0$.
Consequently, if the property is true if $\rho_1$ belongs to the level~$k$ of $T_0$,
it is also true for the trees of the heptagrid which are rooted at a tile of the 
level~$k$+1 of~$T_0$. Which proves the property.
\hfill$\Box$

Below, Figure~\ref{f_hepta_tils} illustrate two different applications of 
Construction~\ref{cons_til}. The left-hand side picture represents an implementation
where at time~0 the initial root is a \GG-tile and, at each time, the father of
$\tau_{m+1}$ is also a \GG-tile for several consecutive values of~$m$ and then, it is
a \YY-tile. In the central and in the right-hand side pictures, we have two views of the
same implementation: an infinite sequence of consecutive \GG-tiles are crossed by a 
line~$\ell$ so that an infinite sequence of consecutive \BB-tiles are crossed by~$\ell$
too.

\vskip 10pt
\vtop{
\ligne{\hfill
\includegraphics[scale=0.3]{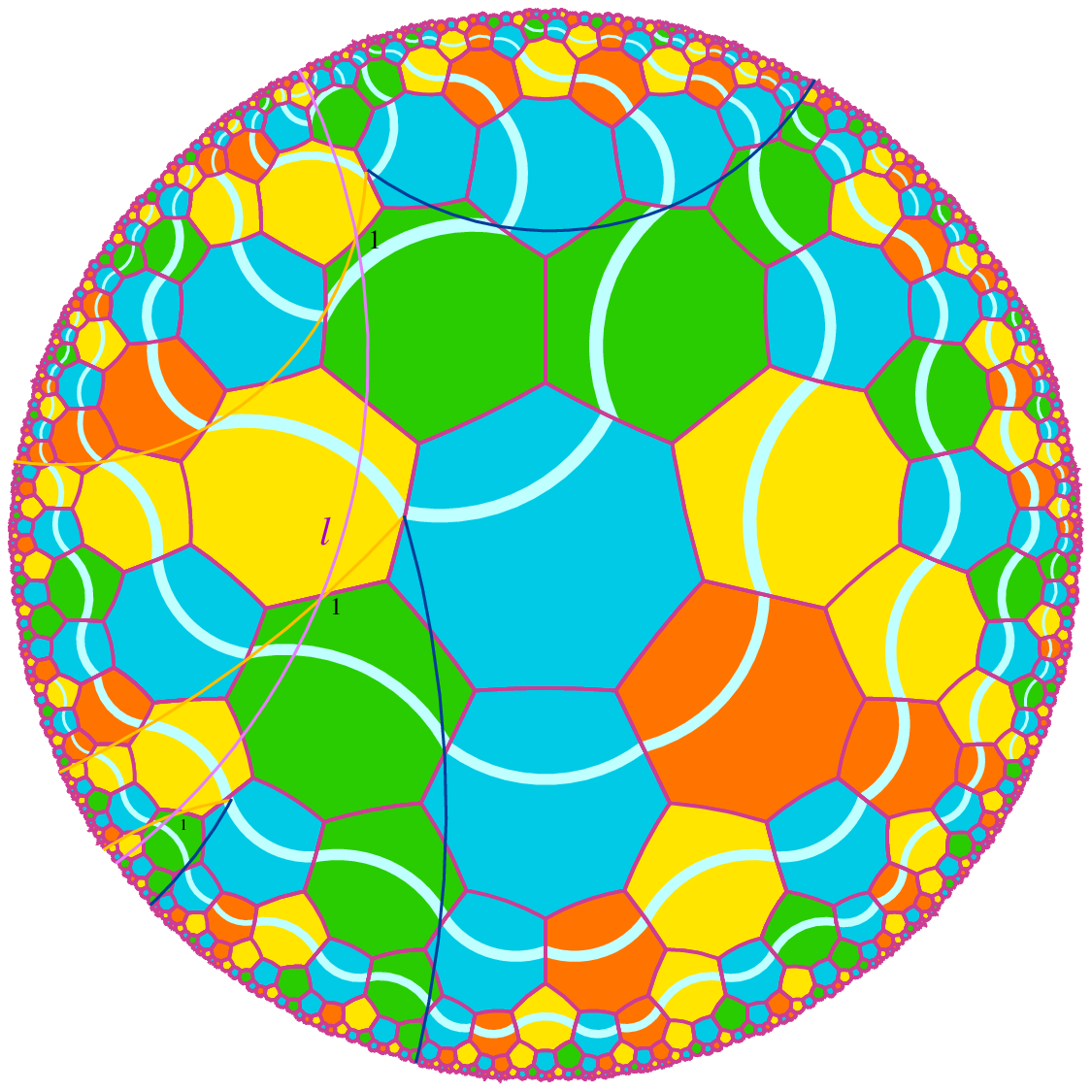}
\includegraphics[scale=0.3]{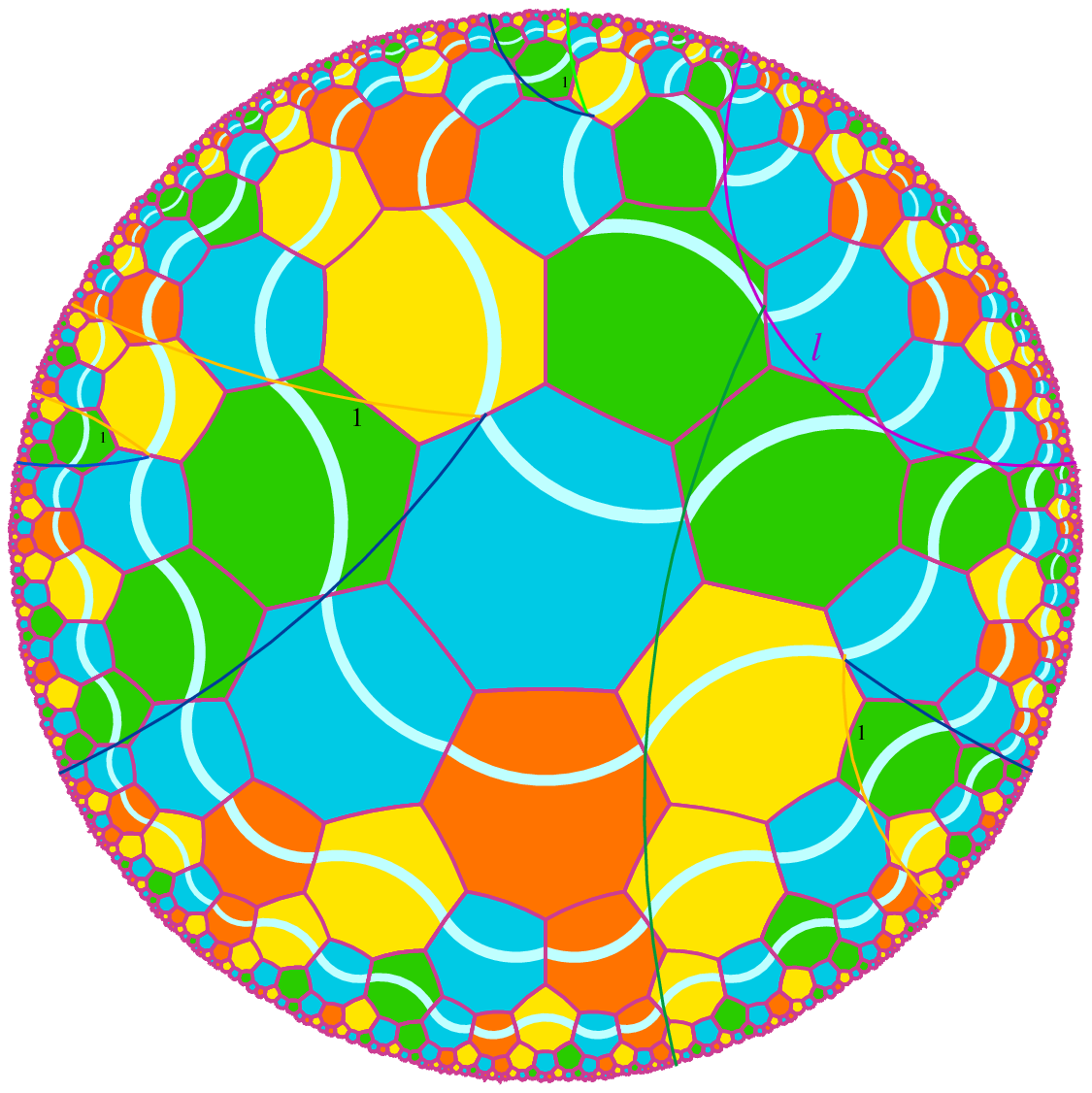}
\includegraphics[scale=0.3]{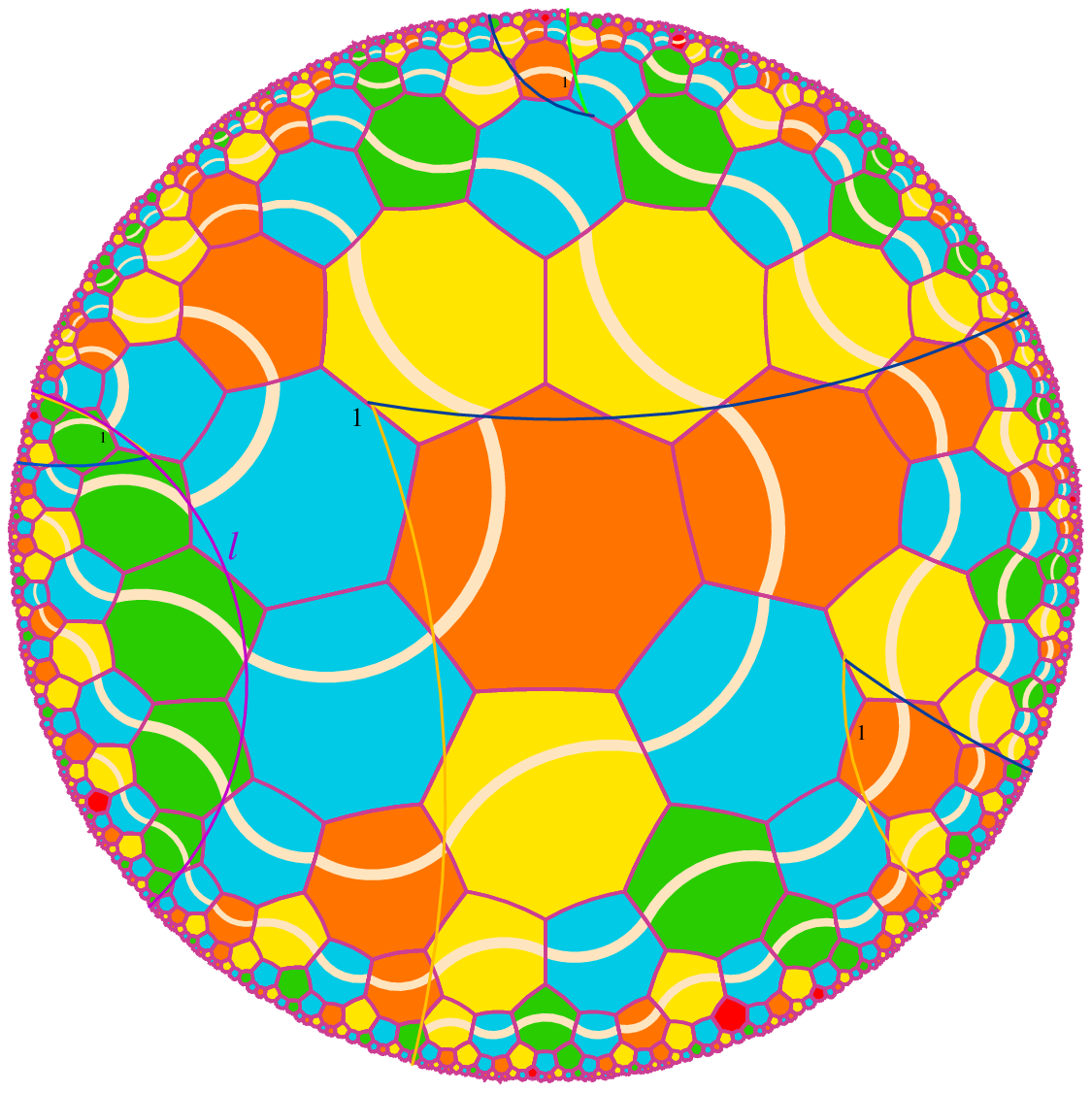}
\hfill}
\begin{fig}\label{f_hepta_tils}
\leurre
Two examples of an implementation of the rules~$(R_0)$ to tile the heptagrid. To left,
a sequence of successive \GG- and \YY-tiles are crossed by the same line. In the centre
and to right, a sequence of \GG-tiles are crossed by the same line.
\end{fig}
}

On those figures, we can notice levels for which Proposition~\ref{psubtree_levels} is of 
course true. But the figures lead us to introduce a definition. From 
construction~\ref{cons_til} and from Proposition~\ref{psubtree_levels}, we can see that 
a level of $T(\tau_i)$ is continued in $T(\tau_{i+1})$ and, {\it a fortiori}, in all 
$T(\tau_{i+k})$ for all positive integer~$k$. 

\dprop {pinclusion}
For any tree of the heptagrid $T(\tau)$, if $\mu$ is a non \BB-tile belonging to that
tree, then \hbox{$T(\mu)\subset T(\tau)$}.
\fprop

\noindent
Proof. It is true when $\mu$ are sons of~$\tau$ or if $\mu$ is the \OO-son of the \BB-son
of~$\tau$, see figure~\ref{f_tree}. Let $u$ and $v$ be the left- and right-hand side ray
respectively issued from the origin $A$ of $T(\tau)$.

In each of those cases, $T(\mu)$ is the image of $T(\tau)$ under an appropriate shift:
a shift along $u$, $v$ when $\mu$ is the \YY-, \GG-son respectively of $\tau$; a shift
along the left-hand side ray~$w$ defining $T(\mu)$ when $\mu$ is the \OO-son of the
\BB-son of~$\mu$. Accordingly, the proposition follows for any tile~$\nu$ whose status is 
not \BB{} by induction on the level of~$\nu$.\hfill$\Box$

\vskip 10pt
\vtop{
\ligne{\hfill
\includegraphics[scale=0.6]{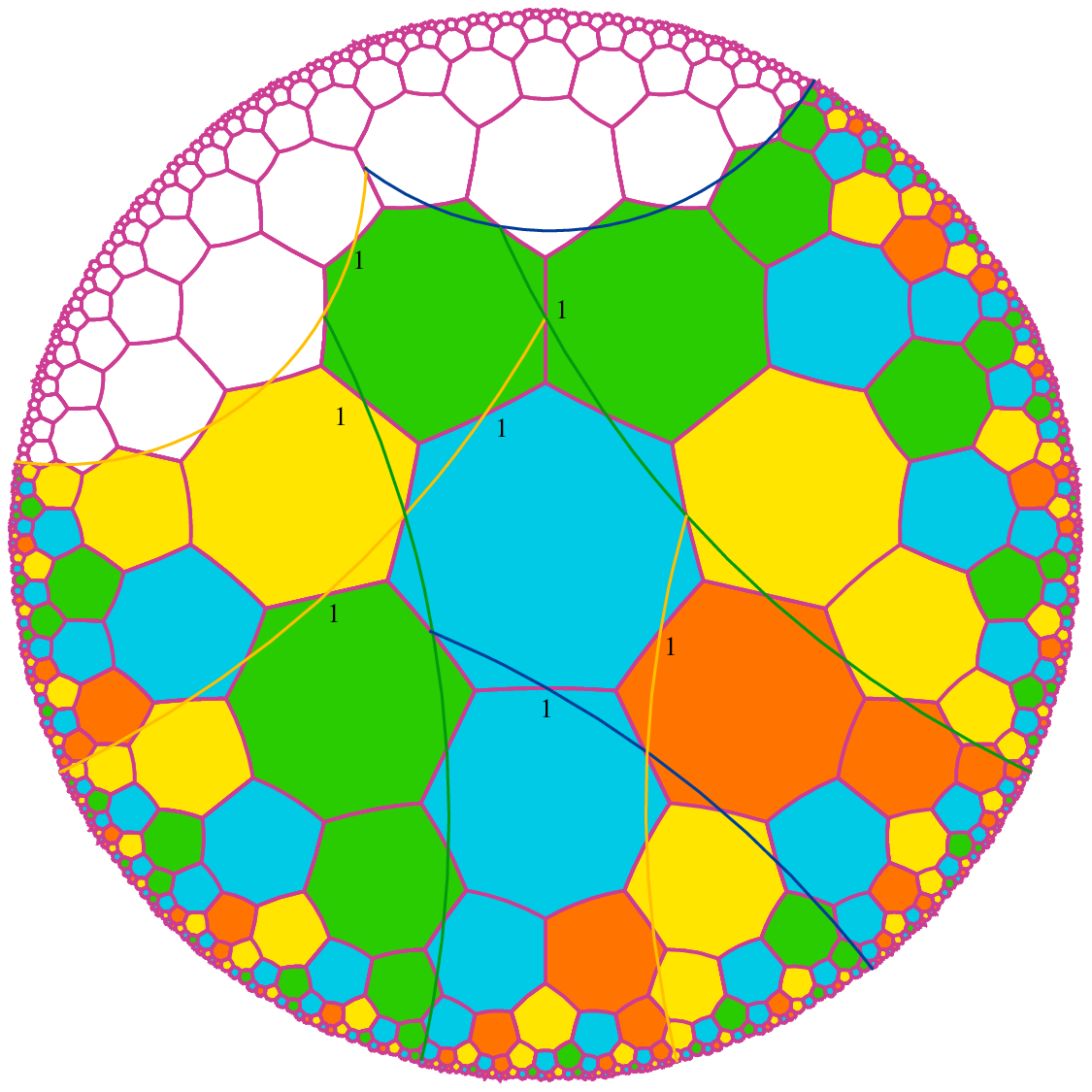}
\hfill}
\begin{fig}\label{f_tree}
\leurre
A tree~$\mathcal T$ of the heptagrid with two sub-trees of $\mathcal T$. One of them
is generated by the \GG-son of the \YY-son of the root of $\mathcal T$, the other
is generated by the \GG-son of the root of $\mathcal T$. Side~$1$ in each tile is defined according to
Table~$(N)$.
\end{fig}
}

We remind the reader that in $T(\tau)$, whatever the status of~$\tau$ which is assumed to
be not \BB, the number of tiles belonging to the level~$m$ of that tree is
$f_{2m+1}$, where $\{f_n\}_{n\in\mathbb N}$ is the Fibonacci sequence satisfying
\hbox{$f_0=f_1=1$}. If the tiles $\mu$ and $\nu$ belong to the same level of $T(\tau)$,
we call {\bf appartness} between $\mu$ and $\nu$ the number $n$ of tiles $\omega_i$ with
$i\in\{0..n\}$ such that those $\omega_i$ belong to the same level and such that 
$\omega_0=\mu$, $\omega_n=\nu$ and that for $i$ with \hbox{$0\leq i<n$}, we have that
$\omega_i$ and $\omega_{i+1}$ share a side. Accordingly, denoting the appartness between
$\mu$ and $\nu$ by $appart(\mu,\nu)$, we get that if those tiles belong to the level~$m$,
then \hbox{$appart(\mu,\nu)\leq f_{2m+1}$}. From Proposition~\ref{psubtree_levels}, it is
plain that the definition of the appartness of two tiles does not depend on the tree of
the heptagrid to which they belong, provided that then, they belong to the same tree.

Let us closer look at the two implementations of Construction~\ref{cons_til} illustrated
by Figure~\ref{f_hepta_tils}. Fix a \GG-tile~$\tau$ of the heptagrid. Fix a mid-point~$A$
of a side of another tile sharing a vertex~$V$ only with~$\tau$. From~$A$, draw two rays
issued from~$A$, one of them~$u$ passes through the mid-point of one of the sides 
of~$\tau$ sharing~$V$ while~$v$ passes through the mid-point of the other side of~$\tau$
sharing~$V$. The ray~$u$ and~$v$ allow us to define a sector of the heptagrid 
pointed by~$A$. Applying the rule~$(R_0)$ to~$\tau$, to its sons and, recursively to the
sons of its sons, we define a tree of the heptagrid. Let $v$ be the ray issued from~$A$
which also crosses the \GG-son of~$\tau$. From the rules $(R_0)$ and from 
Figure~\ref{f_proto_0}, it is not difficult to establish that in $T(\tau)$, $v$ crosses
only \GG-tiles. We can notice that time $m$+1 of Construction~\ref{cons_til}, 
gives us two possibilities only to define the father of a root: indeed, such a father 
cannot be neither a \BB-tile nor an \OO-one, as far as \GG-tiles are never sons 
of either a \BB-tile or an \OO-one. Figure~\ref{f_inc_conf} shows us what may happen
when a father~$\varphi$ is appended to a \YY- or a \GG-tile which is the central tile~$\kappa$ in the 
pictures of the figure, the stress being put on the corresponding trees of the heptagrid
generated in that way. In the leftmost column of the figure, two pictures illustrate 
what happens if $\varphi$ has the same status as~$\kappa$. In the central column $\varphi$ has the other 
status with respect to~$\kappa$. In the rightmost column a new 
father~$\psi$ is appended to~$\varphi$ with, again, a change in the status. Note that if $\psi$ would have
the same status as~$\varphi$, the obtained tree would be the same but later appendings could change the 
situation. Let us call the situation illustrated by the rightmost column an {\bf alternation}. We 
distinguish \YY\GG\YY- and \GG\YY\GG-alternations where the symbols are self-explaining. We also note 
that in an alternation, the rays defining the tree rooted at~$\psi$ are both non secant with respect to 
the rays defined by $\kappa$, so that \hbox{$T(\kappa)\subset T(\psi)$}.

\vskip 10pt
\vtop{
\ligne{\hfill
\includegraphics[scale=0.3]{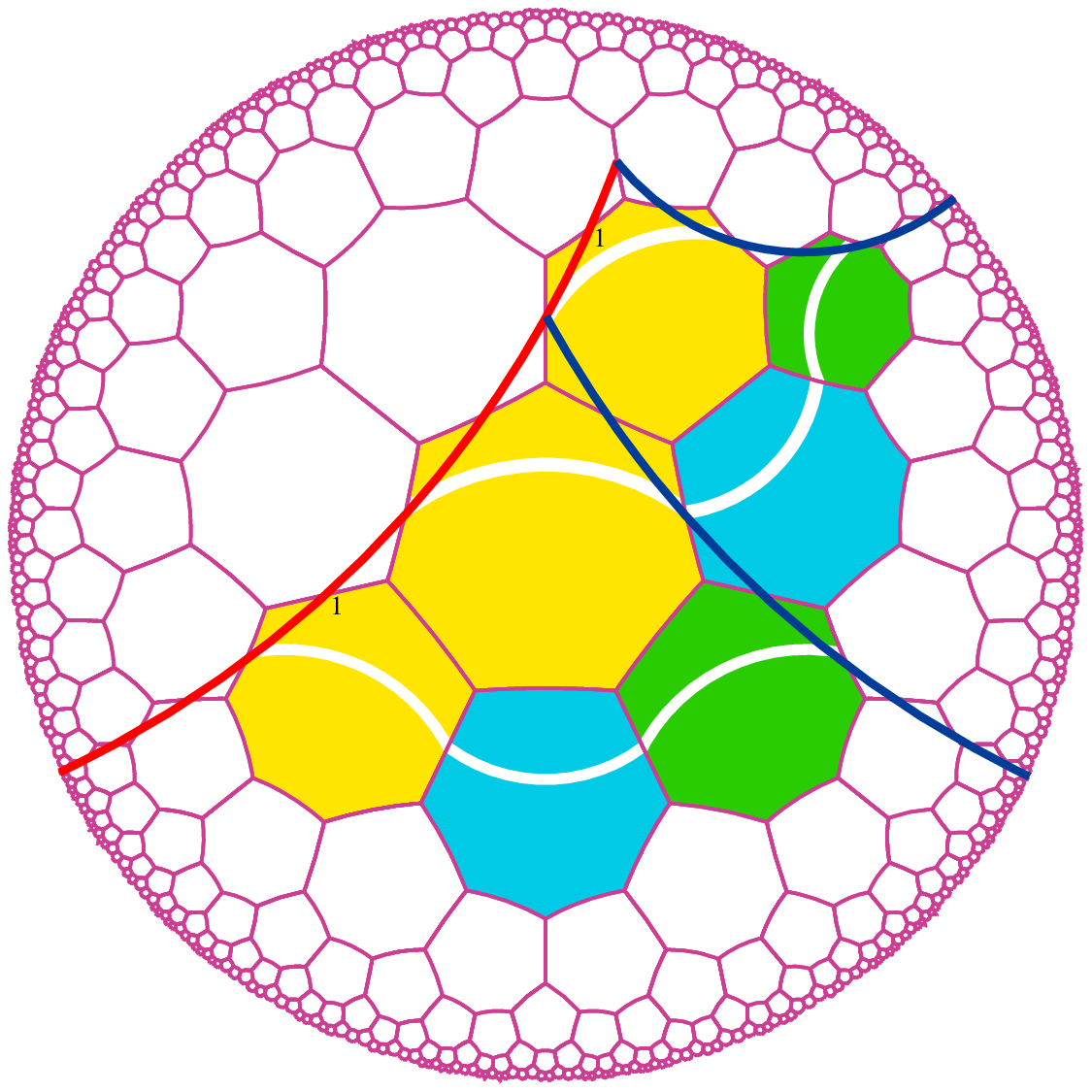}
\includegraphics[scale=0.3]{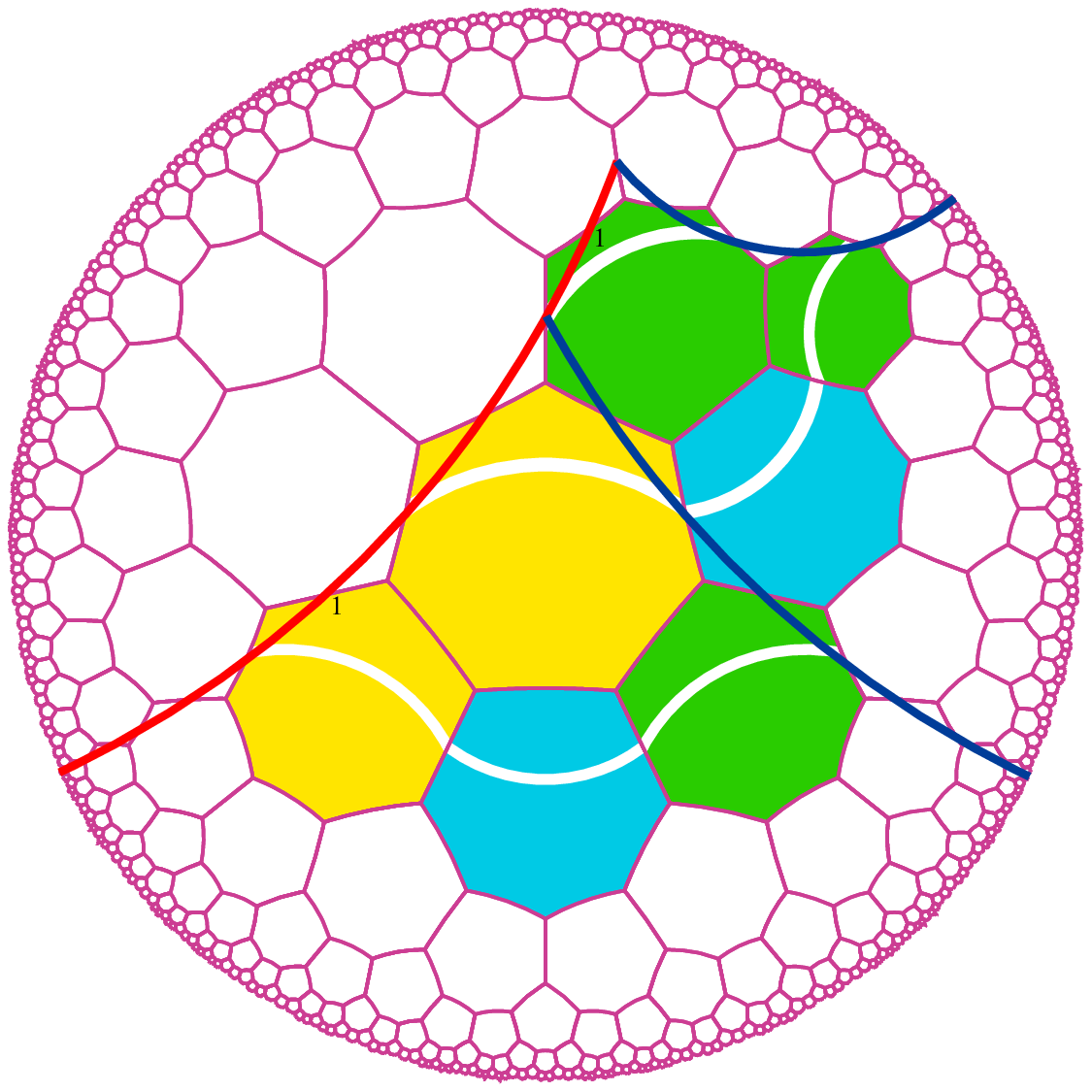}
\includegraphics[scale=0.3]{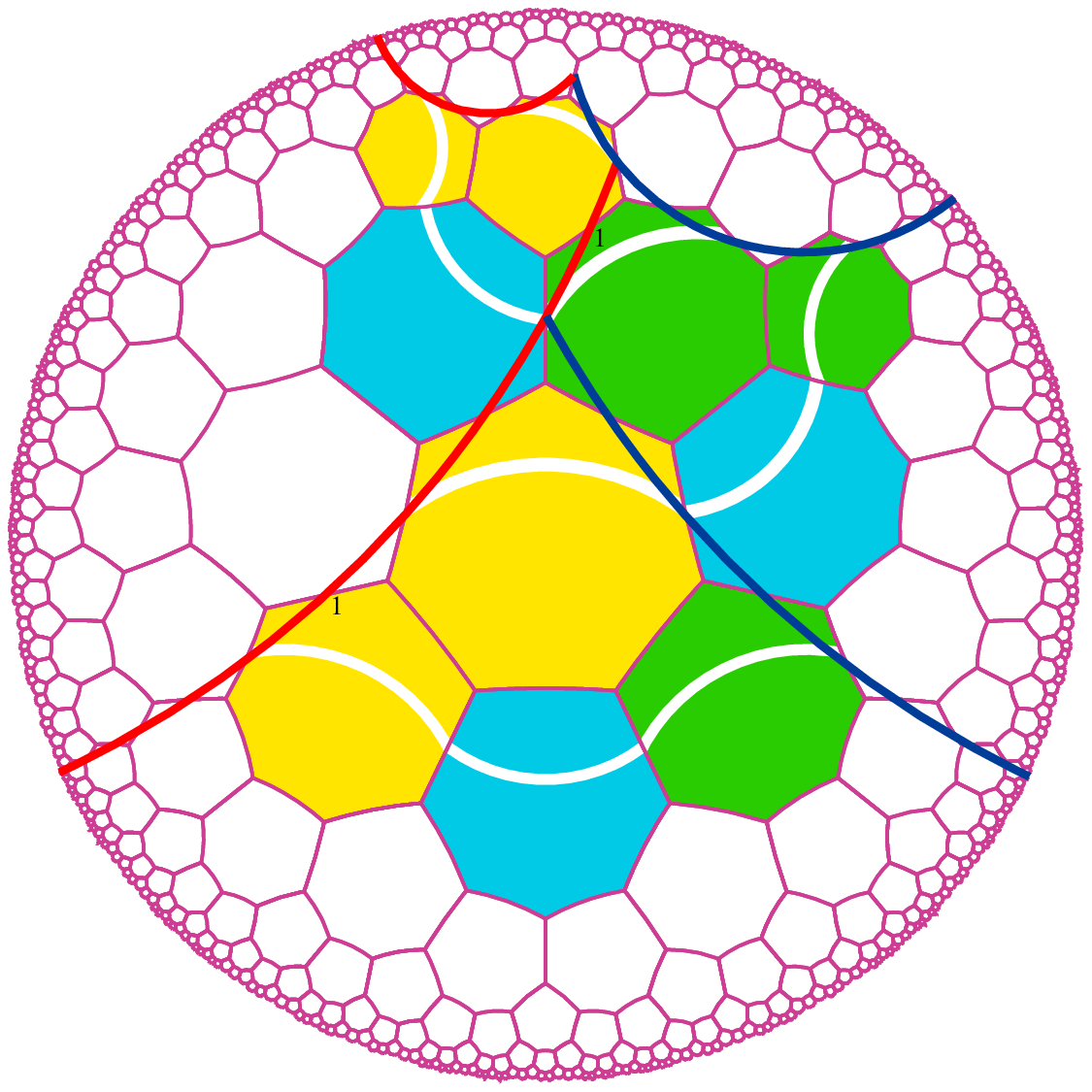}
\hfill}
\ligne{\hfill
\includegraphics[scale=0.3]{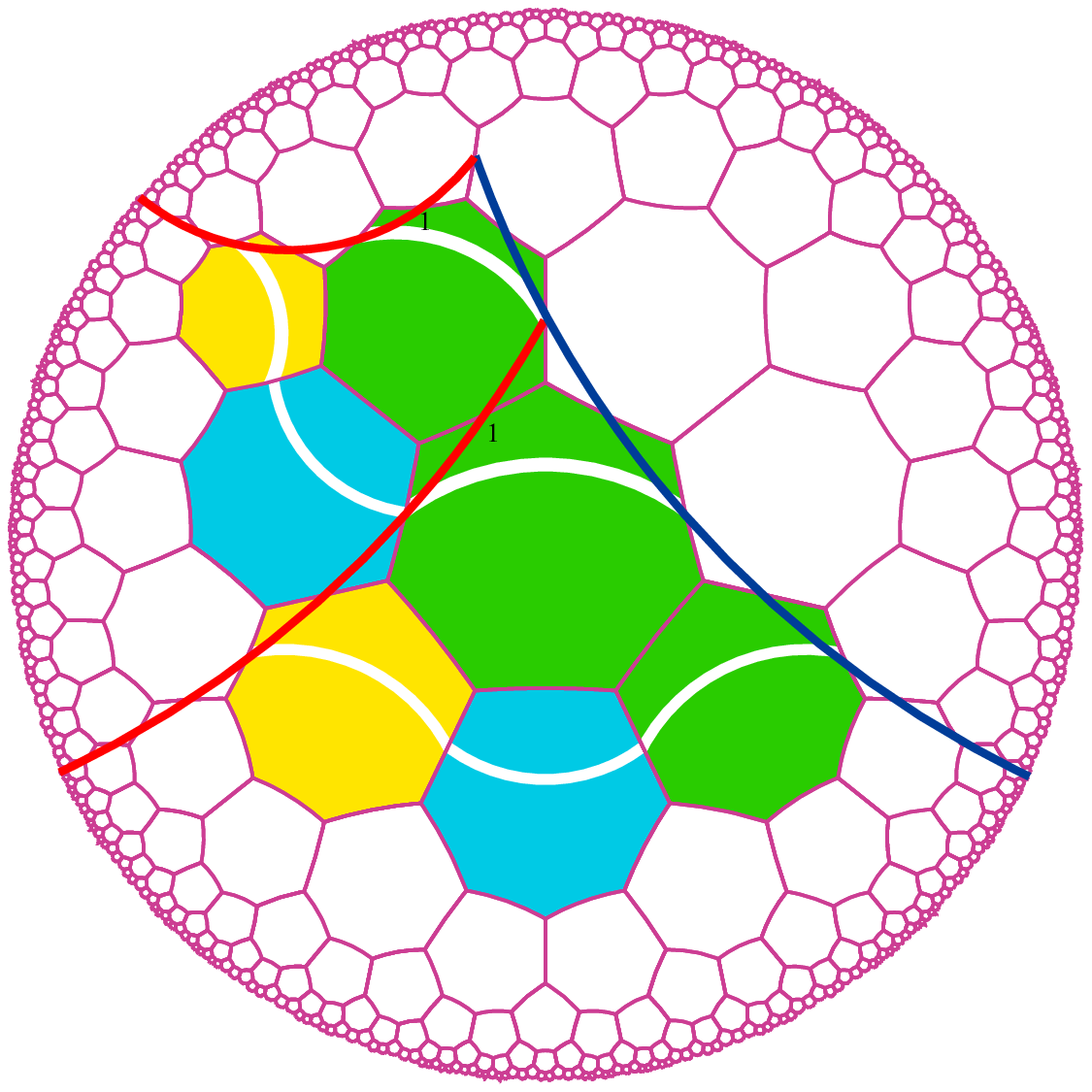}
\includegraphics[scale=0.3]{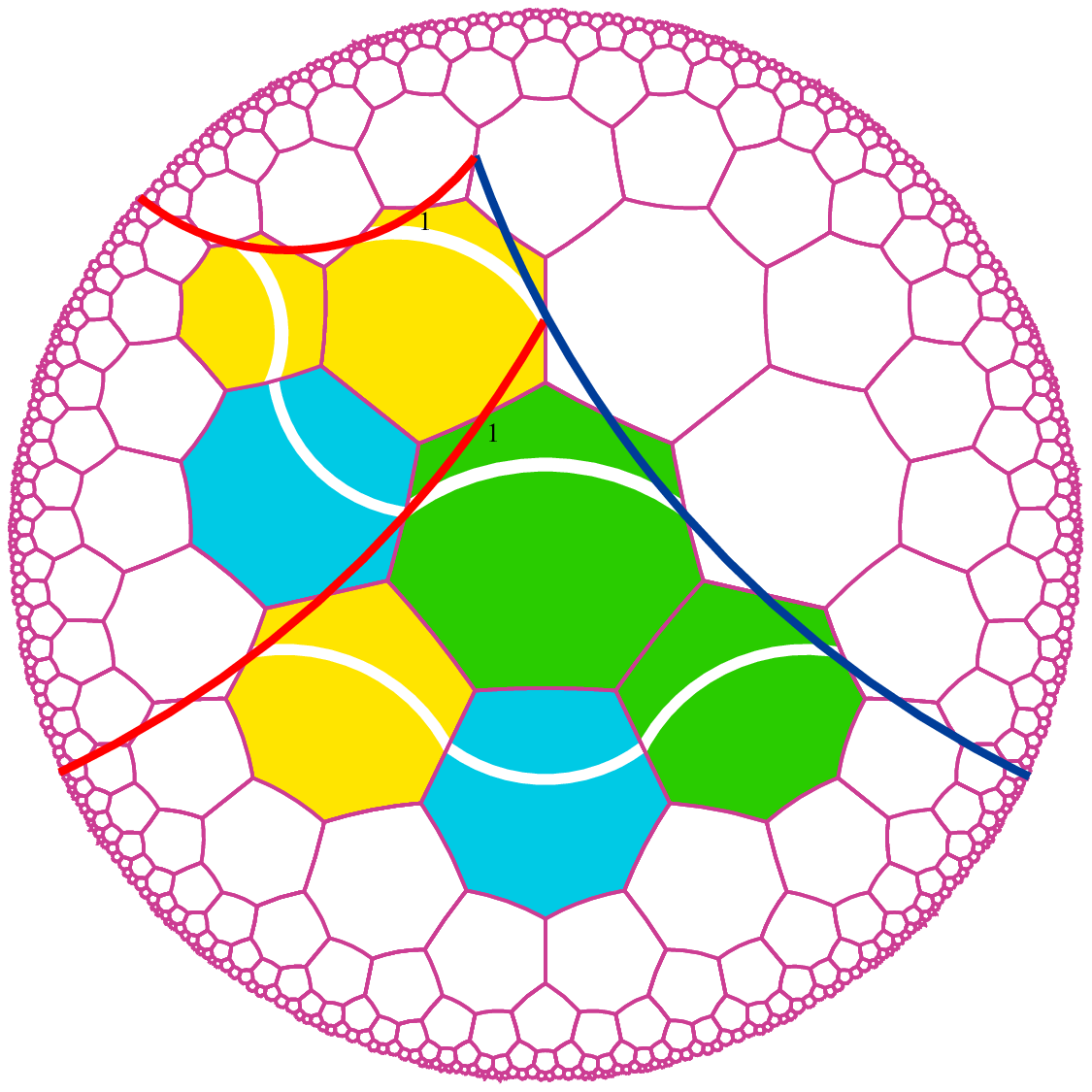}
\includegraphics[scale=0.3]{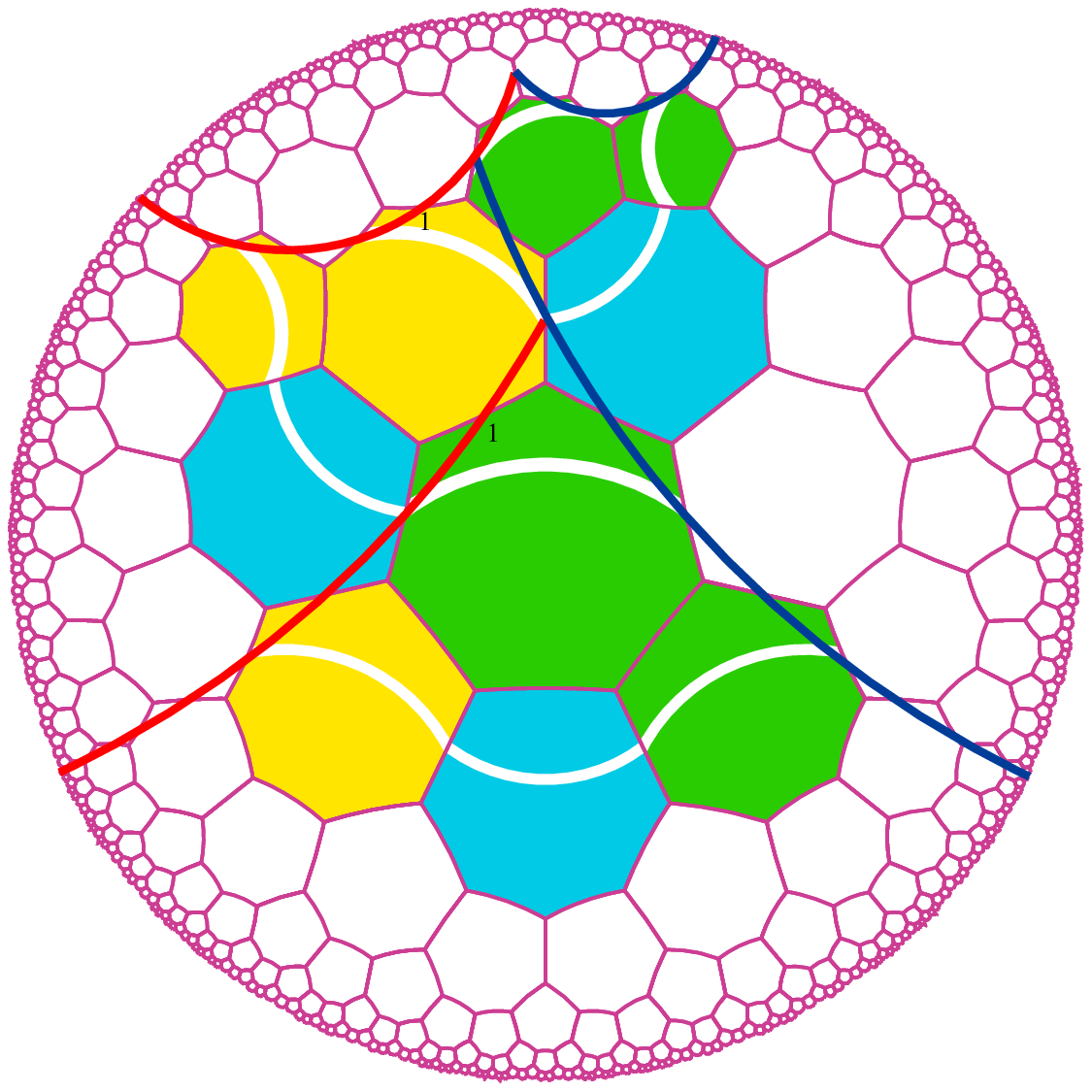}
\hfill}
\begin{fig}\label{f_inc_conf}
\leurre
Appending a father to the root \YY{} or \GG{} of a tree of the heptagrid: first line,
the root is a \YY-tile; second line: it is a \GG-tile. The rightmost column illustrates
the possible alternations.
\end{fig}
}

From those observations, we conclude that there are two basic situations: either,
starting from a time $k$, the father appended at time~$k$+$i$ has always the 
same status as the root at time~$k$ or, there are infinitely many times $i_j$, with
$i_j>k$, such that the situation at times~$k$+$i_j$, $k$+$i_j$+1 and $k$+$i_j$+2 is an 
alternation.

Consider the first case. There are two sub-cases: starting from a time~$k$ the appended
father is always a \GG-, \YY-tile, we call that sub-case the \GG, \YY{\bf -ultimate 
configuration} respectively. 

First, let us study the case of a \GG-ultimate configuration. Accordingly, we may 
infer that the ray issued from the origin of~$T(\tau_k)$ is supported by a line~$\ell$ 
which also supports the right-hand side ray delimiting $T(\tau_{k+i})$ for 
$i\in\mathbb N$. Let $A_j$ be the origin of $T(\tau_j)$.  
It is not difficult to see that for $i\in\mathbb N$, $T(\tau_{k+i+1})$ is the image of
$(\tau_{k+i})$ under the shift along~$\ell$ which transforms $A_k$ into $A_{k+1}$.  
Accordingly, the left-hand side $u_{i+1}$ ray defining $T(\tau_{k+i+1})$ and the 
left-hand side ray $u_i$ defining $T(\tau_{k+i})$ are non-secant. So that the rays $u_i$
define a partition of the half-plane $\pi_\ell$ defined by~$\ell$ which 
contains $T(\tau_k)$. As a consequence, any tile $\mu$ belonging to $\pi_\ell$ falls 
within one of the $T(\tau_{m+i})$'s for some $i$. From that, we infer that there is a 
level~$\lambda$ of $T(\tau_{m+i})$ which contains $\mu$. 

What happens on the other half-plane~$\pi_r$ also defined by~$\ell$? It is not
difficult to see, as shown by the central and the left-hand side pictures of
Figure~\ref{f_hepta_tils}, that $\ell$ also crosses a sequence of consecutive \BB-cells
which lie in $\pi_r$. Let $\beta_i$ be the \BB-son of the \GG-tile $\tau_{k+i}$ for
$i\in\mathbb N$ and let $\omega_i$ be the \OO-son of $\beta_i$. Then, if $u_i$ and $v_i$ 
are the left- and right-hand side ray respectively defining $T(\omega_i)$, it can be 
seen that \hbox{$u_{i+1}=v_i$} for all $i$, $i\geq k$. Indeed, the shift which transforms 
$\tau_{k+i}$ into $\tau_{k+i+1}$ also transforms $\beta_i$ into $\beta_{i+1}$
and, accordingly, it also transforms the line supporting $u_i$ into that supporting 
$u_{i+1}$ which is $v_i$. From that property, it easily follows that 
\hbox{$T(\tau_{mi+i+1})\cap T(\tau_{m+i})=\emptyset$} when $i\geq k$. Accordingly,
any tile $\mu$ of $\pi_r$ which is not a \BB-tile having two sides crossed by $\ell$
falls within $T(\tau_k)$ or within $T(\tau_{k+i})$ for some~$i$ in $\mathbb N$.

Consider the second sub-case when, starting from some~$k$ all $\tau_{k+i}$ is a \YY-tile 
for all~$i$, $i\in\mathbb N$. Call that case the {\bf \YY-ultimate configuration}. This 
time all those tiles are crossed by the line~$\ell$ which supports the left-hand side ray 
defining $T(\tau_k)$. Similarly, a shift along $\ell$
transforms each $T(\tau_{k+i})$ into $T(\tau_{k+i+1})$. Let $u_i$ be the left-hand side 
ray defining $T(\tau_{k+i})$. The same shift also transforms $u_i$ into $u_{i+1}$ so
that those rays define a partition of the half-plane $\pi_1$ defined by $\ell$ which 
contains the $\tau_{k+i}$'s. Accordingly, for any tile~$\mu$ in $\pi_1$, there is
an integer~$i$ in $\mathbb N$ such that $\mu\in T(\tau_{k+i})$. 

Let $\pi_2$ be the other half-plane defined by~$\ell$.  From Figure~\ref{f_proto_0}, we 
know that $\ell$ also crosses a sequence of \OO-tiles: on the level of~$\tau_{k+i+1}$, 
there is an \OO-tile which shares a side with that node and whose \OO-son shares a side 
with $\tau_{k+i}$, being on the other side of~$\ell$ with respect to $\tau_{k+i}$. Let 
$\omega_{k+i}$ be the \OO-tile sharing a side with both $\tau_{k+i+1}$ and $\tau_{k+i}$.
From what we said, the $\omega_{k+i}$'s belong to $\pi_2$. 
The same shift as that which transforms $T(\tau_{k+i})$ into $T(\tau_{k+i+1})$ transforms
$T(\omega_{k+i})$ into $T(\omega_{k+i+1})$. We can note that the $T(\tau_{k+i})$ have their
right-hand side ray supported by~$\ell$ and the left-hand side ray $u_{k+i}$ defining 
that tree is such that $u_{k+i+1}$ is the image of $u_{k+i}$ under the shift which 
transforms $T(\tau_{k+i})$ into $T(\tau{k+i+1})$. The $u_{k+i}$'s define a partition of
$\pi_2$ so that each tile~$\nu$ in $\pi_2$ belongs to some $T(\tau_{k+i})$. In fact, as
far as \hbox{$T(\tau_{k+i})\subset T(\tau_{k+i+1}$}, once $\nu$ is in $T(\tau_{k+i})$
it is also in all $T(\tau_{k+j})$ with $j>i$.

\def\reunion{\mathop{\cup}}
Consider the second case. In each $\tau_i$ defined in Construction~\ref{cons_til},
we define a segment joining the centre of that tile to the center of~$\tau_{i+1}$.
The union of those segments constitute an infinite broken line which splits the complement
$\mathcal C$ in the hyperbolic plane of the points contained in the tiles of 
$T(\tau_0)$ into two parts. In the second case, we know that there is a sequence 
$\{i_j\}_{p\in\mathbb N}$ of times, at which we have an alternation. Whether the 
alternation are is \GG\YY\GG, or \YY\GG\YY, we noted that the left-, right-hand side rays
defined by $T(\tau_{i_j+2})$ are non-secant from the left-, right-hand side rays 
respectively defined by $T(\tau_{i_j})$. The consequence is that in both cases the 
rays defined at the alternations constitute a partition of~$\mathcal C$. Accordingly,
if $\cal T$ is the set of tiles of the heptagrid, then
\hbox{${\cal T}=\displaystyle{\reunion\limits_{j\in\mathbb N} {T(\tau_{i_j})}}$}.
So that if $\mu$ is a tile, it belongs to $T(\tau_0)$ or to $T(\tau_{i_j})$ for some~$j$.

   That allows us to prove:

\begin{lem}\label{absorb}
Let $\{\tau_i\}_{i\in\mathbb Z}$ be a sequence of tiles in a tiling constructed with the 
rules~$(R_0)$, such that for all $i$ in $\mathbb Z$ 
\hbox{$T(\tau_i)\subset T(\tau_{i+1})$}. If the sequence is in an alternating 
configuration then, for any tile~$\mu$ there is an index $i$ such that $\mu$ falls
within $T(\tau_i)$. If it is not the case:

If the sequence is in a \YY-ultimate configuration. Let $k$ be the smallest integer such 
that $\tau_i$ is a \YY-tile for all $i\geq k$. Then, let $\omega_i$ be the tile sharing
a side with $\tau_{i+1}$ and an other one with $\tau_i$. Then for any tile $\mu$ there
is either an index $i$ such that $\mu$ falls within $T(\tau_i)$ or there is an index $j$, 
$j\geq k$ such that $\mu$ falls within $T(\omega_j)$.

If the sequence is in a \GG-ultimate configuration. Let $k$ be the smallest integer such
that $\tau_i$ is a \GG-tile for all $i\geq k$. Let $\beta_i$ be the \BB-son of $\tau_i$
for $i\geq k$ and let $\omega_i$ be the \OO-son of $\beta_i$ for $i\geq k$ too. Then for
any tile $\mu$ which is not a $\beta_i$ with $i\geq k$, there is an index $i$ such that 
$\mu$ falls within $T(\tau_i)$ or there is an index $j$, $j\geq k$, such that $\mu$ falls
within $T(\omega_j)$.
\end{lem}

Lemma~\ref{absorb} and Proposition~\ref{psubtree_levels} allow us to call {\bf isocline} 
the bi-infinite extension of any level of a tree $T(\tau_m)$, for any value of 
$m$ in $\mathbb N$. Note that Figure~\ref{f_proto_0} allows us to define isoclines
in a \GG-ultimate configuration by completing the levels for the exceptional \BB-tiles
indicated in the Lemma and by joining the levels of the $T(\omega_i)$ with the
corresponding $T(\tau_i)$.

Note that the pictures of Figure~\ref{f_hepta_tils} also represent the tilings with 
their isoclines.

Let $T$ be a tree of the heptagrid, let $\rho$ be its root and let $\tau$ be a tile
in $T$. Say that a path $\pi=\{\pi_i\}_{i\in\{0..n\}}$ joining $\rho$ to~$\tau$ is
a {\bf tree path} if and only if for each non negative integer $i$, with $i<n$, $\pi_{i+1}$
is a son of $\pi_i$. It is not difficult to see that, in that case, $n$ is the level of 
$\tau$, which can easily be proved by induction on~$n$.

A {\bf branch} in a tree~$T$ of the heptagrid is an infinite sequence 
$\beta=\{\beta_i\}_{i\in\mathbb N}$ such that for each $i$ in $\mathbb N$, $\beta_{i+1}$
is a son of $\beta_i$. Accordingly, a tree path in $T$ is a path from the root of $T$ to
a tile~$\tau$ in $T$ which is contained in the branch of~$T$ which passes through $\tau$.

We can state:

\dprop{psepar_char}
Let $T$ be a tree of the heptagrid. Let $\mu$ and $\nu$ be tiles in $T$. Let $\pi_\mu$,
$\pi_\nu$ be the tree path from the root $\rho$ of $T$ to $\mu$, $\nu$ respectively.
Then, $T(\mu)\subset T(\nu)$ if and only if $\pi_\mu\subset \pi_\nu$ and
$T(\mu)\cap T(\nu) = \emptyset$ is and only if we have both
$\pi_\mu\not\subset \pi_\nu$ and $\pi_\nu\not\subset\pi_\nu$.
\fprop

\noindent
Proof. Assume that $\pi_\mu\subset\pi_\nu$. As far as the rules $(R_0)$ applied to 
$T(\rho)$ have the same result when they are applied in $T(\nu)$ and in $T(\mu)$,
we have that \hbox{$T(\mu)\subset T(\nu)\subset T(\rho)$} and, clearly,from
Proposition~\ref{psubtree_levels} we get that \hbox{$\pi_\mu\subset\pi_\nu$}.

Presently, consider $\mu$ and $\nu$ and let $\pi_\mu$, $\pi_\nu$ be the tree paths from
$\rho$ to $\mu$, $\nu$ respectively. Consider $\pi_\mu\cap\pi_\nu$. There is a 
tile~$\beta$ belonging to $\pi_\mu$ and $\pi_nu$ such that $\pi_\beta$, the tree path from
$\rho$ to~$\beta$ satisfies $\pi_\beta=\pi_\mu\cap\pi_\nu$. Clearly, $\pi_\beta$ is the
greatest common path of $\pi_\mu\cap\pi_\nu$. Consider the case when $\beta=\nu$.
It means that $\pi_\mu\subset\pi_\nu$. From what we already proved, we have that 
$T(\mu)\subset T(\nu)$.

Consider the case when $\pi_\mu\not\subset\pi_\nu$. In that case, the length of 
$\pi_\beta$ is shorter from both that of $\pi_\mu$ and that of $\pi\nu$. It means that
one son of $\pi_\beta$, say $\beta_\mu$, belongs to $\pi_\mu$ and the other, say 
$\beta_\nu$, belongs to $\pi_\nu$ and $\beta_\mu$, $\beta_\nu$ does not belong to
$\pi_nu$, $\pi_\mu$ respectively by definition of $\beta$. 
Now, $\mu$ and $\nu$ are non \BB-tiles. 

If $\beta$ is a non \BB-tile. We have three cases as far as $\mu$ and $\nu$ can be
exchanged if needed:
\vskip 0pt
$(i)$ $\beta_\mu$ is the \YY-son and $\beta_\nu$ is the \GG- or \OO-son;
\vskip 0pt
$(ii)$ $\beta_\mu$ is the \YY-son and $\beta_\nu$ is the \BB-son;
\vskip 0pt
$(iii)$ $\beta_\mu$ is the \BB-son and $\beta_\nu$ is the \GG- or \OO-son.
\vskip 0pt

In the case $(i)$, Figure~\ref{f_tree} shows us that the right-hand side ray~$u$ of 
$T(\beta_\mu)$ and the left-hand side ray~$v$ of $T(\beta_\nu)$ have a common 
perpendicular which is the line containing the side of the \BB-son of $\beta$ share 
with $\beta$. So that $T(\beta_\mu)\cap T(\beta_\nu)=\emptyset$.

In the case $(ii)$, as far as $\nu$ is a non \BB-tile, there is a tile $\gamma$ in
$\pi_\nu$ which is a descendent of $\beta$ which is the first non \BB-tile on $\pi_\nu$
in between $\beta_\nu$ and $\nu$. It may happen that $\gamma=\nu$. In that case, all tiles
on $\pi_\nu$ after $\beta$ and until $\gamma$ are \BB-tiles. As can be seen from
Figure~\ref{f_tree}, those \BB-tiles are crossed by~$u$. We have that $\gamma$ is an 
\OO-tile and if $w$ is the left-hand side ray of $T(\gamma)$, $u$ and $w$ have a common 
perpendicular as can be seen on Figure~\ref{f_tree} for the \OO-son of the \BB-son
of the tree which is there represented. Accordingly, 
$T(\beta_\mu)\cap T(\gamma)=\emptyset$. As far as $T(\nu)\subset T(\gamma)$ we again
get that $T(\mu)\cap T(\nu)=\emptyset$.

In the case $(iii)$, we can argue in a similar way. This time, let $y$ be the left-hand 
side ray of $T(\beta_\nu)$. If $\gamma$ is the \OO-son of $\beta_\mu$, Figure~\ref{f_tree}
shows us that $T(\gamma)\cap T(\beta_\nu)=\emptyset$ as far as $y$ is the right-hand
side ray of $T(\gamma)$. Now, if $\pi_\mu$ does not pass through $\gamma$ it passes
outside the left-hand side ray $z$ of $T(\gamma)$. Accordingly, 
$T(\mu)\cap T(\beta_\nu)=\emptyset$, so that, all the more, 
\hbox{$T(\mu)\cap T(\nu)=\emptyset$}.

Assume that $T(\mu)\cap T(\nu)=\emptyset$. Clearly, $\pi_\mu\not\subset\pi_\nu$ 
and also $\pi_\nu\not\subset\pi_\mu$. So that we have the situation depicted with
$\pi_\beta=\pi\mu\cap\pi_\nu$ and both $\pi_\beta\not=\pi\mu$ with 
$\pi_\beta\not=\pi_\nu$. Consequently, if both $\pi_\nu\not\subset\pi_\mu$ and
$\pi_\mu\not\subset\pi_\nu$ do not hold then necessarily $\pi_\nu\subset\pi_\mu$
or $\pi_\mu\subset\pi_\nu$ so that $T(\mu)\subset T(\nu)$ or $T(\nu)\subset T(\mu)$
holds.\hfill$\Box$

We have an important property:

\begin{lem}\label{separ_trees}
Two distinct trees of the heptagrid are either disjoint or one of them contains the other.
\end{lem}

The lemma is an immediate corollary of Proposition~\ref{psepar_char}. Moreover, from
the proof of that proposition, we clearly get the following result:

\dprop {ptreepath}
Let $T(\tau)$ be a tree of the heptagrid. Let $T(\mu)$ be another tree of the heptagrid
with $\mu$ within $T(\tau)$. Let $\pi_\mu$ be a tree path from the root of $T(\tau)$ to
$\mu$. Then $\pi_\mu$ contains at least one tile $\nu$ which is not a \BB-tile.
Moreover, for any non \BB-tile $\omega$ in $\pi_\mu$, we have
\hbox{$T(\omega)\subset T(\tau)$}.
\fprop

\def\ww{{\bf w}}
\def\bb{{\bf b}}
   Note that in Figure~\ref{f_proto_0}, the curves representing the isoclines are
constituted by two kinds of segments defined as follows. Those segments join the 
mid-points of two different sides of a tile: one kind, denoted by \ww, is defined by 
joining two sides which are separated by one side, namely joining side~2 and side~7; the other kind, 
denoted by \bb, is defined by joining two sides which are separated by two contiguous sides, namely joining
side~2 and side~6 or joining side~3 and side~7. Call these marks on a tile its {\bf level mark}. The 
distribution of the level marks obeys the following rules:
\vskip 5pt
\ligne{\hfill \ww $\rightarrow$ \bb\ww\ww\hskip 30pt \bb $\rightarrow$ \bb\ww,\hfill$(S)$
\hskip 10pt}
\vskip 10pt
\noindent

\begin{lem}\label{levels}
   It is not difficult to tile the heptagrid with the prototiles \YY, \GG, \BB{} and \OO{}
by applying the rules~$(R_0)$ so that the rules of~$(S)$ also apply if we put \ww{} marks
on \BB-, and \OO-tiles only and \bb{} marks on \YY- and \GG-tiles only.
\end{lem}

\noindent
Proof. Inside a tree of the heptagrid, the result follows by induction on the levels.
If we consider two trees of the heptagrid where one of them contains the other, the
result follows from Proposition~\ref{psubtree_levels} which tells us that the levels
of a sub-tree in a tree of the heptagrid are contained in levels of the tree. From
Lemma~\ref{absorb}, it is possible to construct a sequence of trees of the heptagrid
$\{T(\tau_i)\}_{i\in \mathbb N}$ such that 
$\displaystyle{\reunion\limits_{i\in\mathbb N} T(\tau_i)}$ covers the whole hyperbolic
plane, so that the lemma follows.
\hfill$\Box$

   Note that in \ww-tiles, sides~2 and~7 are joined by the mark while in \YY-tiles it is the cas for 
sides~3and~7 while in \GG-tiles it is the case for sides~2 and~6.

   Construction~\ref{cons_til} allow us to tile the whole hyperbolic plane 
in infinitely many ways. The number of such tilings is uncountable as far as at each time
we have a choice between two possibilities and that the number of steps is infinite.

   It can be argued that any construction of a tiling which, by definition, starts with
any tile, is in some sense described by Construction~\ref{cons_til}. Indeed, whatever
the starting tile, we find at some point a \GG-tile as far as in a tiling, there is
a \GG-tile at a distance at most 3 of any tile~$\mu$. That distance can be observed for 
a \BB-tile: its \OO-son has a \YY-son which to its turn has a \GG-son.


\subsubsection{The trees of the tiling}\label{sss_treetil}

   From now on, we introduce two new colours for the tiles, mauve and red which we denote
by \MM{} and \RR{} respectively. We decide that \MM-tiles duplicate the \BB-tiles when 
they are sons of a \GG-tile and only in that case and that an \RR-tile duplicates 
the \OO-son of an \MM-tile, so that the rules $(R_0)$ are replaced by the following ones:

\vskip 5pt
\ligne{\hfill
$\vcenter{\vtop{\leftskip 0pt\parindent 0pt\hsize 280pt
\ligne{\hfill
\GG{} $\rightarrow$ \YY\MM\GG,\hskip 20pt
\BB{} $\rightarrow$ \BB\OO,\hskip 20pt
\YY{} $\rightarrow$ \YY\BB\GG,\hskip 20pt
\OO{} $\rightarrow$ \YY\BB\OO
\hfill}
\ligne{\hfill
\RR{} $\rightarrow$ \YY\BB\OO,\hskip 20pt
\MM{} $\rightarrow$ \BB\RR,
\hfill}
}}$
\hfill$(R_1)$\hskip 10pt}
\vskip 5pt
   As previously, the status of a tile is, by definition, its colour.

   Figure~\ref{f_proto} illustrates the application of the rules~$(R_1)$ by giving
what they induce for the neighbours of a tile given its status and the status of its
father. Alike what is done in Figure~\ref{f_proto_0}, we also define levels by using
the rules~$(S)$. As far as a \BB-tile or an \OO{} is \ww, \MM- and \RR-tiles are also
\ww. As in Figure~\ref{f_proto_0}, the number of central tiles associated to a same
status is the number of occurrences of the status in the right-hand side parts of the
rules~$(R_1)$.
 
   We call {\bf tree of the tiling} any $T(\nu)$ where $\nu$ is an \RR-tile. We repeat 
that a tree of the tiling is a set of tiles, not the set of points contained in those
tiles. We also here indicate that, as far as \MM-, \RR-tiles behave like \BB-, \OO-tiles
respectively, we later refer to \BB-, \OO-tiles only unless the specificity of \MM-, 
\RR-tiles is required.

   From Lemma~\ref{separ_trees} we can state:

\begin{lem}\label{par_trees}
Let ${\mathcal T}_1$ and ${\mathcal T}_2$ be two trees of the tiling. Either those trees
are disjoint or one of them contains the other. Moreover, a ray which delimits one of 
those trees does not intersect any of the rays delimiting the other tree. The same also
applies to the rightmost and the leftmost branches of those trees.
\end{lem}

\ligne{\hfill
\vtop{\leftskip 0pt\parindent 0pt\hsize=300pt
\ligne{\hfill
\includegraphics[scale=0.5]{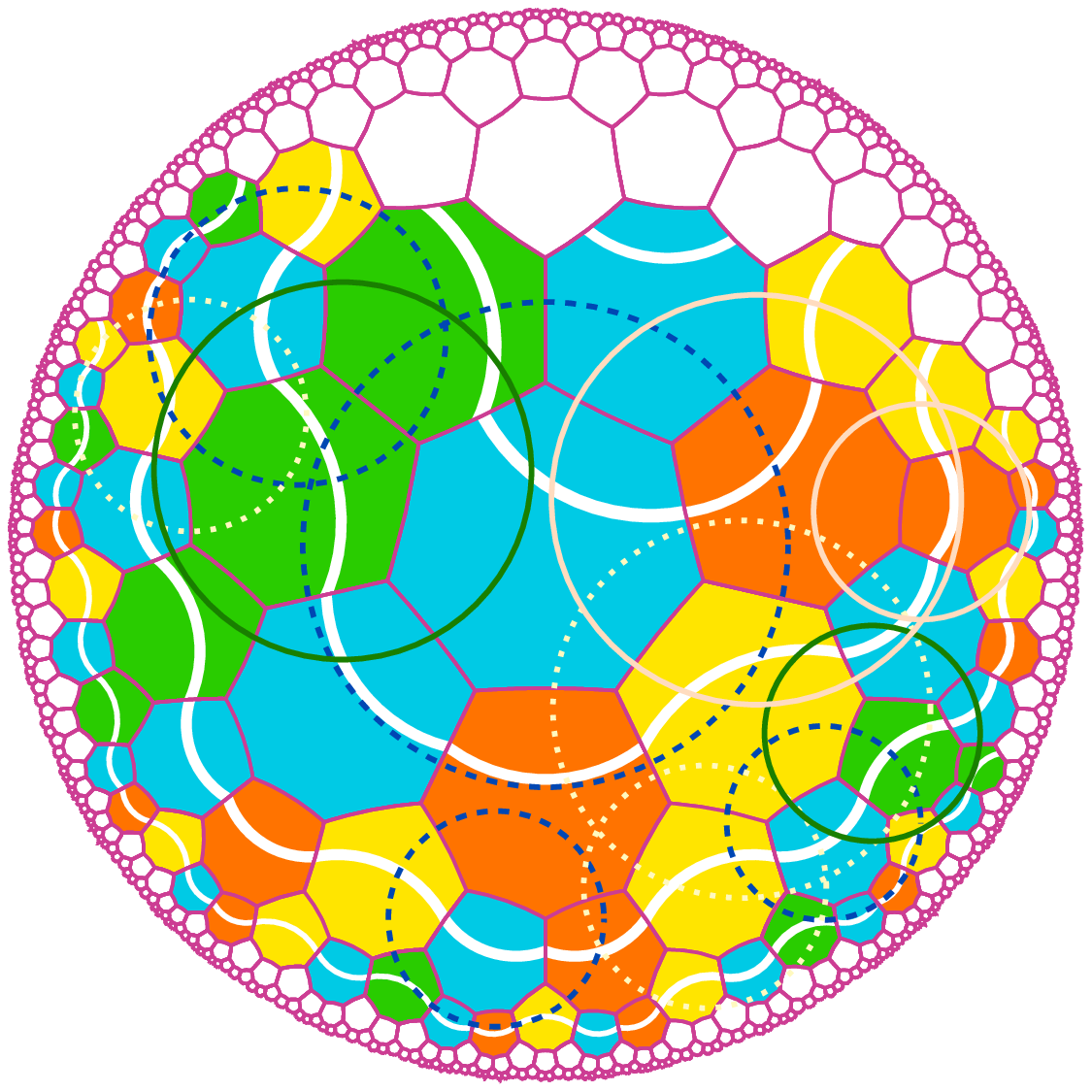}
\hfill}
\begin{fig}\label{f_neigh}
\leurre
Figure for proving the Figures~\ref{f_proto_0} and \ref{f_proto}. The circles crosses the neighbours of the tiles which their centres. The circles have the colour of the tiles
containing their centre.
\end{fig}
}	
\hfill}
\vskip 10pt
\noindent
Proof. Immediate from the proof of Proposition~\ref{psepar_char} when the tree are 
disjoint: the rays do not intersect and the property follows for the borders
as far as they are inside the considered trees.
We have to look at the situation when \hbox{$T(\tau_1)\subset T(\tau_0)$}, where
$T(\tau_0)$ and $T(\tau_1)$ are two trees of the tiling. Consider the tree path $\pi$ in 
$T(\tau_0)$ joining $\tau_0$ to $\tau_1$.  When going on~$\pi$ from $\tau_0$ to $\tau_1$,
let $\nu$ be the last non \BB-tile of~$\pi$ different of $\tau_1$. From
Proposition~\ref{ptreepath}, we know that \hbox{$T(\tau_1)\subset T(\nu)$}. Let 
$u$ and~$v$ be the rays delimiting $T(\nu)$ and let $u_1$ and $v_1$ be those delimiting
$T(\tau_1)$. We repeat here the discussion of cases $(i)$ up to $(iii)$ in the proof of
Proposition~\ref{psepar_char}. We have seen there that the same observation about the
rays hold so that it extends to the borders as far as they are inside the
trees and as far as two rays cannot both cross a border. The proof of
Lemma~\ref{par_trees} is completed.\hfill$\Box$

    Figure~\ref{f_til_trees} illustrates Lemma~\ref{par_trees}. Note that the figure does
not mention all trees of the tiling which can be drawn within the limits of that figure.

    Let us go back to the process described by Construction~\ref{cons_til}. The process
leads us to introduce the following notion:

\vskip 10pt
\vtop{
\ligne{\hfill
\includegraphics[scale=0.35]{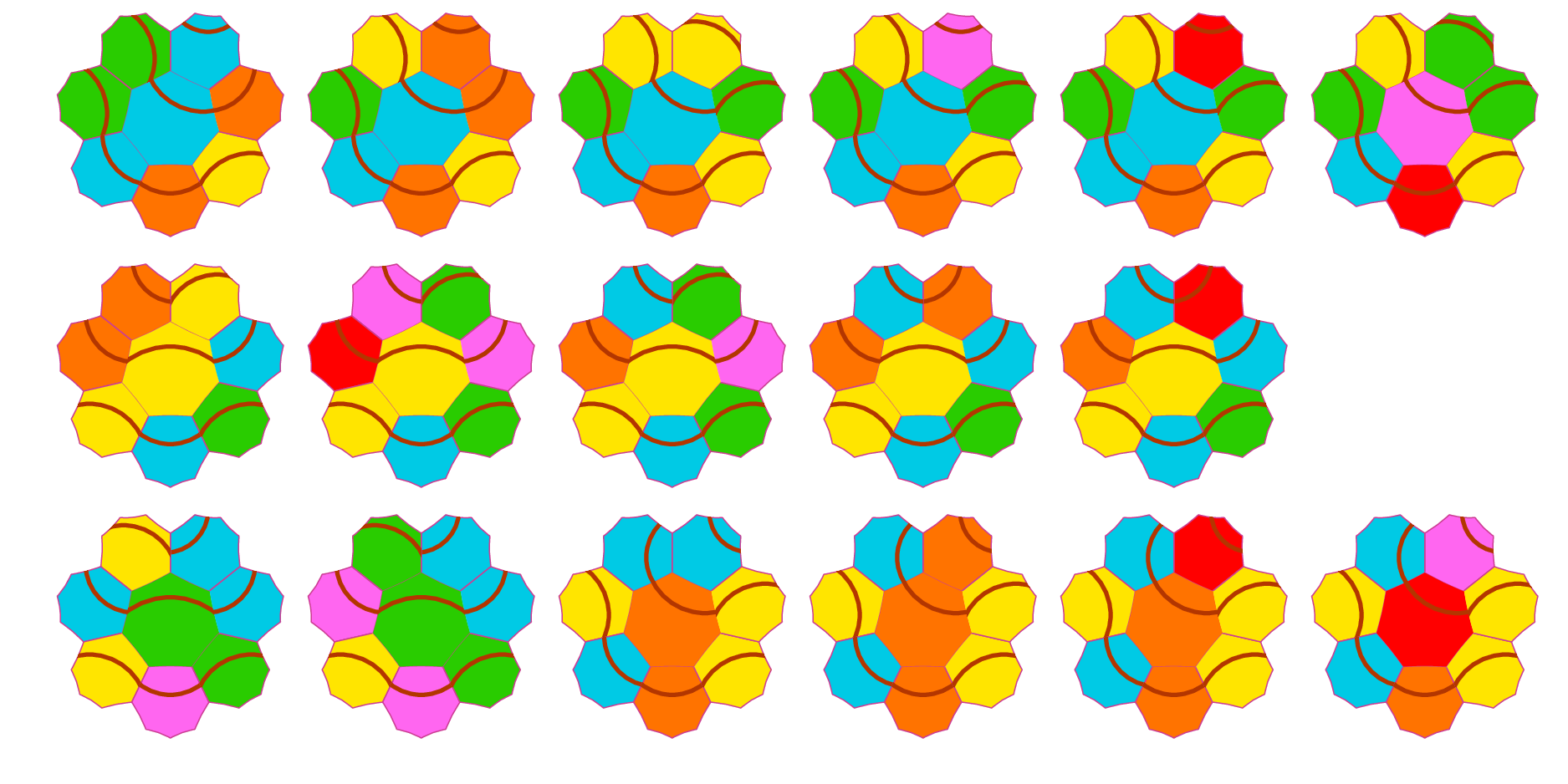}
\hfill}
\vspace{-10pt}
\begin{fig}\label{f_proto}
\leurre
The prototiles generating the tiling: we describe all cases for the neighbourhood of
a tile, whatever it is: \BB, \YY, \OO, \GG, \MM{} or \RR. The neighbourhoods around a 
tile of the same colour correspond to the different occurrences of that colour in the 
right-hand side part of rules~$(R_1)$. 
\end{fig}
}

\vskip 10pt
\vtop{
\ligne{\hfill
\includegraphics[scale=0.6]{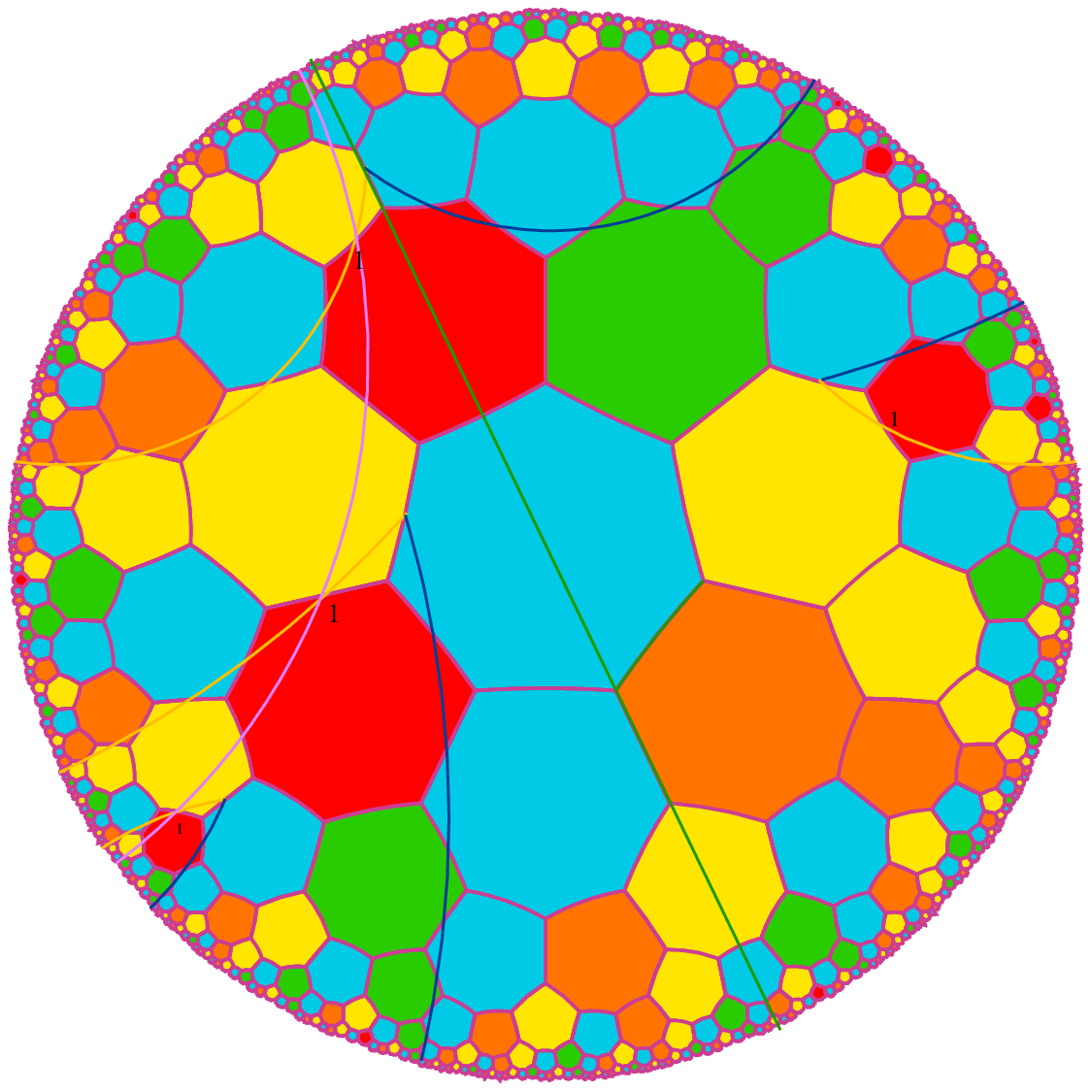}
\hfill}
\begin{fig}\label{f_til_trees}
\leurre
A tree~$\mathcal T$ of the tiling with three sub-trees of the tiling contained in
$\mathcal T$. One of them is contained in another one while two of those trees of the tiling inside $\mathcal T$ are disjoint.
\end{fig}
}

\begin{defn}\label{thread}
A {\bf thread} is a set $\cal F$ of trees of the tiling such that:
\vskip 2pt
{\leftskip 20pt\parindent 0pt
$(i)$ if $A_1,A_2\in{\cal F}$, then either $A_1\subset A_2$ or
$A_2\subset A_1$; 

$(ii)$ if $A\in{\cal F}$, then there is $B\in{\cal F}$ with $B\subset A$,
the inclusion being proper.

$(iii)$ if $A_1,A_2\in{\cal F}$ with $A_1\subset A_2$ and if $A$ is a tree of
the tiling with $A_1\subset A$, and $A\subset A_2$, then $A\in {\cal F}$. 
\par}
\end{defn}

   Said in words, a thread is a set of trees of the tiling on which the inclusion defines
a linear ordered, which has no smaller element and which contains all trees of the tiling
which belong to a segment of the set, according to the order defined by inclusion.

\begin{defn}\label{ultrathread}
A thread $\cal F$ of the tiling is called an {\bf ultra-thread} if it 
possesses the following additional property:
\vskip 2pt
{\leftskip 20pt\parindent 0pt
$(iv)$ there is no $A\in{\cal F}$ such that for all $B\in{\cal F}$,
$B\subset A$.
\par}
\end{defn}

\begin{lem}\label{ultrathread_defs}
A set $\cal F$ of trees of the tiling is an ultra-tread if and only if it
possesses properties~$(i)$, $(ii)$ and~$(iii)$ of definition~{\rm\ref{thread}} 
together with the following:
\vskip 2pt 
{\leftskip 20pt\parindent 0pt
$(v)$ for any $A\in{\cal F}$ and for any tree $B$ of the tiling,
if $A\subset B$, then $B\in{\cal F}$.
\par}
\end{lem}

For proving the theorem, we need a kind of converse of Proposition~\ref{ptreepath}.

\dprop{inpath}
Let $A$ and $B$ be two trees of the tiling with $A\subset B$. Let $\rho$ be the root 
of~$B$ and let $\pi$ be the tree path from $\rho$ to the root~$\tau$ of~$A$. Let $C$ be 
a tree of the tiling such that \hbox{$A\subset C\subset B$}. Then there is a tile~$\nu$ 
of~$\pi$ such that $C=T(\nu)$.
\fprop

\noindent
Proof of the Lemma. Indeed, an ultra-thread satisfies $(v)$. Assume the opposite.
There are  $A\in{\mathcal F}$ and $B$ be a tree of the tiling such that $A\subset B$
and $B\not\in\mathcal F$. Let $C$ be another tree of $\mathcal F$. Then
$B\not\subset C$,otherwise, from~$(iii)$ we get $B\in\mathcal F$. 
From Lemma~\ref{par_trees}, $C\subset B$. As far as $C\in\mathcal F$, we may apply to~$C$ 
the argument we applied to~$A$. We get a sequence $C_i$ of elements of $\mathcal F$
such that \hbox{$A\subset C_i\subset C_{i+1}\subset B$}. From Proposition~\ref{inpath}
the roots of $C_i$ belong to the path from the root of~$B$ to that of~$A$ which is 
contained in a branch of~$B$. Accordingly, as such a path is finite, the sequence of $C_i$
is also finite. Let $C_m$ be the biggest tree of the sequence of $C_i$. Repeating
the argument applied to~$A$ by applying it to~$C_m$, we get that for any $C$ in 
$\mathcal F$, $C\subset C_m$ which is a contradiction with $(iv)$. So that an ultra-thread
satisfies $(v)$.

Conversely. Assume that a thread $\cal F$ satisfies $(v)$. Then, it obviously satisfies
$(iv)$.
\hfill$\Box$

\noindent
Proof of Proposition~\ref{inpath}. Let, $A$, $B$ and $C$ as in the statement of the 
proposition. Let $\rho$, $\tau$, $\nu$ be the roots of $B$, $A$, $C$ respectively. Let 
$\pi_A$, $\pi_C$ be the tree path from $\rho$ to~$\tau$,$\nu$ respectively.
Let $\omega$ be the tile on both $\pi_A$ and $\pi_C$ which is the farthest from~$\rho$
and assume that $\pi_A$ and $\pi_C$ go through different sons of~$\omega$. From 
Proposition~\ref{psepar_char} we get that \hbox{$A\cap C=\emptyset$} which is impossible.
So that necessarily $\omega=C$, which means that $\omega\in\pi_A$.\hfill$\Box$

   Accordingly, an ultra-thread is a maximal thread with respect to the
inclusion. A thread $\mathcal F$ which is not an ultra-thread is said {\bf bounded} and there is a tree 
$A$ in $\mathcal F$ such that for each $B$ in $\mathcal F$, we get \hbox{$B\subset A$}. In that
case, $A$ is called the {\bf bound} of $\mathcal F$.

Consider the following construction:
\begin{const}\label{cultra}
{\leftskip 20pt\parindent 0pt\rm
\phantom{construction}\\
- time~0: fix an \RR-son $\rho$ of a~\MM-tile which is itself the son 
of a \GG-tile; let $F_0$ be $T(\rho)$; 

- time~1: at the level~3 of~$F_0$, and on its left-hand 
side border, there is another \RR-tile $\rho_{-1}$:  let $F_{-1}$ be
$T(\rho_{-1})$; clearly, \hbox{$F_{-1}\subset F_0$}.

Repeating that process by induction, we produce
a sequence $\{F\}_{n\leq0}$ of trees of the tiling such that $F_i$ is 
contained in $F_{i-1}$ for all negative~$i$; denote by $\rho_i$ the
root of~$F_i$. 
If $\tau_{i+1}$ is the son of a tile~$\tau_i$, we say that $T(\tau_i)$ {\bf completes}
$T(\tau_{i+1})$.
 
- time~$2n$+1, $n\geq0$: complete~$T(\rho_{2n})$ by 
$T(\rho_{2n+1})$, where $\rho_{2n+1}$ is an \MM-tile which is the son of a \GG-tile
$\omega_{2n+1}$ which we take as the \GG-son of a \YY-tile $\xi_{2n+1}$;

- time $2n$+2: complete $T(\xi_{2n+1})$ by $T(\rho_{2n+2})$, where $\rho_{2n+2}$ is an 
\RR-tile whose \YY-son is $\xi_{2n+1}$; let $F_{n+1}$ be $T(\rho_{2n+2})$ which
contains~$F_n$.
\par}
\end{const}
\vskip 2pt
\dprop{pultra}
The sequence constituted by the $F_n$, $n\in\mathbb Z$ of Construction~{\rm\ref{cultra}} 
is an ultra-thread.
\fprop

\noindent
Proof. By construction, \hbox{$F_n\subset F_{n+1}$}. Moreover, as a consequence of
Lemma~\ref{par_trees}, we know that the rays delimiting $F_{n+1}$ do not meet those
delimiting $F_n$. The linear order follows from the construction itself. Clearly, the 
property~$(ii)$ is also satisfied. By construction, in between $F_n$ and $F_{n+1}$ there
is no tree of the tiling: there are two trees of the heptagrid whose roots are not 
\RR-tiles. So that the sequence constitute a thread. Clearly, property~$(iv)$ is also
satisfied by the construction.\hfill$\Box$

\subsection{Isoclines}
\label{subsubsec_iso}

   We go back to the rules~$(S)$ defined in the Subsection~\ref{ss_pre_tiles}.
We proved there that it is possible to tile the plane with the rules~$(R_0)$ so that
the rules~$(S)$ also apply provided that \ww-marks are put on \BB- and \OO-tiles
exactly and that \bb-marks are put on \YY- and \GG-tiles exactly. In fact, we can prove 
more:

\begin{lem}\label{les_iso}
Consider a tiling of the heptagrid with \YY, \GG, \BB{} and \OO{} as prototiles obtained
by applying the rules~$(R_0)$. Then, defining \ww-marks on \BB- and \OO-tiles exactly and
\bb-marks on \YY- and \GG-tiles exactly, then the \bb- and \ww-marks obey the rules 
of~$(S)$.
\end{lem}

Proof. According to the assumption, around any tile of the tiling we have the
configurations of Figure~\ref{f_proto_0}. Now, the pictures of the figure satisfy the
rules of~$(S)$ if we put \bb- and \ww-marks as stated in the assumption of the lemma.
Accordingly, the tiling also obey the rules of $(S)$ if we consider \bb- and \ww-marks 
only.\hfill$\Box$

Figure~\ref{f_isoc} illustrates the property that the levels of a tree of the tiling 
coincide with those of its sub-trees which are also trees of the tiling. We already 
noticed that property for the trees of the heptagrid, see 
Proposition~\ref{psubtree_levels}. That allows us to continue the levels to infinity on 
both sides of a tree of the tiling. We call {\bf isoclines} the curves obtained by 
continuing the levels of trees in~$\cal T$.

   In the sequel, it will be important to mark the path of some isoclines on 
each tile of the tiling. The isoclines are unchanged if some \BB- and \OO-tiles are
replaced by \MM- and \RR-ones respectively provided that rules $(R_1)$ are applied. 
Figure~\ref{f_proto} shows us that the levels are also defined in the same way.

\vskip 10pt
\vtop{
\ligne{\hfill
\includegraphics[scale=0.6]{nov_til_isoclines.ps}
\hfill}
\begin{fig}\label{f_isoc}
\leurre
Illustration of the levels in the tiling. Seven of them are indicated in the figure. 
Four trees of the tiling are shown with the rays defining the corresponding tree of the 
tiling.
\end{fig}
}
\vskip 5pt
\noindent



\subsection{Constructing an aperiodic tiling}\label{ss_aperiod}

    We remind the reader that in the heptagrid, a tiling is periodic if there is
a shift~$\tau$ such the tiling is globally invariant under the application of~$\tau$.
The goal of the present subsection is to prove:
    
\begin{thm}\label{th_aperiod}
There is a tiling of the heptagrid which is not periodic. It can be constructed with
$157$ prototiles.
\end{thm}

    We presently turn to the construction which will be reused to prove 
Theorem~\ref{undec}.

    The idea is the following: with trees of the tiling, we define an infinite family 
of triangles which do not intersect and which are bigger and bigger. That property 
entails that the tiling cannot be periodic.

    The construction is performed as follows.

    We define two families of {\bf trilaterals}, a red one and a blue one. In each family, we have 
{\bf triangles} and {\bf phantoms}, so that we have blue and red triangles and also blue and red
phantoms. We call them the {\bf interwoven} triangles as far as the blue trilaterals generate the red ones
which, to their turn generate the blue trilaterals. Blue and red are the {\bf colours} of the trilateral.
Blue and red are said {\bf opposite colours}. Triangle or phantom is the {\bf attribute} of a trilateral.
Triangle and phantom are said of {\bf opposite attributes}.

    The first steps of the construction are represented by Figure~\ref{f_interwov}. Although
the figure is drawn in the Euclidean plane, it can be implemented in the heptagrid.

    We require that triangles of the same colour do not intersect each other. They will be implemented by 
following trees of the tiling as far as the borders of such a tree do not intersect those of another one. 
The legs of a triangle or those of a phantom will follow the borders of a tree $T(\tau)$ of the tiling. 
The basis of the triangle or of the phantom will follow a level of $T(\tau)$. 

For properties shared by both triangles and phantoms whichever the colour, we shall speak of trilaterals. 
For the set of all trilaterals, we shall speak of the {\bf interwoven triangles}.

   For the construction, we consider a sequence of \RR-tiles $\{\rho_i\}_{i\in \mathbb N}$
such that for each $i$ in $\mathbb N$, \hbox{$T(\rho_{i+1})\subset T(\rho_i)$}, and such
that $\rho_{i+1}$ is the \RR-son of an \MM-tile which is the \MM-son of a \GG-tile which 
is the \YY-son of $\rho_i$. We say that the pattern \YY\GG\MM\RR{} joins $\rho_i$ to~$\rho_{i+1}$. Now, 
we require that $\rho_0$ belongs to an isocline, chosen at random and which we call {\bf iscoline 0}.
We number the isoclines with number in $\mathbb Z$. Each isocline $8n$, $n\in\mathbb N$ is said {bf green}
and each isocline $n$ with \hbox{$n = 4 (mod 8)$} is said {\bf orange}.  Under that condition, the sequence
of the $\rho_i$ is called a {\bf wire}. For any $\rho_i$ we say that $i$ is its {\bf absissa}. We say that
$\rho_{2i+1}$ is the {\bf mid-point} between $\rho_{2i}$ and $\rho_{2i+2}$. Note that, by construction, 
the mid-point lies on an orange isocline and each $\rho_{2i}$ lies on a green isocline.  

   The role of the green isoclines is to construct the generation~0 of the trilaterals whose colour is
blue. Each \RR-tile on a green isocline, it is called a {\bf primary seed} triggers a trilateral, moreover,
for each $i$ in $\mathbb N$, the trilaterals raised at $\rho_{2i}$ and $\rho_{2(i+1)}$ have the same 
colour and opposite attributes. The \RR-tiles on an orange isocline raise the {\bf principal seeds} which 
trigger a blue or a red trilateral. 
 
\begin{const}\label{c_interwov}
\phantom{construction}\\
along each wire $\{\rho_i\}_{i\in\mathbb N}$ of the tiling: 

step $0$ defines the trilaterals of generation~$0$ which are blue; $\rho_{2i}$ emit legs of a trilateral
$T_0$ which are stopped by the isocline passing through $\rho_{2i+2}$; $\rho_{2i+2}$ emit legs of a 
trilateral $T_1$ which has the same colour as $T_0$ but the opposite attribute with respect to~$T_0$;
the $\rho_{2i+1}$ which lies inside a triangle of generation~0 emits a red trilateral; let $T_1$ and $T_2$
be the trilaterals raised at that $\rho_{2i+1}$ and at $\rho_{2i+5}$ respectively for the same $i$;
$T_1$ and $T_2$ are both red and they have opposite attributes; accordingly, the basis of $T_1$ is raised 
at $\rho_{2i+5}$; the seeds at $\rho_{2i+1}$ also emit a mauve signal along their orange isocline from
side to side; 

$-$ step $n$$+$$1$, $n\in\mathbb N$: for each trilateral $T$ of the generation~$n$, let $\rho_i$ be its
vertex and let $\rho_j$ emit its basis; then $\rho_k$ is its {\bf mid-point} where $k$ satisfies
\hbox{$2k=i+j$}; also, $j$$-$$i$ is the {\bf height} of~$T$; the isocline passing through $\rho_k$ is 
said the {\bf mid-line} of~$T$; then for each triangle~$T_0$ of the generation~$n$, its mid-point emits 
the vertex of a trilateral~$T_1$ and the basis of a trilateral~$T_2$; $T_1$ and $T_2$ have opposite 
attributes and both have the opposite colour with respect to~$T_0$; 
when the mauve signal~$\mu$ emitted at step~$0$ is accompanied by the basis of a phantom, it is stopped 
by the legs of the first triangle~$T$ which it meets and the isocline of~$\mu$ is the mid-line of $T$; 
when the mauve signal is accompanied by the basis~$\beta$ of a triangle~$T$, it is stopped by the first 
legs of the same colour as~$\beta$, which completes the construction of~$T$; the trilaterals of the 
generation~$n$$+$$1$ are the trilaterals whose vertex is raised at the mid-point of a triangle of the 
generation~$n$.
\end{const}

   The construction is illustrated by Figure~\ref{f_interwov}.

\dprop{tri_indices}
The trilaterals of the odd generations are red, the even generations are blue. If $h_n$
is the height of a trilateral of the generation~$n$ we have $h_n=2^{n+1}$. The absissa $\xi_{n,m}$ of the 
vertex of the $m^{\rm th}$ trilateral of the generation~$n$, $m\in \mathbb N$, is given by 
\hbox{$\xi_{n,m} = 2^n-1+m.2^{n+1}$}, assuming that $\xi_{0,0}=0$.
\fprop

\noindent
Proof. As far as the trilaterals of generation~0 are blue, the trilaterals of generation~1 are red and 
those of generation~2 are blue so that, by induction, the trilaterals of an odd generation are red and 
those of an even generation are blue. By construction, the absissas of the mid-points of the trilaterals of
generation~0 are $2m+1$, $m\in\mathbb N$. As far as $h_0=2$ trivially holds, the formula is true for 
generation~0. We also have that absissas of
the heads of the trilaterals of generation~0 are $2m$, $\in\mathbb N$ which also satisfies the formula
of the proposition. From Construction~\ref{c_interwov}, as far as $\xi_{0,0}=0$, we can see that
$\xi_{1,0}=1$. From Construction~\ref{c_interwov}, we can see that $\xi_{n+1,0}=\xi_{n,0}$+$h_n$ as far as
$h_{n+1}=2h_n$. As far as $\xi_{0,0}=0$, we get that $\xi_{n,0}=2^n-1$, from which we obtain the formula
of the proposition.  \hfill$\Box$

Denote by $\mu_{n,m}$ the mid-point of the $m^{\rm th}$ trilateral of the generation~$n$. From the proof 
of the proposition, we note that $\mu_{n,m}=(m$+$1).2^{n+1}-1=(2m$+$2).2^n-1$ which 
means $\mu_{n,m}$ is also the mid-point of a trilateral of the previous generation. In fact, each second 
mid-point of trilaterals of the previous generations is still the mid-point of a trilateral of the 
generation~$n$. The other mid-points are mid-points of triangles so that they emit vertices of trilaterals
of the generation~$n$. That proves the construction too. Note that the proof is illustrated by 
Figure~\ref{f_interwov}. The reader is referred to the Appendix for other pictures
illustrating the first five steps of the construction.

Together with Proposition~\ref{tri_indices}, we have additional properties:

\begin{lem}\label{inter_gene}
A trilateral~$T$ of the generation~$n$$+$$1$ contains a single phantom~$P$ of the generation~$n$ and 
there are two triangles
$T_0$, $T_1$ of the generation~$n$ such that $T_0$, $T_1$ contains the vertex, 
the basis of~$T$ respectively, in both cases on their mid-line. Moreover, $T$ and $P$ have the same
mid-line. A trilateral~$T$ of the generation~$n$$+$$2$
contains three trilaterals which are of the same colour of the generation~$n$ when $n\geq 1$, two of 
them being triangles and, in between them, a phantom~$P$, the third one. Also, $T$ and $P$ have the same 
mid-point.
\end{lem}

\noindent
Proof. The lemma is an easy consequence of Proposition~\ref{tri_indices}.

Taking the notations of the proposition, note that 
\vskip 0pt
$\xi_{n+1,m+1}=2^{n+1}-1+(m$+$1).2^{n+2}=\xi_{n+1,m}+2^{n+1}=\xi_{n+1,m}+2^n+2^{n+1}+2^n$

Moreover, $\xi_{n+1,m}=2^{n+1}-1+m.2^{n+2}=2^n-1+2m.2^{n+1}+2^{n+1}+2^n$\\
$=\xi_{n,2m}+2^n$,

\noindent
which proves the first assertion of the lemma, as far as $2^n$ is half the height of $T_1$. Note that the
absissa of the mid-point of~$T$ is $\xi_{n+1,m}+2^{n+1}$ while that of $T_1$ is $\xi_{n,2m+1}+2^n$. An easy
computation shows as that the two absissas are equal.

Similarly,\\
$\xi_{n+2,m+1}= 2^{n+2}-1+(m$+$1).2^{n+3}=\xi_{n+2,m}+4.2^{n+1}=\xi_{n+2,m}+2^n+3.2^{n+1}+2^n$,\\

\noindent
which proves the last part of the lemma. For what is the mid-points, the proof follows from the latter
computation and from two applications of the first part of the lemma. 

\ifnum 1=0 {
Assume that it is true for the generation~$n$. By Construction~\ref{c_interwov}, the vertex
of a trilateral of the generation~$n$+1 is a mid-point of a trilateral $T_0$ of the 
generation~$n$ of another colour or of another attribute. The legs raised from that tile cross a first 
orange signal in the next trilateral~$T_1$ and they cross the basis of that trilateral at the second 
orange signal issued from the mid-point of the trilateral~$T_2$, the next one after
$T_1$. As far as in between $T_i$ and $T_{i+1}$ with \hbox{$i\in\{0,1\}$}, there is a trilateral, the 
trilateral contains two triangles of the generation~$n$ exactly, which proves the first assertion of the 
lemma. If the trilateral of the generation~$n$+1 is a phantom~$\varphi$, its mid-point~$\mu$ 
corresponds to the first orange signal met by its legs after its vertex. By construction,
$\mu$ cannot be in a triangle as far as the legs of a triangle stop the orange signal running 
along an orange isocline. And so, $\mu$ is the mid-point of a phantom~$\psi_n$ of the 
generation~$n$ and \hbox{$\psi_n\subset\varphi$}. By induction, that proves the second 
assertion of the lemma.
} \fi
\vskip 10pt
\vtop{\leftskip 0pt\parindent 0pt
\ligne{\hfill
\includegraphics[scale=0.5]{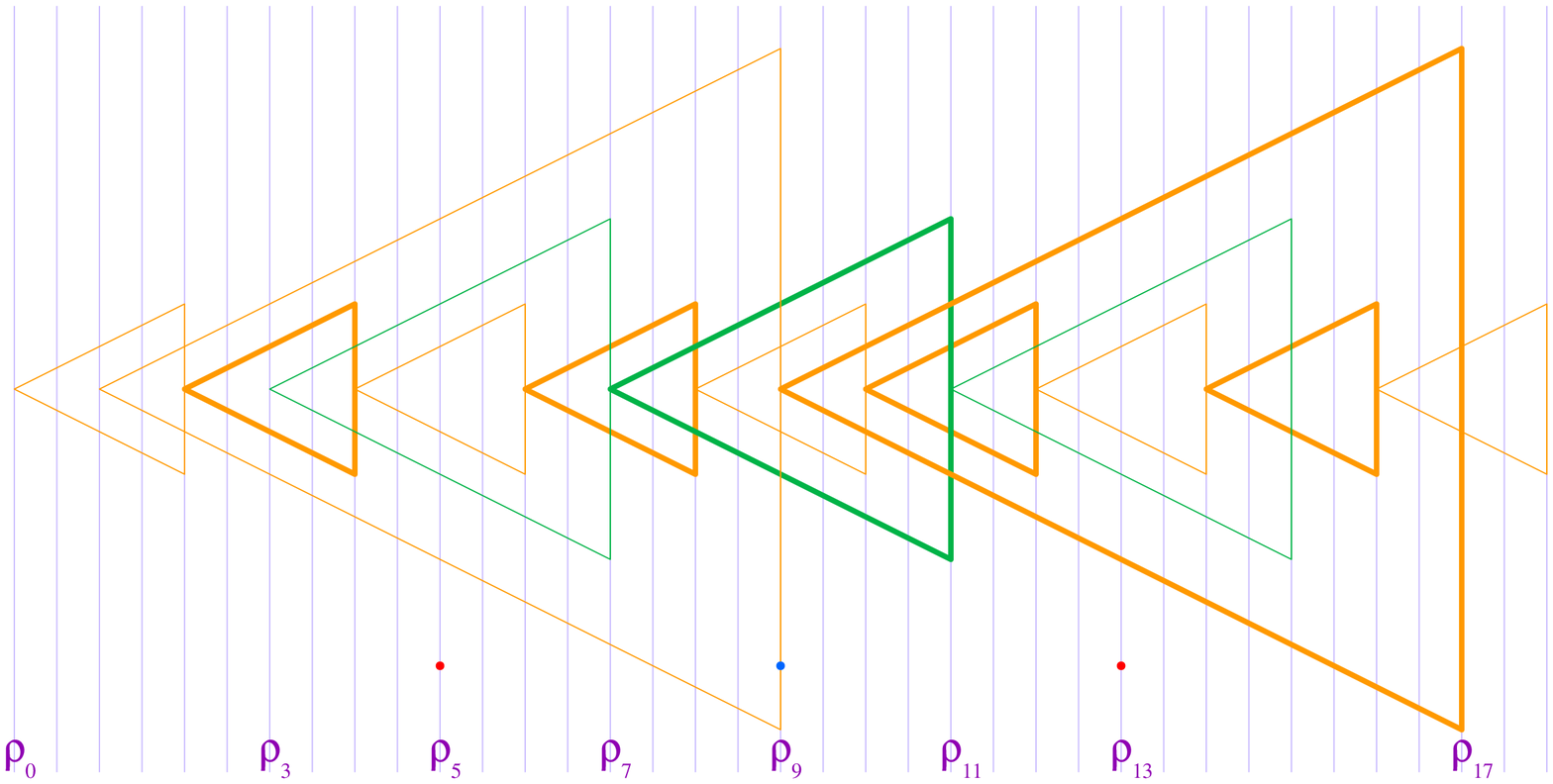}
\hfill}
\begin{fig}\label{f_interwov}
\leurre
Illustrating the construction of the interwoven triangles. 
We can see how to construct a triangle of the generation~$n$+1 from triangles of the generation~$n$.
\end{fig}
}

We remain with the proof that the attributes of the trilaterals we have found in the above compoutations
are those given by the lemma.
For what is the connection between a trilateral of the generation~$n$+1 and the trilaterals mentionned in
the lemma, we know from construction~\ref{c_interwov} that $T_0$ and $T_1$ are triangles. Consequently,
the trilateral contained in~$T$ is a phantom as far as the vertex of $P$ belongs to the basis of~$T_0$ 
and the basis of~$P$ contains the vertex of~$T_1$. Consider a trilateral of the generation~$n$+2. From what
we just proved, it contains a phantom~$P_0$ of generation~$n$+1. Applying the lemma to~$P_0$, we get 
that $P_0$ contains a phantom~$P$ of the generation~$n$ and two triangles $T_0$ and~$T_1$ such that the
basis of $T_0$ contains the vertex of~$P$ and the vertex of~$T_1$ belongs to the basis of~$P$. Now, the 
above computations show us that $T_0$ and~$T_1$ are contained in $T$. That completes the proof of the
lemma. \hfill$\Box$

\dprop{no_cross}
The legs of a trilateral do not intersect the legs of another one, whichever its colour, whichever its 
attribute. Moreover, two triangles of the same colour are either disjoint or one of them is embedded in the
other one.
\fprop

\noindent 
Proof. Immediate corollary of Lemma~\ref{par_trees} and of Proposition~\ref{tri_indices}.

The colour, the attribute and the generation of a trilateral constitute its {\bf characteristics}.

\vskip 10pt
\ligne{\hskip 10pt
\vtop{\leftskip 0pt\parindent 0pt\hsize=300pt
\ligne{\hskip 0pt
\includegraphics[scale=0.45]{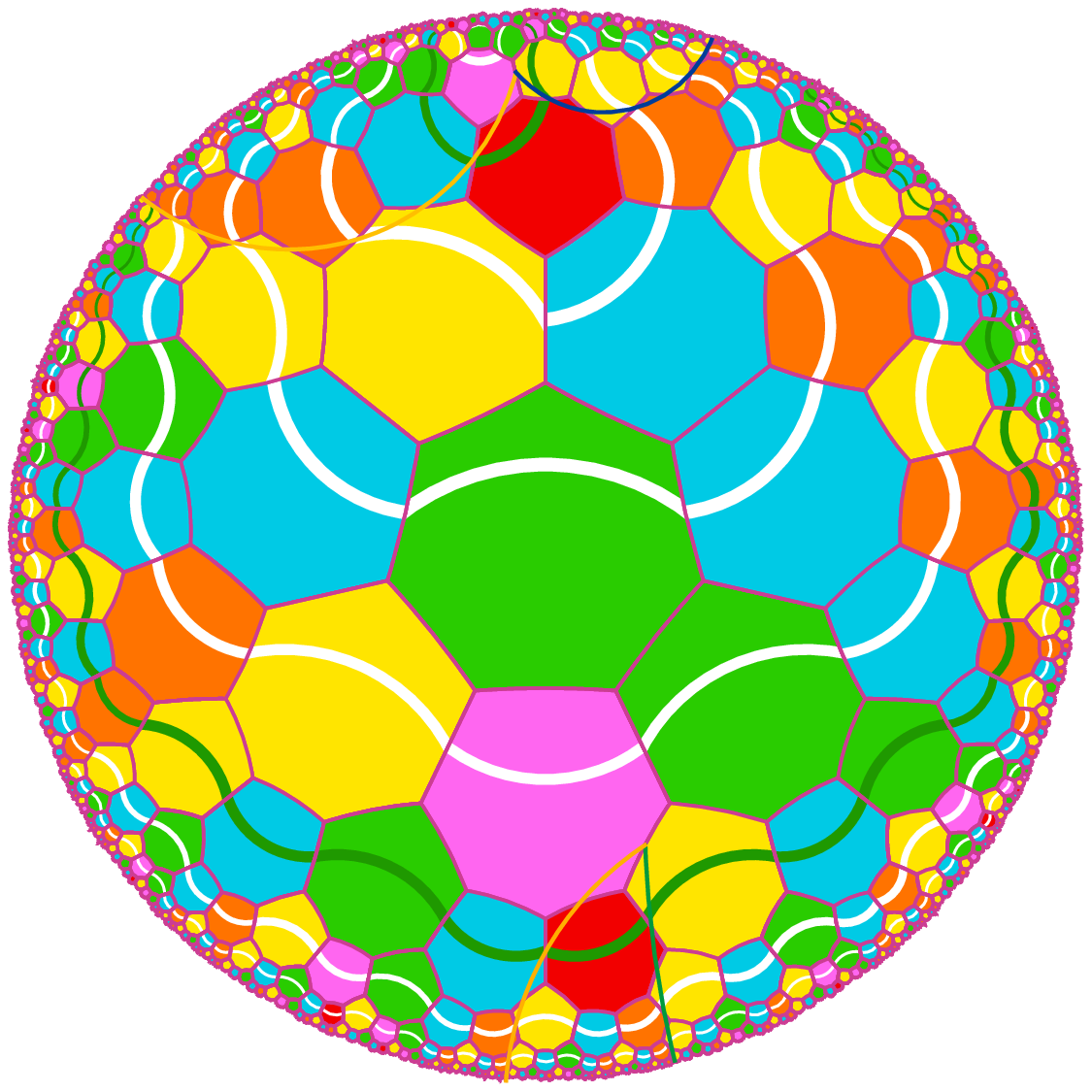}
\includegraphics[scale=0.45]{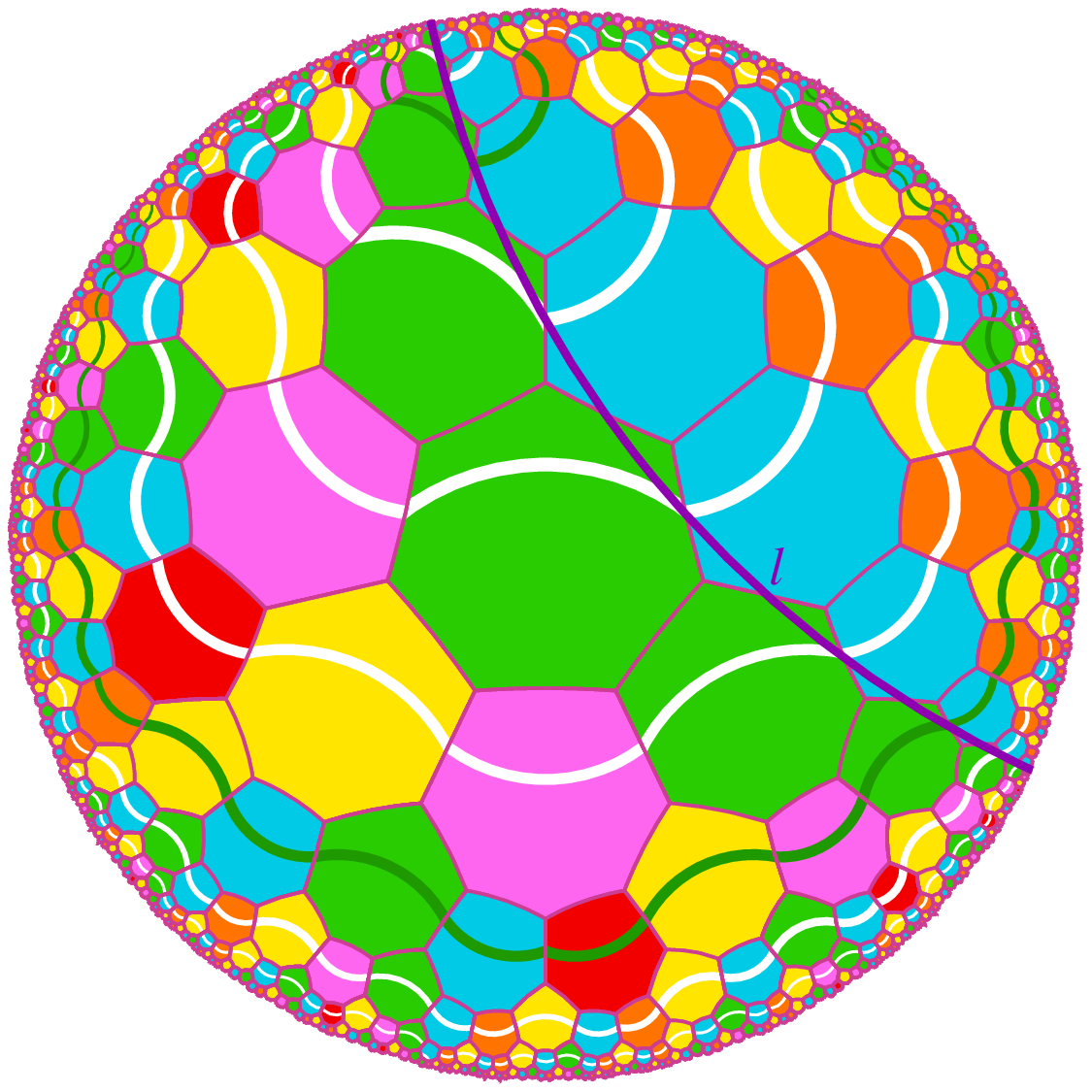}
\hfill}
\begin{fig}\label{f_ultra}
\leurre
Representation of seeds and isoclines in two tilings of the heptagrid. To left, two trees 
of the tiling are illustrated, both rooted at an \RR-tile. They belong to an ultra-thread.
To right, the tiling has no ultra-threads, threads only.
\end{fig}
}
\hfill}

   The trilaterals we defined in Section~\ref{s_aperiod} can be embedded in the 
tiling of the hyperbolic plane illustrated by Figure~\ref{f_ultra}. In the figure, we
can see that the periodic numbering of the isoclines from~0 up to~4 is implemented with
the help of three colours used to materialise the isoclines: blue, green and orange. 

\dprop{mid-lines}
In each triangle~$T$ of the generation~$n$, $n\geq 1$, its mid-line $\mu$ crosses $n$ phantoms~$P_m$ of 
the generation~$m$ with \hbox{$0\leq m < n$}. Moreover, $\mu$ is also a mid-line for each $P_m$,
where \hbox{$0\leq m < n$}.
\fprop

\noindent
Proof. Apply Lemma~\ref{inter_gene}: $T$ contains a phantom $P_{n-1}$ of the generation~$n$$-$1. If $n>1$,
the lemma also applies to $P_{n-1}$ giving rise to a phantom $P_{n-2}$ of the generation $n$$-$2. By 
induction, we prove the proposition which is clearly true for generation~1 which follows from the previous
argument. The statement about $\mu$ also follows by induction from the computation performed in the
proof of Proposition~\ref{tri_indices}. \hfill$\Box$

\subsection{Application to isoclines and to threads}\label{ss_iso_thr}

   Going back to isoclines, we already noticed that they allow to define levels in the 
whole hyperbolic plane. As shown by Figures~\ref{f_proto_0} and~\ref{f_proto}, isoclines
do not intersect and above an isocline there is always an isocline, so that isoclines
constitute a partition of the tiling. We need some additional information to 
Construction~\ref{cultra}. In that construction, we defined a {\bf wire}, denoted by
$\mathcal Q$, as the sequence of tiles joining all $\rho_i$, $i\in \mathbb N$, in the 
tree. The sequence of $T(\rho_i)$, $i\in\mathbb N$ defines a thread. Remember that from 
$\rho_i$ to $\rho_{i+1}$, both included, we have the statuses \RR, \YY, \GG, \MM{} and 
\RR{} for the elements of~$\mathcal Q$. Those five tiles belong to five consecutive 
isoclines. Note that the number of tiles of~$T(\rho_0)$ which are at distance~$n$ from
$\rho_0$ is $f_{2n+1}$.

\begin{lem}\label{iso_apart}
Let $I_n$ be the elements of $T(\rho_0)$ belonging to the isocline~$n$ where 
$n\in \mathbb N$. Let $u_n$, $v_n$ be the leftmost, rightmost element of~$I_n$ 
respectively. Let $y_n$ be the element of $\mathcal Q$ belonging to the isocline~$n$. Then
\vskip 5pt
\ligne{\hfill
$\vcenter{\vtop{\leftskip 0pt\parindent 0pt\hsize=250pt
\ligne{\hfill
$appart(u_n,y_n)\geq f_{2n-3}$, $appart(y_n,v_n)\geq f_{2n-1}$,\hfill}
\ligne{\hfill $appart(u_n,y_n), appart(y_n,v_n) \leq f_{2n+1}$\hfill}
}}$
\hfill$(A)$\hskip 20pt}
\end{lem}
   
\noindent
Proof. It is plain, from Figure~\ref{f_proto}, that if $\rho_i$, \hbox{$i\in\{1,2,3\}$},
the status of $\rho_i$, $i$ from 1 to~3, are \YY, \BB{} and \OO{} respectively.
From Lemma~\ref{separ_trees}, we have \hbox{$T(\rho_1)\cap T(\rho_3)=\emptyset$} and
${\mathcal Q}\backslash \{\rho_0\} \subset T(\rho_1)$. Now, consider $\eta_j$, 
\hbox{$j\in\{1,2,3\}$} the sons of~$\rho_1$. As far as the statuses of those sons are, 
in the order of~$j$, \YY, \BB{} and \GG{} respectively, Lemma~\ref{separ_trees} tells
us that \hbox{$T(\eta_1)\cap T(\eta_3)=\emptyset$}. Moreover, it is plain that
\hbox{${\mathcal Q}\backslash\{\rho_0,\rho_1\}\subset T(\eta_3)$} and we know that
\hbox{$T(\eta_3)\subset T(\rho_1)$}. Accordingly, we may conclude that 
\hbox{${\mathcal Q}\backslash\{\rho_0,\rho_1\}\cap T(\eta_1)=\emptyset$}. In $T(\eta_1)$
the level of the isocline~$n$ is $n$$-$2 and in $T(\rho_3)$, the same isocline contains
the level $n$$-$1 of that tree. From that, we get the estimates of~$(A)$.
\hfill$\Box$

   Let us remember that Construction~\ref{cultra} defines an ultra-thread 
${\mathcal F}_{i\in\mathbb Z}$, where each $\mathcal F$ is $T(\rho_i)$ where
the tiles joining $\rho_i$ to $\rho_{i+1}$ have the statuses \RR, \YY, \GG, \MM{}
and \RR{} in that order. As far as $i$ runs over $\mathbb Z$ we may, in that case,
define $\mathcal Q$ as a sequence of tiles indexed in $\mathbb Z$ with the property
that ${\mathcal Q}_{4i}$ is exactly $\rho_i$. We again call that new sequence the
quasi-axis of that ultra-thread. Then, it is possible to prove:

\begin{lem}\label{bigtree}
Let $\{T(\rho_i)\}_{i\in\mathbb Z}$ be the sequence of trees of the tiling defined by
Construction~{\rm\ref{cultra}}. Then, for each tile~$\tau$ of the heptagrid, there
is $i\in\mathbb Z$ such that $\tau\in T(\rho_i)$. Accordingly, for any tile $\tau$ of
the heptagrid which is not a \BB-tile, there is $i\in \mathbb Z$ such that
\hbox{$T(\tau)\subset T(\rho_i)$}. 
\end{lem}

\noindent
Proof. There is an index $n$ such that $\tau$ belongs to the isocline~$n$. Let $y_n$ be
the tile of $\mathcal Q$ belonging to the isocline~$n$ too. Let $\rho_i$ be the closest
$\rho_j$ such that $y_n\in T(\rho_j)$ as far as, clearly, ${\mathcal Q}_m$ belong to
$T(\rho_j)$ for any $j$ and any $m$ starting from some value.
Now, we can find $j<i$, $j\in \mathbb Z$, such that $(A)$ should be satisfied 
with $u_j$ and $v_j$ being the leftmost and rightmost tiles respectively of the trace of 
$T(\rho_j)$ on the isocline $n$. Taking $\tau$ a tile of 
status different from \BB{} and from \MM, as we can find $j\in\mathbb Z$ such that
$\tau\in T(\rho_j)$, from Proposition~\ref{ptreepath} we conclude that
\hbox{$T(\tau)\subset T(\rho_j)$}.\hfill$\Box$

As a corollary of Lemma~\ref{bigtree}, we can deduce the following property of
the ultra-thread obtained from Construction~\ref{cultra}:

\begin{lem}\label{ultra_ultra}
Let $\mathcal F$ be the ultra-thread given by Construction~{\rm\ref{cultra}} and let
$\mathcal G$ be another ultra-thread. Then, for each tree of the tiling $G$ in
$\mathcal G$, there is a tree of the tiling $F$ belonging to~$\mathcal F$ such that
\hbox{$G\subset F$}.
\end{lem}

\noindent
Proof. Immediate.

\dprop{p_level_ultra}
Let $\mathcal F=\displaystyle{\reunion_{n\in\mathbb Z}F_n}$ be an ultra-thread and let 
$\tau_n$ be the root of $F_n$. Consider the set of levels of the tiling. For each 
level~$m$ in $\mathbb Z$, there is an $n$ in $\mathbb Z$ such that the level of~$\tau_n$
is higher than~$m$. Moreover, let $I_m$ be the set of tiles on the level~$m$
belonging to~$F_n$. Let $\ell_m$, $r_m$ be the leftmost, rightmost tile respectively 
in $I_m$. Let $u$ in $\mathbb Z$, $u>n$. Then 
\hbox{$I_u\subset I_m$} and \hbox{$appart(\ell_m,\ell_u).appart(r_m,r_u)>0$}.
\fprop

\noindent
Proof. Let $h_k$ be the level of $\tau_k$, $k$ in $\mathbb Z$. Let $\tau_\ell$ with
$\ell>k$. Then, by definition of $\mathcal F$, we have 
\hbox{$F_k=T(\tau_k)\subset T(\tau_\ell)=F_\ell$}, so that $h_\ell>h_k$.

The relations concerning $I_m$, $I_u$ and their respective extremal tiles comes from
the fact the borders of trees of the tiling do not meet.\hfill$\Box$

\begin{lem}\label{bigtree_ultra}
Let $\mathcal F=\displaystyle{\reunion_{n\in\mathbb Z}F_n}$ be an ultra-thread and let 
$\tau$ be a tile. Then there is $m$ in $\mathbb Z$ such that $\tau\in F_m$.
\end{lem}

\noindent
Proof. Consider a broken line $\mathcal B$ which joins the centers of each $\tau_n$, 
$n\in\mathbb Z$, where $\tau_n$ is the root of $F_n$. Let $k$ be the level of~$\tau$. 
That level meets $\mathcal B$ at some tile~$\nu$. From Proposition~\ref{p_level_ultra},
there is an $m$ in $\mathbb Z$ such that the level of~$\tau_m$ is higher than $k$. By
construction, $F_m$ contains $\nu$ as far as each $F_n$ contains all the tiles crossed
by $\mathcal B$, starting from its root. Let \hbox{$\delta= appart(\tau,\nu)$}.
Let $I_u$, $\ell_u$ and $r_u$ defined as in the proof of Proposition~\ref{p_level_ultra}.
As far as $appart(\ell_u,\ell_v)>0$ and $appart(r_u,r_v)>0$ if $u<v$ and as far as those
appartnesses are integers, we have that $appart(\ell_u,\tau)\rightarrow\infty$
and \hbox{$appart(r_u,\tau)\rightarrow\infty$} when
$u\rightarrow\infty$. Accordingly, there is $w$ in $\mathbb Z$ such that
$appart(\ell_w,\tau),appart(r_w,\tau)>\delta$, so that $F_w$ contains $\tau$.
\hfill$\Box$

\begin{lem}\label{no_ultra}
There are tilings of the heptagrid with the tiles \YY, \GG, \BB, \OO, \MM{} and \RR{}
and the application of the rules~$(R_1)$ such that all its threads are bounded.
\end{lem}

\noindent 
Proof. Consider a mid-point line $\ell$ of the hyperbolic plane as defined in 
Section~\ref{ss_heptatil}. Assume that $\ell$ crosses all the levels which can be put
by a tiling of the heptagrid. It is possible to assume that $\ell$ crosses \GG-tiles
and that the centres of those tiles lie on the same side of~$\ell$. Each one of those 
tiles generates a tree of the heptagrid whose right-hand side ray is contained in~$\ell$.
That rules out the possibility of an ultra-thread in such a tiling. Otherwise, let 
$\mathcal F=\displaystyle{\reunion_{n\in\mathbb Z} F_n}$ be an ultra-thread. Take
$\tau$ as a \GG-tile crossed by~$\ell$. From Lemma~\ref{bigtree_ultra}, there is $n$
in $\mathbb Z$, such that $\tau$ is contained in~$F_n$. From Lemma~\ref{separ_trees},
$T(\tau)\subset F_n$. But now, $F_{n+1}$ contains $F_n$ but, as its border does not meet
that of~$F_n$, that border should meet that of $T(\nu)$ where $\nu$ is a \GG-tile crossed
by~$\ell$ such that $T(\tau)\subset T(\nu)$. But in that case, we could choose $\nu$ such
that its border meet that of $F_{n+1}$, a contradiction with Lemma~\ref{separ_trees}.
That proves Lemma~\ref{no_ultra}.\hfill$\Box$

   Accordingly, some realisations of the tiling contain ultra-threads, some realisations
of it contain none of them as illustrated by Figure~\ref{f_ultra}.

\subsection{The prototiles for an aperiodic tiling}\label{ss_til_aperiod}

   From now on, we introduce a distinction of the isoclines. Each fourth isecline, starting from one of 
them defined at random, receives a colour: alternatively {\bf green} and {\bf orange}. Considering that a 
green isocline is higher than an orange one, that defines the directions {\bf up} and {\bf down} in the 
hyperbolic plane. The isoclines also allow us to define the directions {\bf to left} and {\bf to right}. 

From now on, an \RR-tile will be called a {\bf seed}. The seeds which are crossed by a green isocline
are the vertices of a trilateral of the generation~0, so that they trigger the construction of that
trilateral. A seed which sits on an orange isocline is the vertex of a trilateral of the generation~$n$
with $n\geq 1$. An isocline which is neither green nor orange is said to be {\bf blue}.

    In the present subsection, we implement Construction~\ref{c_interwov} as a tiling. To that 
purpose, we define a set of {\bf prototiles}: the tiles of the tiling are copies of prototiles.
By copy we mean an isometric image which place a tile from an isocline onto another one such that
left, right up and down of the former place coincide with those directions on the new isocline. 
Figure~\ref{f_nproto_a_i} defines the tiles required for the implementation of the isoclines. 
\vskip 10pt
\ligne{\hfill
\vtop{\leftskip 0pt\parindent 0pt\hsize=300pt
\ligne{\hfill
\includegraphics[scale=0.25]{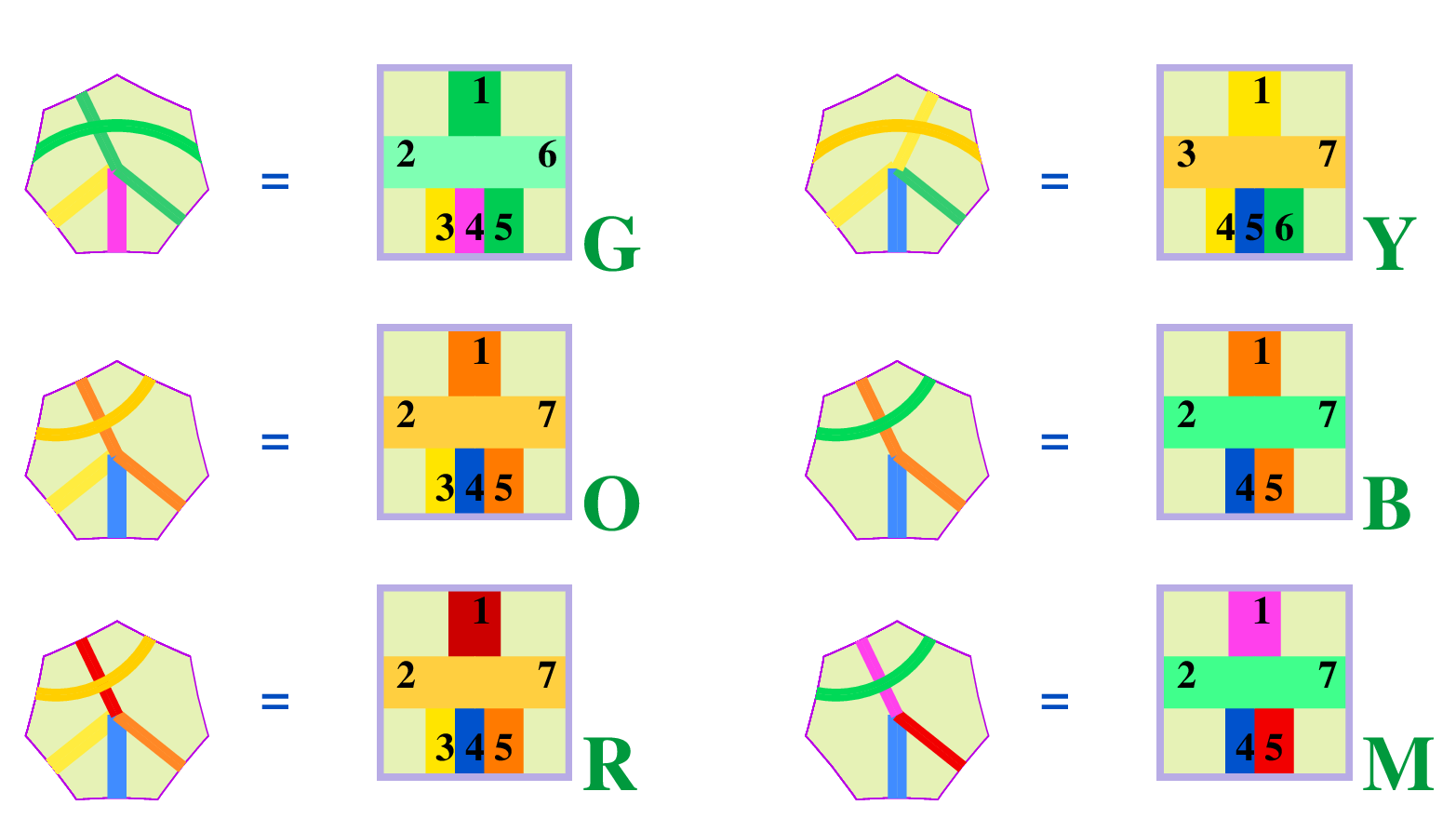}
\hfill}
\ligne{\hskip-10pt
\includegraphics[scale=0.3]{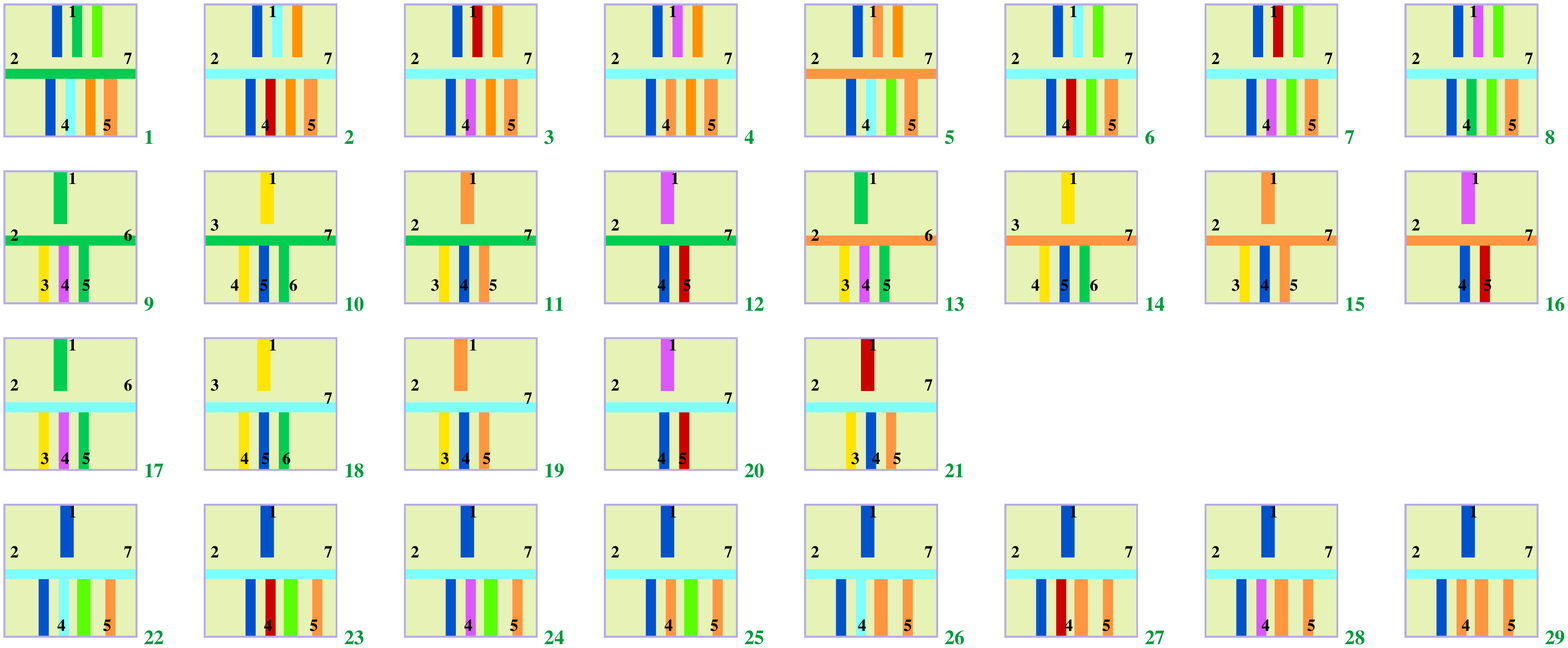}
\hfill}
\begin{fig}\label{f_nproto_a_i}
\leurre
Implementation of the rules~$(R_1)$ and of the isoclines by 29 prototiles. Note the convention for
representing heptagonal tiles by squares.
\end{fig}	
}
\hfill}

To force the succession of green and orange isoclines, we use the \BB-tiles as far as they occur 
in most rules in $(R_1)$. Tiles 1 up to~8 of the figure illustrate how we define that succession. 
Tiles 9-12, 13-16 and 17-21 illustrate the other tiles than \BB-ones 
when they receive a green isocline, an orange one and a blue one respectively.
Note that \RR-tiles are mentioned with blue isoclines only as far as they are seeds when sitting 
on a green or an orange isocline. The last row illustrates the tiles needed to start marking the 
\BB-tiles: it is the case for the son of a \YY-, an \OO{} or an \RR-tile. Depending on the 
isocline of the father of such a \BB-tile and the surrounding isoclines, we have the appropriate 
tile to be synchronised with \BB-tiles on the same isocline as far as green and orange isoclines 
split the hyperbolic plane into two halves.

Consider a tree of the tiling ${\mathcal T} = T(\tau)$, rooted at $\tau$. Define 
\hbox{$\{\beta_i\}_{i\in\mathbb N}$} to be the branch of $\mathcal T$ as follows. If $\tau$ is 
a \GG-tile, then $\beta_0 =\tau$. Otherwise, $\beta_0$ is the \BB-son of the \MM-son of $\tau$. 
Then, for any non negative integer $n$, $\beta_{n+1}$ is the \BB-son of $\beta_n$. Call that 
branch the {\bf $\beta$-branch} from $\tau$. We say that the branch consists of $\tau$ and of its 
recursive \BB-off-springs. The interest of that definition is that the $\beta$-branch of 
$\mathcal T$ does not intersect any tree of the tiling contained in $\mathcal T$. 

It can easily be 
seen from Figure~\ref{f_proto} that the prototiles of the figure can tile $T(\tau)$. 
The first three rows of the figure indicate the convention we use to represent the heptagonal
prototiles by square ones. The convention is based on the fact that we have mainly a top down 
direction and a left right one given by the isoclines. The top number indicates 1, the side to 
the father. At the bottom side of the square we have the numbers of the sides to the sons of the 
tile. On the left- and right-hand side edges, we have the number of the sides crossed by the 
isocline on which the tile sits.

\def\gg{{\bf g}}
As an example, it is not difficult to see that a \ww-tile cannot abut another \ww-tile
and that, similarly, a \bb-tile cannot abut another \bb-tile. From that and similar 
considerations we leave to the reader, the tiles with a blue isocline can build the
pictures of Figure~\ref{f_proto} and only them. As far as besides the isocline the 
tiles with a green isocline of Figure~\ref{f_nproto_a_i} look like those with a
blue isocline, we obtain the pictures of Figure~\ref{f_proto} and only them with the
tiles of Figure~\ref{f_nproto_a_i}. We also clearly obtain that green and blue
isocline do not mix and do not cross each other. The first row of Figure~\ref{f_nproto_a_i}
allow us to build a $\beta$-branch in any tree of the tiling. But the first row alone
generates a $\beta$-branch whose root is rejected at infinity. For a true $\beta$-branch
rooted at a tree of the heptagrid, we need the tiles of the last row of 
Figure~\ref{f_nproto_a_i}: the father of a \BB-tile is either a \YY-, an \OO- or an 
\RR-tile. In each case, the father may be on a green or on a blue isocline while the
\BB-tile may be on a blue or on a green one. Of course, if the \BB-tile, its father is 
on a green tile then its father, the \BB-tile respectively, is on a blue one.

Note that the green, orange isocline defined by a first tile~1, 5 respectively impose the position 
of all other green, orange isoclines by the fact that the tiles bearing a green, orange isocline 
can only abut on the same level tiles also bearing a green, orange isocline respectively. 

In order to define the prototiles to construct the trilaterals, we need another property which can 
be deduced from Proposition~\ref{tri_indices} and Lemma~\ref{inter_gene}:

\dprop{legs_and_bases}
The legs of a trilateral~$T$ of the generation~$n$$+$$1$ are cut once by the basis~$B$ of the 
triangle~$T_0$ of the generation~$n$ whose mid-point is the vertex~$V$ of~$T$. That 
isocline~$\beta$ which contains $B$ is issued from $\rho_j$ where $\rho_j$ is the mid-point 
between $V$ and the mid-point of~$T$. In between $V$ and $\beta$, the legs of $T$ are cut by bases 
of phantoms only. In between $\beta$ and the basis of~$T$, the legs of $T$ are not cut by any 
trilateral of whichever generation.
\fprop

\noindent
Proof. From Lemma~\ref{inter_gene}, we know that the vertex of~$T$ is the mid-point of a 
triangle~$T_0$ of the generation~$n$ and that the basis of~$T$ is issued from the mid-point of a 
triangle $T_1$ of the generation~$n$ too. As far as the height of~$T_0$ is the half of that of~$T$ 
the basis of~$T_0$ satisfies the statement of the proposition. Some trilaterals of generation~$m$, 
with $m\leq n$ whose vertices are contained in~$T$ cut the basis of~$T$, but their basis does not 
cut the legs of~$T$. For what are trilaterals of generation higher than $n$+1, either they 
contain~$T$ or they are disjoint from~$T$ so that in both cases, their basis do not cut the legs 
of~$T$.\hfill$\Box$

\vskip 10pt
\ligne{\hfill
\vtop{\leftskip 0pt\parindent 0pt\hsize=300pt
\ligne{\hskip-15pt
\includegraphics[scale=0.4]{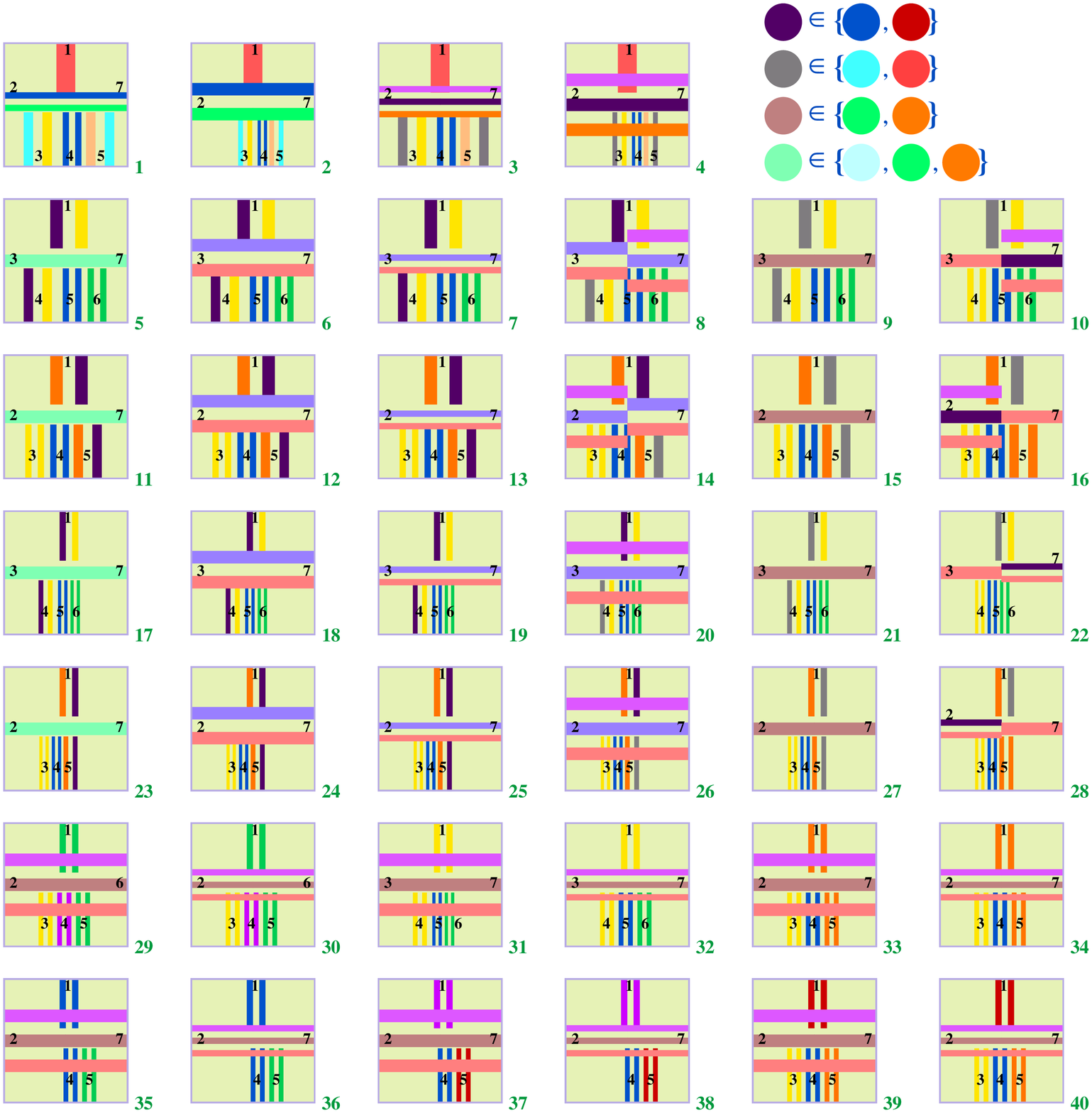}
\hfill}
\vspace{-10pt}
\begin{fig}\label{f_nproto_tri}
\leurre
The 128 prototiles for constructing the trilaterals. Among those prototiles, 28 of them represent 
a red or a blue trilateral. Note the conventions of colours in order to restrict the number of 
pictures downto 40 of them.
\end{fig}	
}
\hfill}

   Figure~\ref{f_nproto_tri} gives the prototiles for constructing the trilaterals which
have to be appended to those of Figure~\ref{f_nproto_a_i}. Note that the prototiles~1 to~4 of
Figure~\ref{f_nproto_tri} complete the prototiles of Figure~\ref{f_nproto_a_i} for what are the 
seeds on a green or an orange isocline. 

The first row of the second part of Figure~\ref{f_nproto_tri} illustrates prototiles to trigger 
the construction of the legs of a trilateral. Note that both left- and right-hand side legs are 
represented. Moreover, as indicated by the figure itself, we use a few grey colours to be replaced 
by various hues of blue and red. It is the reason why the 128 prototiles are illustrated by only 
40 ones. In fact, as indicated in the propositions and lemmas devoted to the trilaterals, the legs 
can be uniformly dealt with. Note the fact that for a leg, whichever its side is, we use two hues 
of the same colour: a dark version for the first half of the leg starting from the vertex, and a 
lighter version for the second half. The rightmost part of the first row in the second part of
the figure illustrates thee conventions we use for the hues which represent two or three colours.
\vskip 10pt
\vtop{
\ligne{\hfill
\includegraphics[scale=0.35]{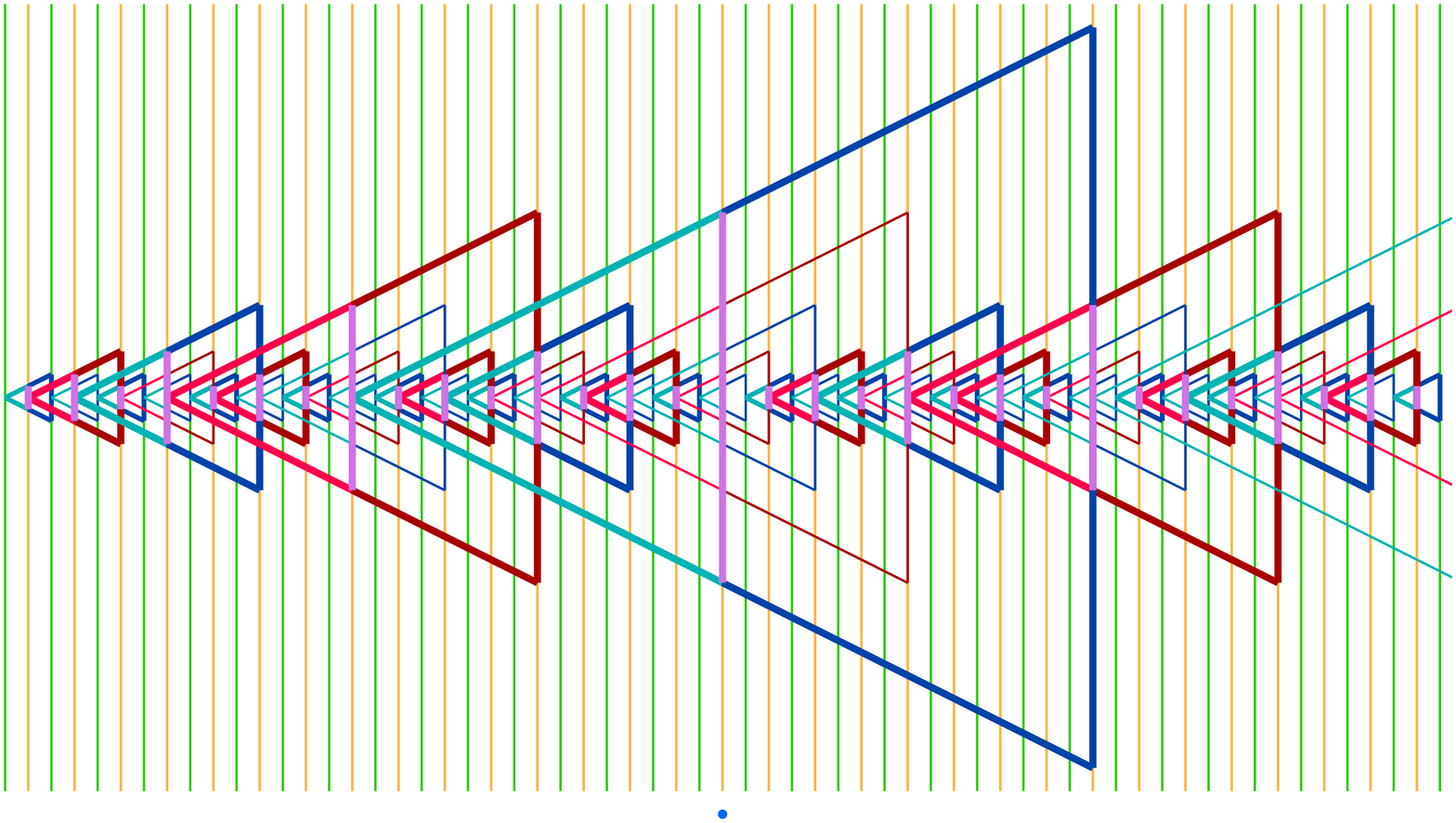}
\hfill}
\vspace{-10pt}
\begin{fig}\label{f_tttt_interwov}
\leurre
The construction of an aperiodic tiling.
\end{fig}
}

Consider a {\bf triangle} $T$ whose vertex is denoted by~$V$ and its mid-point by $\omega$. From
Proposition~\ref{legs_and_bases}, the first basis of a {\it triangle} $T_0$ cutting a leg of~$T$ 
cuts the path from $V$ to the basis of~$T$ at the mid-point~$\nu$ between $V$ and~$\omega$. Such a 
basis met by a leg when running over it from~$V$ occurs at the fourth of the leg. From 
Lemma~\ref{inter_gene}, the other bases cutting the leg of~$T$ in between $V$ and $\nu$ are bases of
trilaterals of lower generations. Clearly, the mauve signals running on the isoclines of thoses 
bases meet triangles of a generation lower than that of~$T_0$, so that when those bases cut the 
leg of $T$ there is no mauve signal with them. So that the first time a leg of~$T$ meets a mauve 
signal, it is on the isocline passing through~$\omega$. Accordingly, the change of coulour for the 
leg of~$T$ occurs at that moment. Later, there is no meeting of a basis of a trilateral, the basis 
of~$T$ being excepted. When it is the case, the basis does not contain a mauve signal at that 
meeting. 

   Accordingly, the proof of Theorem~\ref{th_aperiod} is completed. The number of needed prototiles
is the sum of the numbers indicated in Figures~\ref{f_nproto_a_i} and~\ref{f_nproto_tri}.      
\hfill$\Box$

Figure~\ref{f_tttt_interwov} illustrates the proof of Theorem~\ref{th_aperiod}.

\section{Completing the proof of Theorem~\ref{undec}}
\label{s_proof}

   Presently, we shall see how to obtain the prototiles we need in order to prove
Theorem~\ref{undec}. It is important to see that the theorem will be proved only when
we have produced the set of prototiles.

   From the previous construction, we know that we have bigger and bigger triangles, so
that if we take red triangles as the frame of the simulation of a Turing machine, it is
a possible solution. It is enough that the set of prototiles is adapted to a given
Turing machine $M$ in order to perform its computation in any triangle. If $M$ does not
halt, the computation is stopped when the computing signal meets the basis of the 
triangle and it will be the case in all triangles. If $M$ halts, the halting will be
observed in some triangle. It is easy to implement the halting state by a prototile
one side of which cannot be abut by any prototile.

   However, that program can be fulfilled if we can perform the computation in a red triangle.
The scenario is the following. The initial configuration is displayed along the 
right-hand leg $\ell_r$ of the red triangle~$T$. That leg consists of \OO-tiles, the vertex 
of~$T$ being excepted: it is the tile~\ref{f_nproto_tri}.3, an \RR-tile sitting on an orange 
isocline. From the \OO-tiles, we consider the path in $T$ which goes from a tile~$\tau$ of 
$\ell_r$ to a tile of the basis of~$T$ by following the \YY-son of~$\tau$ and, recursively, 
the \YY-sons of those \YY-sons. Call such a path the {\bf \YY-path from~$\tau$} and say that $\tau$ 
is its {\bf source}. From Lemma~\ref{separ_trees}, we know that a \YY-path from a tile $\ell_r$ 
does not meet the legs of a triangle contained in~$T$. The role of a \YY-path from~$\tau$ is to 
convey the content of the square of the tape of~$M$ which lies in $\tau$. The \YY-path updates 
that content as soon as the apropriate state is seen, so that the \YY-path records the history of 
the computation on the square represented by its source. A computing signal $\xi$ starts from the 
root $\rho$ of $T$ and it visits the \YY-paths according to the program of~$M$. In order to go 
from one \YY-path to the next one, the $\xi$ travels on a level of~$T(\rho)$. That signal conveys 
the current state~$\eta$ of~$M$. When $\xi$ meets a \YY-path conveying the current content 
$\sigma$ of the square of the tape which is the source of that \YY-path, $\xi$ performs the 
instruction associated to $\eta$ and $\sigma$ in the program of~$M$. The \YY-path convey the new 
letter contained by the square at the source of the \YY-path. It also conveys the new state of~$M$ 
as well as the direction~$\delta$ towards the \YY-path whose source is a neighbour 
of the source from which the previous \YY-path originated. To that goal, $\xi$ goes to the next 
level along the \YY-path it met and, on that level, goes to the new \YY-path in the direction 
given by~$\delta$.

   As far as $T$ may contain other red triangles in which the same computation of $M$ is
performed, those computations should not interfere with each other. We already know that
the \YY-paths generated in $T$ do not meet those of a triangle inside~$T$. It is also
necessary that the levels on which $\xi$ travels in~$T$ are not those on which a similar
signal travels in a triangle contained in~$T$. Accordingly, we have to deal with that
point.

   Call {\bf free row} of a trilateral~$T$, the intersection with~$T$ of an orange or a green 
isocline which does not meet the legs of a triangle contained in~$T$. Note that the notion might be 
applied to red phantoms as well but we reserve it for red triangles. We deal with that problem in 
Subsection~\ref{ss_freerows}.

   We also notice from Figure~\ref{f_nproto_tri} that active seeds trigger both the construction 
of legs of a trilateral~$T$ and the construction of the basis of a trilateral whose status is 
opposite to that of~$T$. However, as indicated by Figure~\ref{f_ultra}, it may happen that the 
basis triggered by an active seed will not meet legs of an appropriate triangle. It is the case if 
the tree of the tiling raised by the active seed is the bound of a thread. That raises another 
problem dealt with in Subsection~\ref{ss_synchro}.
  
\subsection{Free rows in red triangles}\label{ss_freerows}

Before considering how to detect the free rows in a red triangle, it is important to know whether 
there are enough of them for the computation purpose.

\begin{lem}\label{free_rows}
In a red trilateral of the generation~$2n$$+$$1$ there are $2^{n+1}$$+$$1$ free rows. 
\end{lem}

\noindent
Proof. The smallest generation for a red triangle is generation~1 and such a triangle contains two 
green isoclines and one green one. Accordingly, such a triangle contains 3 free rows. From 
Lemma~\ref{inter_gene}, a red triangle~$T$ of the generation $2n$+3 contains two red triangles 
$T_0$ and $T_1$ of the generation~$2n$$-$1 with, in between them, a phantom of that generation 
which contains $\varphi_{2n-1}$ free rows. The height of $T$ is four times that of $T_0$. It is 
not difficult to see that the vertex of~$T$ is contained in a phantom~$P_0$ of the 
generation~$2n$$-$1 too and that the vertex of~$T$ is the mid-point of~$P_0$. Similarly, the basis 
of~$T$ is contained in the mid-line of a phantom~$P_1$ of the generation~$2n$$-$1 too. Accordingly,
\vskip 0pt
\ligne{\hfill $\varphi_{2n+1} -1 = 2(\varphi_{2n-1}-1)$
\hfill\hskip 20pt}

\noindent
which gives us
\vskip 0pt
\ligne{\hfill $\varphi_{2n+1} = 2\varphi_{2n-1}-1$
\hfill$(\ast)$\hskip 20pt}

\noindent
as far as the mid-line $T$ is counted twice if we consider $2\varphi_{2n-1}$. An easy induction from 
$(\ast)$ shows us that $\varphi_{2n+1}=2^{n+1}+1$. We again find that $\varphi_1=3$. We can check on
Figure~\ref{f_interwov_4}, see the Appendix, that $\varphi_3=5$.\hfill$\Box$

Note that a red triangle of the generation~$2n$+1 is crossed by $4^{2n+2}=8^{n+1}$ isoclines.
Accordingly, if the number of free rows of a red triangle of the generation~$2n$+1 is very small 
with respect to its height, it still tends to infinity as $n$ tends to infinity.

   It is worth noticing that if we choosed the blue triangles instead of the red ones in order to
simulate the computation of the Turing machine, using a similar definition for free rows with
the help of blue signals instead of the red ones, we would obtain that in each blue triangle 
there is a single free row, the mid-line of the triangle, see \cite{mmundecTCS} for the proof.
The reason is that generation~0 consists of blue trilaterals in which there is a single free row
while in a red triangle of generation~1 there are three free rows.

   Accordingly, it is worth dealing with the dectection of the free rows in red triangles. To that 
aim we proceed as follows. The legs of a red triangle~$T$ send a red signal inside~$T$ along a green
or an orange isocline. If the signal meets a leg of the same side and of the same colour, the signal
goes on.  It goes on too if it meets the legs of a blue triangle or the legs of a phantom, whichever
the colour. If the signal meets a leg of an opposite side, we prevent tiles to implement such a 
meeting. We also give the leg the possibility to send a yellow signal inside~$T$ along a such an 
isocline. If the signal meets a leg of an opposite side, it is the other leg of $T$ and the signal 
is established: a free row is detected. If the yellow signal meets a leg of the same side,it means 
that it must be replaced by a red signal as far as no tile implements the crossing of the yellow 
signal by a leg of a red triangle. However, the yellow signal may freely cross legs of blue 
triangles and of phantoms whatever their colour. Accordingly, each green or orange isocline 
inside~$T$ conveys a signal: a red one if on that isocline the signal meets the leg of a red 
triangle inside $T$, a yellow one if on that isocline the signal does not meet a red triangle 
inside~$T$. In Subsection~\ref{ss_tiles} we shall see how the problem is solved.

\subsection{Synchronisation}\label{ss_synchro}

   We already noted the problem of possible active seeds which are the origin of a bound for some
bounded thread.

  Another problem arises: as far as on the same green or orange isocline there might be several 
seeds, it is important that the red signals raised in a triangle occurs on the same isocline as a 
yellow signal inside another triangle. Call {\bf latitude} a finite set of consecutive green and 
orange isoclines. The {\bf latitude of a trilateral} is the set of green and orange isoclines from 
the isocline of its vertex to that of its basis. 

Note that the lateral red signals give rise to signals which may travel along an isocline far away 
from the legs of any triangle. Those signals of opposite laterality may meet in between two red 
triangles and outside them: in that case, a left-hand side signal coming from right meets a 
right-hand side signal coming from left. It is important that the latitude of a trilateral coincide 
with that of trilaterals of the same characteristics belonging to different threads. The red 
signals used for detecting the free rows are not enough for that property.

   To better see what is involved, we need the following notion. Consider a trilateral~$T$ of 
generation~$n$+1. If the vertex~$V$ of~$T$ is inside a triangle~$T_1$, we say that $T_1$ is the 
{\bf father} of~$T$. Note that a trilateral may have no father: it is the case in a wire defined 
by a bounded thread. If $T$ has a father of $T_1$ we may define the father of $T_1$ if that later 
exists. Accordingly, for any trilateral~$T$, we construct a sequence $\{T_k\}_{k\in [0..h]}$ 
such that $T_0=T$ and for each $k$ in $[0..h$$-$$1]$, $T_{k+1}$ is the father of~$T_k$ and $T_h$ 
has no father. Each $T_k$ with $k$ in $[0..h]$ is called an {\bf ancestor} of~$T$ and $T_h$ is 
called the {\bf remotest ancestor} of $T$. Note that the generation of the remotest ancestor of a 
trilateral~$T$ depends on the wire to which $T$ belongs.

We append two kinds of signals. We consider a special signal for blue trilaterals: the vertex of a
blue trilateral as well as the ends of its basis trigger a {\bf blue} signal, the same one whichever
the laterality of the end emitting it, whichever the attribute of the trilateral. Such a signal is 
important due to the fact that a thread may not cross the latitude of a given trilateral. Also, to 
distinguish latitudes of red trilaterals we need to mark the isoclines passing through the heads 
of red triangles. We call that latter mark the {\bf silver} signal. It is raised by the vertex~$V$ 
of a red triangle and it travels on the orange isocline which passes through~$V$.

The silver and the blue signals allow us to prove the following property: 

\begin{lem}\label{remote}
Let $T$ be a trilateral belonging to a wire $\mathcal W$. Let $S$ be a trilateral whose 
characteristics are those of~$T$, $S$ belonging to a wire~$\mathcal V$, with 
${\mathcal V}\not={\mathcal W}$. Then $T$ and $S$ has the same latitude if and only if $T$, $S$ 
have an ancestor $X$, $Y$ respectively, such that the vertex of~$X$ and that of~$Y$ lie on the same
isocline.
\end{lem}

   Note that the lemma mentions an ancestor within the same latitude and it says nothing of the 
remotest ancestors of~$T$ and~$S$ which may belong to different latitudes. As a consequence of the 
lemma, we can say that the latitudes of red triangles of the generation~$2n$$+$$1$ are the same 
whatever the wire giving rise to the interwoven triangles and for any $n\in\mathbb N$. 

\noindent
Proof of Lemma~\ref{remote}. We prove that property by induction on~$n$, for $n\geq 1$. Note that 
for any trilateral~$T$ of generation~0, its ancestors are~$T$ itself.
Consider a triangle~$T$ of generation~0{} in $\mathcal W$. Its vertex~$W$ sits on a green 
isocline $\omega$. If $\omega$ meets a seed~$V$ belonging to $\mathcal V$, $V$ receives the blue 
signal emitted by~$W$ as far as that signal cannot leave~$\omega$: that signal may be interrupted by
a basis lying on $\omega$, but that very basis restores the signal starting from its ends. 
Accordingly, $W$ triggers a trilateral~$S$ of generation~0. The colour of~$S$ is the same as that 
of~$T$, we remain with its attribute. What we have seen up to now shows us that the trilaterals of 
generation~0 lie within the same latitudes both for $\mathcal W$ and for $\mathcal V$. We have to 
see that the attributes are the same for the same lattitude. Let $A$ be the mid-point of~$T$. It 
triggers a trilateral~$Q$ of generation~1. If $Q$ is not a triangle, its basis defines a seed $B$ 
of~$\mathcal W$ which triggers a triangle $H$. It is not difficult to see that $B$ belongs to a 
triangle $J$ of generation~0 whose vertex is on the basis of the phantom whose vertex is on the 
basis of~$T$. Accordingly, we may assume that $Q$ is a red triangle. Now $Q$ emits a silver signal 
which travels on its orange isocline~$\varpi$ which meets a seed $C$ of $\mathcal V$. As far as 
isoclines cannot cut each other, the distance from $W$ to~$A$ is the same as the distance from $V$ 
to~$C$. And so, $C$ is the mid-point of~$S$ which raises a red triangle. Consequently, $S$ is a 
triangle too. We proved the lemma for generation~0. The argument of the silver signal to identify 
the attribute of~$S$ show us that the lemma is also true for generation~1.

Assume that the lemma is true for the generation~$n$. Consider a trilateral~$T$ of the 
generation~$n$+1 whose vertex is in $\mathcal W$. We may assume that $T$ is a triangle: if it were a
phantom $P$ we would consider as $T$ the triangle triggered by the seed of $\mathcal W$ lying on 
the basis of~$P$. Let $W$ be the vertex of~$T$ and let~$N$ be its mid-point. We know that on the 
same wire, there is a seed $Q$ which also triggers a triangle of the same colour as~$T$ and the 
distance of $Q$ from $W$ is twice the height of~$T$. Let $\omega$, $\varpi$ and $\varphi$ be the 
isoclines passing through~$W$, $N$ and $Q$ respectively. Let $V$, $M$ and $P$ be the seeds of 
$\mathcal V$ lying on the isoclines $\omega$, $\varpi$ and $\varphi$ respectively. The distance 
from~$V$ to $P$ is that from $W$ to~$Q$ which means that the trilateral issued from $V$ has the 
same height as $T$ so that its generation is the same and its colour is also the same. Also, $M$ 
is the mid-point of~$S$. Clearly, if $T$ is a red triangle, so is $S$ as far as the silver signal 
emitted by $W$ passes also through~$V$. If $T$ is blue, $N$ triggers a red trilateral and, arguing 
like we did for generation~1, we may assume that $N$ triggers a red triangle. Accordingly, $M$ also
triggers a red triangle as far as it receives the silver signal emitted by $N$. Now, a triangle is 
triggered at the mid-point of another triangle, so that $S$ is a triangle too as far $M$ is its 
mid-point. Consequently, we proved that for the generation~$n$+1 the latitudes are the same for
trilaterals with the same attributes, provided that the trilateral are both present in $\mathcal W$
and in $\mathcal V$.

From that, the lemma follows. If $S$ and $T$ have the same latitude, their attributes are the same
and their ancestors lie within the same latitudes as long as they are present in both wires. 
If $T$ and $S$ have an ancestor whose vertices are on the same isocline, clearly, those ancestors 
have the same latitude and, step by step, their successive sons occupy the same latitudes, so that 
it is the case for $S$ and~$T$ too.  \hfill$\Box$ 

Later on, we refer to Lemma~\ref{remote} when we say that the silver and the blue signals allow us 
to synchronise all wires of the tiling.

   The problem raised by possible bounds of threads is dealt with as follows. The blue signal 
emitted by a the basis of a trilateral of a wire may meet the basis emitted on the same isocline 
by a blue trilateral of another wire. Such a meeting is permitted: it solves the problem of 
possibly missing trialterals in a bounded thread.
\vskip 10pt
\vtop{
\ligne{\hfill
\includegraphics[scale=0.35]{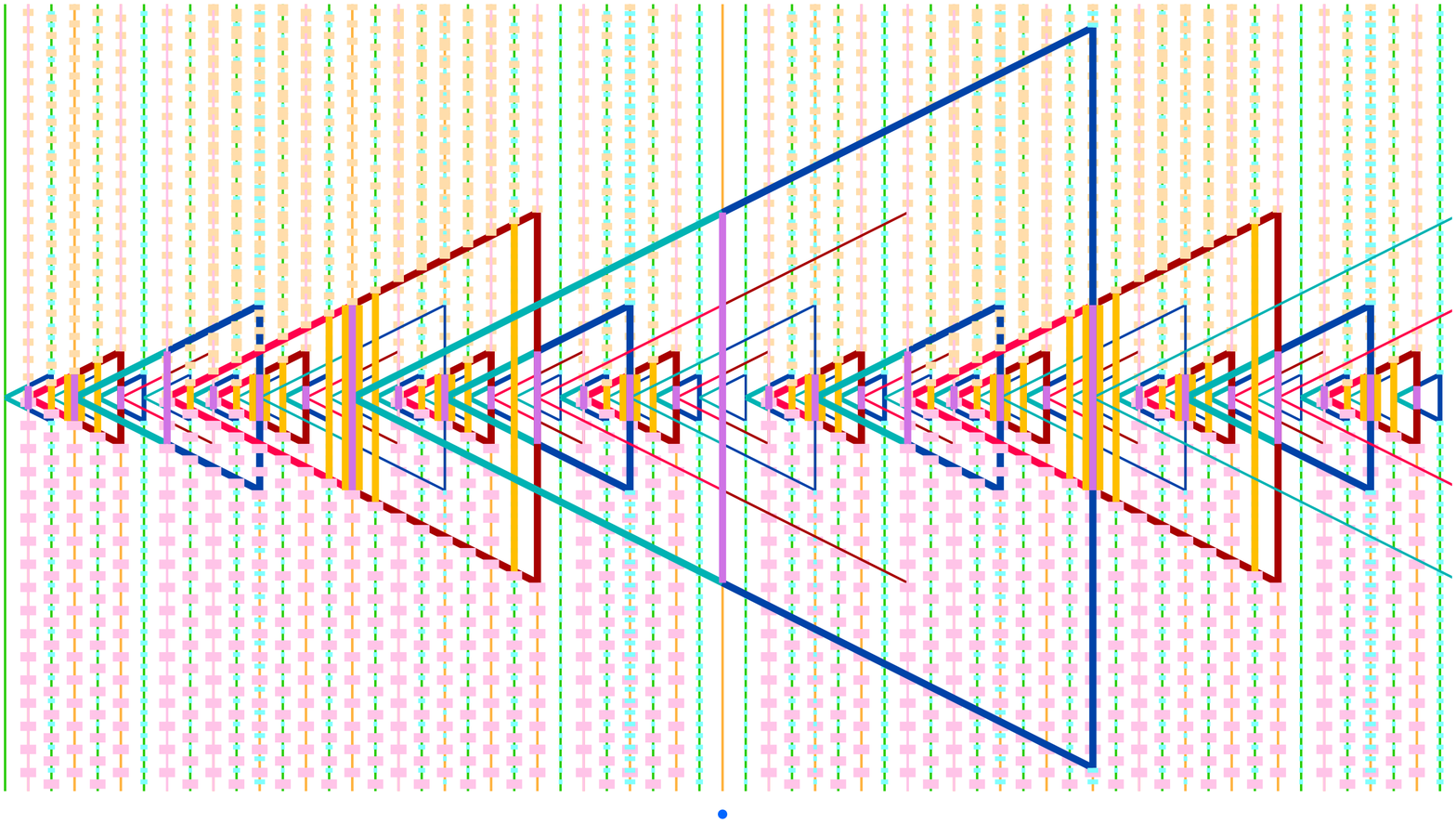}
\hfill}
\vspace{-10pt}
\begin{fig}\label{f_freerows}
\leurre
The free rows in the red triangles. They are in yellow in the figure. Note that the yellow signal is
supperposed with the mauve one on the mid-line of red triangles.
\end{fig}
}

From our description of the signals emitted by the legs of a triangle in order to detect free rows 
inside them, we can see that such signals must cross legs of the same laterality. It is the reason 
why we consider that instead, the legs of a red triangle emit a red signal of their laterality 
outside the triangle. As illustrated by Figure~\ref{f_freerows}, we can see that a red signal 
emitted by a red triangle $T_0$ included into another red triangle~$T_1$ also cuts the legs 
of~$T_1$. Those red signals are similar to the blue signals above defined for both trilaterals. 
The difference is here that they concern red triangles only and that they are not emitted by the 
vertex and the basis only: they are emitted on each green or orange isocline crossing the leg. We a
lso decide that right-, left-hand side red signals coming from left, from right respectively may 
match with a red basis coming from right, from left respectively.

Note that appending the silver signal means just changing a bit the tiles conveying an orange 
isocline, but it also requires to append five tiles as far as there are tiles outside legs and bases
of trilaterals which convey an orange isocline with no signal at all.

We remain with the condition meant by the {\it general tiling problem}. We borrow the next 
subsection to paper~\cite{mmundecTCS} with a few changes.

\subsection{The general tiling problem}\label{ss_til_issue}

In the proofs of the general tiling problem in the Euclidean plane 
by Berger and by Robinson, there is an assumption which is implicit and 
which was, most probably, considered as obvious at that time.

   Consider a finite set~$S$ of {\bf prototiles}. We call {\bf solution} of the
tiling of the plane by~$S$ a partition~$\mathcal P$ such that the closure of any
element of~$\mathcal P$ is a copy of an element of~$S$. We notice that the
definition contains the traditional condition on matching signs in the case when
the elements of~$S$ possess signs. We also notice that a copy means an
isometric image. In that problem, we assume that only shifts are allowed and we
exclude rotations. Note that, in the Euclidean case, rotations are also ruled out. 
Here rotations have to be explicitly ruled out as far as the shifts leaving the 
tiling globally invariant also generate the rotations which leave the tiling globally 
invariant. In fact we accept isometries and only those such that a tile marked \ww{} or \bb{} on a 
given isocline is transformed into a tile marked \ww{} or \bb{} respectively on an isocline, the 
same one or another one.

   Note that the general tiling problem can be formalized as follows:
\vskip 3pt
\ligne{\hfill$\forall S\ \ (\exists\,{\cal P}\ sol({\cal P},S)\vee%
              \neg(\exists\,{\cal P}\ sol({\cal P},S))),$
\hfill}
\vskip 3pt
\noindent
where $sol({\cal P},S)$ means that $\cal P$~is a solution of~$S$ and where
$\vee$~is interpreted in a constructive way: there is an algorithm
which, applied to~$S$ provides us with~'yes' if there is a solution and~'no'
if there is none.

   The origin-constrained problem can be formalized in a similar way by:
\vskip 3pt
\ligne{\hfill$\forall (S,a)\ \ (\exists\,{\cal P}\ sol({\cal P},S,a)\vee%
              \neg(\exists\,{\cal P}\ sol({\cal P},S,a))),$
\hfill}
\vskip 3pt
\noindent
where $a\in S$, with the same algorithmic interpretation of~$\vee$ and where the formula 
$sol({\cal P},S,a)$ means that $\cal P$~is a solution of~$S$ which starts with~$a$. Note that if 
$a$ is a blocking tile, {\it i.e.} a tile which cannot abut any one in $S$, then we may face a 
situation where we cannot tile the plane because $a$ was chosen at random while it is possible to 
tile the plane. A solution is to exclude $a$ from the choice. The other one is to allow the 
occurrence of contradictions because a wrong tile was chosen while the appropriate one would raise 
no contradiction. Of course, there must be a restriction: such a change should occur finitely many 
times at most for the same place.

Obviously, if we have a solution of the general tiling problem for the considered instance, we 
also have a solution of the origin-constrained problem, with the facility that we may choose the 
first tile. To prove that the general tiling problem, in the considered instance, has no solution, 
we have to prove that, whatever the initial tile, except the blocking one, the corresponding 
origin-constrained problem has no solution either.

   In Berger's and Robinson's proofs the construction starts with a special tile, called the 
{\bf origin}. Their proof holds for the general problem as far as they force the tiling to have a 
dense subset of origins. In the construction, the origins start the simulation of the space-time i
diagram of the computation of a Turing machine~$M$.  All origins compute the same machine~$M$ with i
the same initial configuration of~$M$. The origins define infinitely many domains of computation 
of infinitely many sizes. If the machine does not halt, starting from an origin, it is possible to 
tile the plane. If the machine halts, whatever the initial tile, we nearby find an origin and, from
this one, we shall eventually fall into a domain which contains the halting of the machine: at that 
point, it is easy to prevent the tiling to go on.

   The present construction aims at the same goal.
   
   From Proposition~\ref{tri_indices}, we know that the trilaterals are bigger and bigger
once their generation is triggered along a wire. Consequently, what we suggested with
the \YY-paths and the free rows answer positively the possibility to simulate any Turing
machine working on a semi-infinite tape which, as well known, does not alter the 
generality. We remain with the way to force such computations.

\subsection{The seeds}\label{ss_sides}

We establish that there are enough seeds for starting the computation of a Turing machine in the 
interwoven triangles. 

We have the important property:

\begin{lem}
\label{iso_seeds}
Let the root of a tree of the tiling~$T$ be on a green or an orange isocline. Then, there is a seed
in the tiles of~$T$ on the next orange or green isocline respectively, downwards. Starting from that
last isocline, there are seeds, downwards, on all the isoclines.
\end{lem}

   We have a very important density property:

\begin{lem}\label{density}
For any tile~$\tau$ in a tiling, fitted with the isoclines, there is a seed on a green or an orange
isocline within a ball around~$\tau$ of radius~$8$.
\end{lem}

\noindent
Proof.  From Figure~\ref{f_proto} we can see that if we consider a \GG-tile $\tau$, there
is a seed at a distance~2 from~$\tau$. As far as a \YY-tile has a \GG-son, there is a
seed at a distance at most~3 from a \YY-tile. By construction, there is a seed at 
distance~1 from an \MM-tile. Figure~\ref{f_proto} shows us that there is a \GG-tile at 
distance~1 from a \BB-tile. Accordingly, there is a seed at distance at most~3 from a 
\BB-tile. There is a \YY-tile at distance~1 from an \OO-tile so that there is a seed at 
distance at most 4 from an \OO-tile. Accordingly, there is another seed at distance at 
most~4 from a seed. If the seed found in that way is on a green or an orange isocline, four 
isoclines further there is a seed on a green or on an orange isocline at distance at most 8.
\hfill$\Box$

   From Lemma~\ref{density}, we know that around any tile~$\tau$ there is a seed in a 
disc of radius~4. We can say a bit more:

\begin{lem}\label{all_iso}
Assume that there is a seed on a green isocline. Then, there is at least a seed on the next green
isocline and on each further isocline whichever its colour.
\end{lem}

\noindent
Proof. Let $\tau$ be a seed on an isocline. From Lemma~\ref{density} it can be easily found 
everywhere. Say that the isocline to which $\tau$ belongs is isocline~0. The sons of~$\tau$, say
$\tau_i$, $i\in\{1..3\}$ are not seed and none of them is \MM. Accordingly there is no 
seed on the levels~1 and~2 of~$T(\tau)$. On the level~2 the statuses of the sons of the
$\tau_i$, $i\in\{1..3\}$, are \YY, \BB, \GG{} or \OO. Accordingly, there is no seed
on the level~3 but as far as there is a tile~\MM{} on that level, there is a seed on 
level~4. But a \GG-tile always has a \GG-son so that if the \GG-tile on level~2 raises 
a seed on level~4 the \GG-descendants of that tile generate a seed at each level~$n$ with
$n>4$. At least one of those isoclines is green so that we can repeat the argument.
That completes the proof of the lemma. \hfill$\Box$.

From now on, a seed on a green or an orange isocline is called an {\bf active seed}.

In each red triangle, we define a limited grid in which we simulate the execution of the same Turing
machine starting from the same initial finite configuration. Of course, the whole initial 
configuration occurs in a big enough red triangle. If the configuration is not complete in a red 
triangle, the computation halts on the basis of the red triangle. Accordingly as the red triangles 
are bigger and bigger, if the machine does not stop, it is possible to tile the plane. If the 
machine halts, the halting produces a tile which prevents the tiling to be completed. As far as i
the halting problem of Turing machines starting from a finite initial configuration is undecidable, 
that reduction proves that the tiling problem of the hyperbolic plane is also undecidable.


    As far as we know that there are enough active seeds, we have to look at how behave the 
triangles constructed from them.

   The construction performed in Section~\ref{s_aperiod} required the realization of the interwoven
triangles starting from at least one wire as far as that alone entails the construction of bigger i
and bigger triangles which are disjoint from each other. To prove Theorem~\ref{undec} we need that 
the computation is performed more or less similarly in the triangles of each wire. But that 
property is guaranteed by the synchronisation property of the silver and blue signals, as proved 
by Lemma~\ref{remote}.

\subsubsection{The implementation}\label{ss_implem}

   As can immediately be seen, the important feature is not that we have strictly parallel lines, 
and that squares are aligned along lines which are perpendicular to the tapes. What is important 
is that we have a {\bf grid}, which may be a more or less distorted image of the just described 
representation.

We can reinforce Lemma~\ref{density}:

\begin{lem}\label{superdensity}
For each tile~$\tau$, there is an active seed whose distance from~$\tau$ is at most~$12$.
\end{lem}	

\noindent
Proof. In the worst case, assume that $\tau$ is a \BB-tile on an isocline $m$. The
\OO-son of $\tau$ has a \YY-son so, that at distance at most 4 there is a seed~$\rho$
say on the isocline~$n$. As far as there are seeds on levels $n$+4+$k$ for any 
$k\in\mathbb N$, there is at least an active seen on some level $n$+$j$ with $j\leq 8$.
Now, clearly, $n\leq m$+4. That proves the lemma. \hfill$\Box$

On Figure~\ref{f_ultra}, we can see two active seeds and several seeds which are not active. 
Accordingly, most seeds are not active but, the active ones are also dense in the heptagrid. It 
means that if we start to tile the heptagrid from an arbitrary tile, the blocking one being 
excepted, later or sooner we fall upon an active seed. We go back to that topic a bit later and also
when we shall discuss the exact set of prototiles.

   As already mentioned, the legs issued from an active seed~$\sigma$ follow the borders of 
$T(\sigma)$. Note that the active seeds also send signals on the green and the orange isoclines.

  What is important is the thread-structure and Lemma~\ref{inter_gene}. Note that the silver and the
blue signals prevent the occurrence of two active seeds on the same isocline giving rise to 
trilaterals of different characteristics. 

\ifnum 1=0 {
We now turn to Subsection~\ref{ss_freerows} in order to solve the problem of the free rows.
   
\subsection{Detecting the free rows}\label{ss_freerows}

   We remain with a problem about the detection of the free rows.

   We already defined the principle of that detection which gurantees that the free rows lie on the
same isoclines for red triangles of the same latitude. We have simply to define the prototiles which
implement that principle.

   The set of prototiles defined by Figures~\ref{f_nproto_a_i} and~\ref{f_nproto_tri} guarantees the 
definition of green and orange isoclines for the whole heptagrid in such a way that we can number the 
isoclines with integers and in such a way that the green and orange isoclines receive numbers $4n$, 
$n\in\mathbb Z$. The silver signal allows us to synchronise the generation of triangles and to obtain the 
alternation of triangles and phantoms among the trilaterals of the same generation and the altarnation of
colours between the generations: even ones are blue, odd ones are red.

We complete the prototiles of Subsection~\ref{ss_aperiod} in order to signalise
the free rows. When that will be done, we shall turn to the definition of the prototiles
for the computation of the simulated Turing machine.
} \fi

\subsection{The tiles}\label{ss_tiles}

In this sub-section, we shall describe as precisely as possible the tiles
needed for the constructions defined in the previous sub-sections. The
description is split into two parts. 

We revisit the prototiles defined in Subsection~\ref{ss_til_aperiod} with 
Figures~\ref{f_nproto_a_i} and~\ref{f_nproto_tri}. Indeed, we have to implement the 
detection of the free rows, the construction of the red and of the blue signals and 
then the travel of the computing signal $\xi$. That latter is tightly connected with 
the program of the simulated Turing machine~$M$ so that the related prototiles should 
be better called {\bf meta-tiles} as far as they bear variable signs for the content of 
a square of the tape of~$M$, for the state of~$M$ and for the direction~$\delta$ which 
has to be followed in order to meet the next \YY-path. The detection of the free rows 
and the construction of red and blue signals are defined by Sub subsection~\ref{sss_proto}. 
The management of the signal~$\xi$ is performed in Sub subsection~\ref{sss_meta}.

\subsubsection{The proto-tiles}
\label{sss_proto}

   With the silver signal, we fix the implementation of the triangles and 
of the phantom. The actual place of the generation~$n$+1 is fixed by the first choice of
an active seed which is in a free green isocline. If the active seed triggers a triangle,
a phantom, the active seeds of the basis of the trilateral trigger a phantom, a triangle
respectively. Whence the expected alternation which the whole construction is based upon.

   The set of tiles we turn to now is called the set of 
{\bf prototiles}. We subdivide the set into two parts: the construction of the isoclines
and the construction of the trilaterals. A prototile is a pattern. Indeed, a tile 
is the indication of two data: the location of a tile in the heptagrid and the 
indication of a copy of the prototile which is placed at that location. The mark of the 
isoclines indicates which shifts are allowed: from a tile on an isocline to another tile
on another isocline, provided that the marks of the image match with those of the
new isocline. 
 
\vskip 10pt
\ligne{\hfill
\vtop{\leftskip 0pt\parindent 0pt\hsize=300pt
\ligne{\hskip-15pt
\includegraphics[scale=0.3]{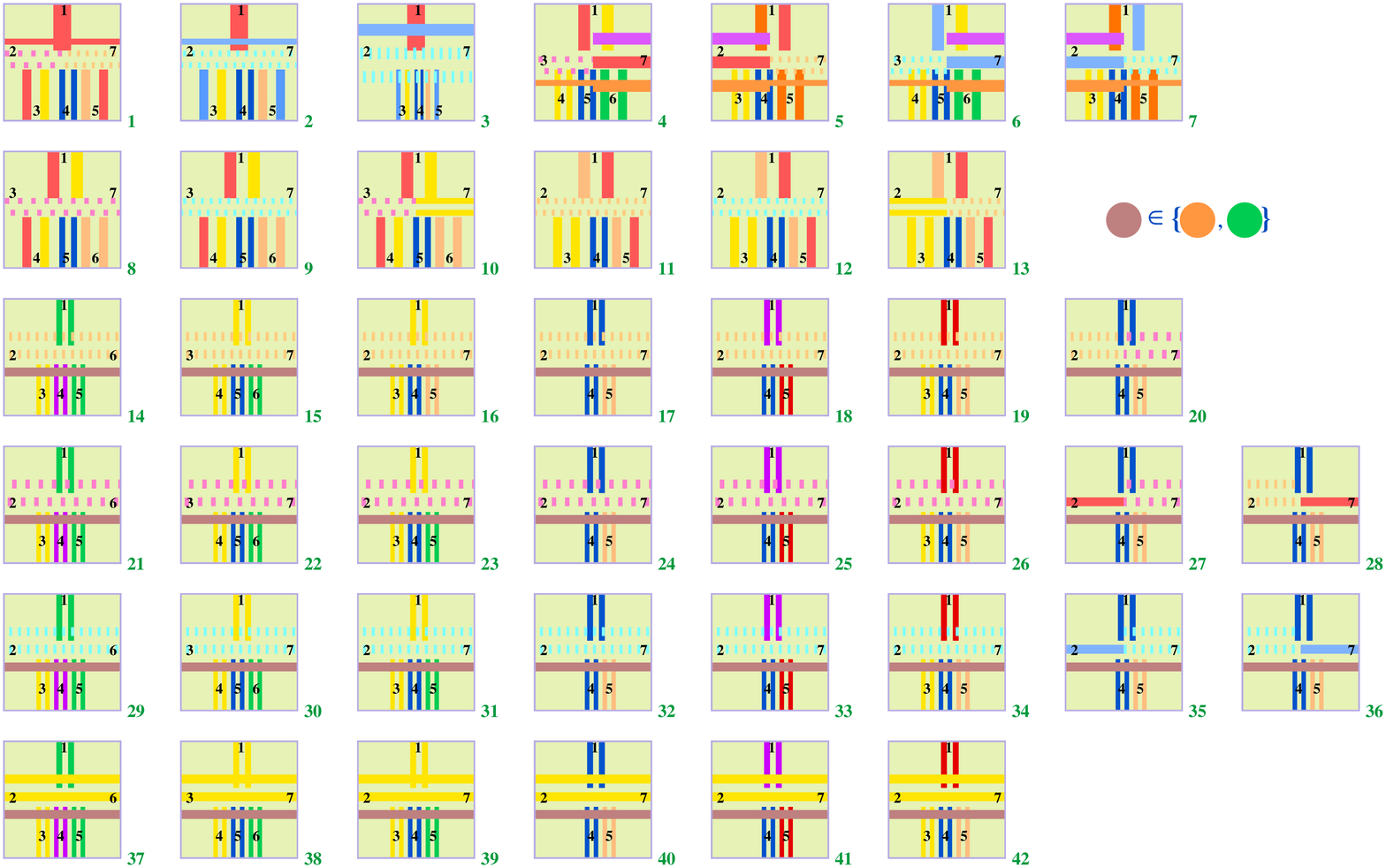}
\hfill}
\vspace{-10pt}
\begin{fig}\label{f_proto_frow}
\leurre
Generation of the red and the yellow signals by the legs of the triangle. Note that
there are two red signals: one for the left-hand side legs and the other for the 
right-hand side ones. Taking into account the possible colours of the isocline, we
get $75$ tiles to be appended for those from Figures~{\rm\ref{f_nproto_a_i}}
and {\rm\ref{f_nproto_tri}}.
\end{fig}
}
\hfill}

   The set of prototiles forces the construction of the tiling with
the isoclines. It also forces the activation of the seeds and the 
consecutive construction of the embedded triangles including the detection
of the free rows in the triangles. We have already two figures to illustrate the 
prototiles: Figures~\ref{f_nproto_a_i} and~\ref{f_nproto_tri}.
Each figure defines marks on the tiles for the construction of the tiling, of the 
isoclines, of the triangles and of the phantoms respectively. We append to
those figures two new ones in order to introduce the new signals we defined: the red and
the yellow ones which allow us to locate the free rows.

Figure~\ref{f_proto_frow} illustrates the generation of the
yellow and the red signals by the legs of a triangle. The figure makes use of meta-tiles
in that meaning that the light mauve colour indicating the isocline can be freely 
replaced either by the orange colour or by the green one depending on which isocline we consider:
remember that the red, blue and yellow signals run on green or orange isoclines only.
Accordingly, that colour represents a variable for colours of the isoclines. 
A part of the tiles of Figure~\ref{f_proto_frow} are already present in Figure~\ref{f_nproto_tri}:
the new red, blue or yellow signs are appended to those of Figure~\ref{f_nproto_tri} for 20 of them.

\ifnum 1=0 {
\vskip 10pt
\ligne{\hfill
\vtop{\leftskip 0pt\parindent 0pt\hsize=300pt
\ligne{\hfill
\includegraphics[scale=0.4]{nov_prototuiles_freerowsII.ps}
\hfill}
\begin{fig}\label{f_proto_frowII}
\leurre
The prototiles conveying the red signals outside the legs of a triangle both inside and
outside the triangle. Note the use of meta-colours for the isoclines and for the signals.
The light blue colour represents both forms of the red signal. The light purple colour
represents the two red signals and the yellow signal. The last picture of the third line 
illustrates the join tiles. We also append the prototiles for conveying the silver 
signal. Taking into account that the isoclines may have two colours, we get $67$ 
prototiles. 
\end{fig}
}
\hfill}
} \fi

Note that the tiles allowing red and blue signals to meet are attached to \BB-tiles: there are 
enough of them on each isocline. The distance between two consecutive \BB-tiles on an isocline is 
at most 5, as can be checked on Figure~\ref{f_ultra}.

From that remark and summing the prototiles defined in Figures~\ref{f_nproto_a_i}, 
\ref{f_nproto_tri} and \ref{f_proto_frow}, we get 232 prototiles.

\ifnum 1=0 {
A last remark is about the vertex of a triangle. From Figure~\ref{f_nproto_tri}, we
can see that the the tile {\bf Rtg0} does not send a red signal but it sends a green 
signal and a silver one together. The silver signal allows us to synchronise the 
construction of the trilaterals as already mentioned and with respect to free rows,
that signal should be considered as a red one.


The figure indicates the prototiles for the tiles outside the legs and the basis of a
triangle, both inside and outside the triangle. Note that we get 67 prototiles. Among 
them, note that we have two join tiles, one for each kind of isoclines. The join tile
of Figure~\ref{f_proto_frowII} allows a right-hand side signal coming from the left to
join a left-hand side signal coming from the right. It means that the red signals emitted
by the legs of two triangles lying on the same isoclines can join outside the
triangles. Moreover, we decided to permit that joining on a \BB-tile only, wich reduces
the number of prototiles but it does not alter the generality of the construction: in 
between legs of such triangles there are always \BB-tiles: the vertex~$\nu$ of a 
triangle is the \RR-son of an \MM-tile which has a \BB-son $\beta$. The \BB-son of 
$\beta$ is on  an isocline which comes just after that of~$\nu$. The figure also
displays the ten prototiles required to convey the silver signal, of course on a green
isocline.

Summing all the numbers obtained in each figure among Figures~\ref{f_proto_a_i}, 
\ref{f_nproto_tri}, \ref{f_proto_frowI} and~\ref{f_proto_frowII}, we obtain:
} \fi

\begin{lem}\label{proto_count} 
There is a set of $232$ prototiles which allow us to construct a tiling of the heptagrid
implementing the embedded triangles with their isoclines together with the detection of 
the free rows in each triangle, the lattitudes of trilaterals with identical attributes being
synchronised.
\end{lem}

Note that the number of free rows in a trilateral is that of Lemma~\ref{free_rows}
as far as the basis of a triangle is not signalised as a free row.

In the appendix, several figures illustrate the construction of the tiling by focusing
each one on one of the pictures belonging to Figure~\ref{f_proto}.

\def\oo{{\bf o}}
\def\yy{{\bf y}}
\def\gg{{\bf g}}
\def\mm{{\bf m}}
\vskip 10pt
\noindent

\subsubsection{The meta-tiles}
\label{sss_meta}

   Let us now examine the construction of prototiles for simulating a Turing machine.
As already mentioned,that part of the prototiles depends upon the Turing machine~$M$ which
is simulated. It also depends on the data $\mathcal D$ to which $M$ is applied. Of course, it would 
be possible to consider Turing machines starting from an empty tape. The consequence would be a huge
complexification of the machine which would store the data into its states. It is simpler to 
consider that $M$ applies to a true data. It is the reason while we call these tiles 
{\bf meta-tiles}.

   As already mentioned, the simulation of the computation of~$M$ is organised in the red
triangles, starting from generation~0. The interest of those infinitely many generations is the fact
that as far as the number $n$ of the generation increases, the number of free rows in the 
corresponding triangles also increases, which gives the tiling more time in the simulation of~$M$. 
Also note that the space also increases as far as the height of a red triangle of the 
generation~$2n$+1 is $8^{n+1}$ according to Lemma~\ref{tri_indices}. As already mentioned, the 
initial configuration is displayed along the rightmost branch of a red triangle~$T$ which, outside 
the head of~$T$, consists of \OO-tiles. A tree of the heptagrid rooted at a tile on the rightmost 
branch of~$T$ has its leftmost branch constituted of \YY-tiles. Now, from Lemma~\ref{par_trees}, 
that borders does not meet the legs of a triangle inside~$T$. Accordingly, the computation 
signal~$\xi$ travels on \YY-tiles only when it goes from a free row of~$T$ to the next one and it 
crosses consecutive tiles when it travels on a free row of~$T$.

   The meta-tiles are illustrated by Figure~\ref{f_nmeta}, where the caption explains the 
meaning of the colours.

\vskip 0pt
\ligne{\hfill
\vtop{\leftskip 0pt\parindent 0pt\hsize=320pt
\ligne{\hskip-10pt
\includegraphics[scale=0.3]{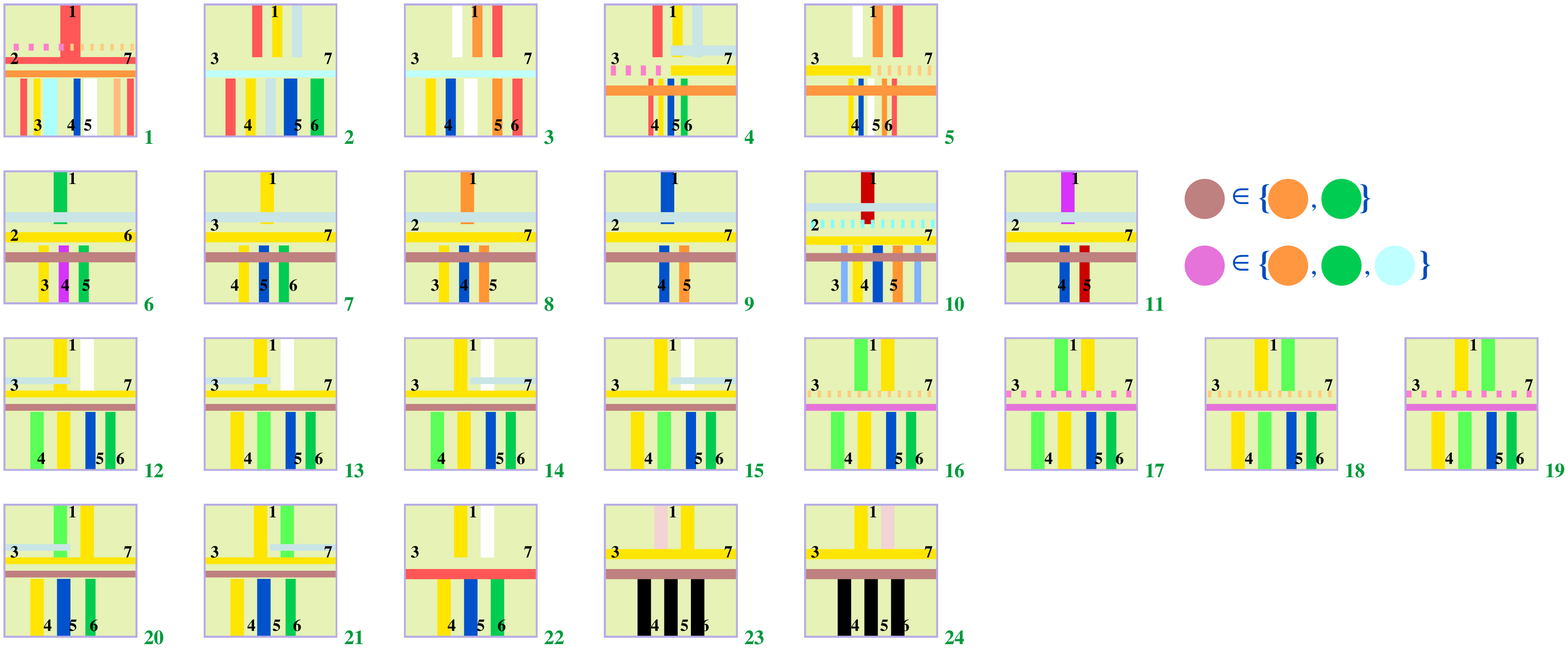}
\hfill}
\vspace{-15pt}
\begin{fig}\label{f_nmeta}
\leurre
The meta-tiles for simulating the execution of the Turing machine~$M$. Tile~1
represents the active seed for a triangle. Tiles~2 up to~5 allow the signals emitted for the
computation to go down until the first orange or green isocline reached by tiles~3 and~4 for the
left- and right-hand side legs repsectively. Tiles~6 up to~11 allow the computing signal to
travel on a free row of the triangle. Tiles 12 up to~15 illustrate the execution of an instruction.
Tiles 16 up to~19 represent the crossing by the \YY-path of the isoclines covered by a red
lateral signal. Tiles 20 and~21 illustrate the junction of the \YY-path with a free row, allowing 
the computing signal to go towards the next \YY-path. Tile~22 illustrates the case when the 
computing signal reaches the basis of the triangle which interrupts the computation. Tiles~23
and~24 indicate that the computation by the Turing machine halts. Those tiles are unique and
cannot abut any other tile of the tiling.
\end{fig}	
}
\hfill}
\vskip 10pt

A few supplementary explanations are needed. 

Meta-tile~\ref{f_nmeta}.1 sends a white signal to the right-hand side leg of the triangle it 
generates. It triggers tile \ref{f_nmeta}.3 which is on that leg in order to represent the squares 
of the Turing tape of~$M$. Meta-tiles \ref{f_nmeta}.6 up to \ref{f_nmeta}.11 illustrate the travel 
of the current state of~$M$ on a free row. Note that if a seed occurs on that isocline, it must be 
active and it is either meta-tile~\ref{f_nmeta}.1 or \ref{f_nmeta}.10 depending on whether the 
tile triggers a red triangle or a blue trilateral respectively. Meta-tiles~\ref{f_nmeta}.12 up 
to \ref{f_nmeta}.15 illustrate the execution of an instruction: the current state travels on the 
free row going to left or to right, meta-tiles~\ref{f_nmeta}.14, \ref{f_nmeta}.15 or 
\ref{f_nmeta}.12, \ref{f_nmeta}.13 respectively. The difference is seen on the \YY-son: if the new 
state goes to left, to right, the mark is put to left, to right respectively of the yellow mark of 
the \YY-son. Meta-tiles \ref{f_nmeta}.16 up to \ref{f_nmeta}.19 allow the new signal following the 
\YY-path to cross non free rows. When it reaches the free row, the new current states goes to left, 
to right, meta-tiles~\ref{f_nmeta}.20, \ref{f_nmeta}.21 respectively depending on the side from 
which the current state came along the \YY-path. Meta-tile~\ref{f_nmeta}.22 is used when the current
state reaches the basis of the red triangle: the computation is stopped as far as there is not 
enough free rows in that triangle for the computation of~$M$. Meta-tiles~\ref{f_nmeta}.23
and \ref{f_nmeta}.24 are used when the current state is the halting state: when it meets the free 
row, the halting of the computation of~$M$ is implemented. Note that those latter tiles cannot 
abut any other tiles in those we defined and cannot abut each other or each one with itself.
And so, we can see that the computation in a triangle always stops. Either it happens by the meeting
of the computing signal along a \YY-path with the basis of the triangle, or it happens by the
halting of~$M$ itself. Meta-tiles \ref{f_nmeta}.22 up to \ref{f_nmeta}.24 illustrate those 
situations.

The number of meta-tiles depends upon the number~$s$ of states and the number~$\ell$ of 
letters of~$M$. From Figure~\ref{f_nmeta}, we can made the following counting:

{\leftskip 30pt\parindent 0pt
in Figure~\ref{f_nmeta}, tiles:
1 $-$ 5: 5 tiles;\\
6 $-$ 11: 12$\times\ell$ tiles: two possible isoclines and $\ell$ possible states;\\
12 $-$ 15: 2$\times I$ tiles; $I$: number of instructions of $M$;\\
16 $-$ 19: 3$\times I$ tiles: the four tiles together, three possible isoclines and 
$I_\ell$ possible instructions;\\
20 $-$ 21:  2$\times I$ tiles; two possible isoclines, $I_\ell$ possible instructions;\\
22 $-$ 24: 3 tiles.
\par}

\noindent
where $I$ is the number of instructions of the program of~$M$, $\ell$ is the number of states and 
$s$ is the number of letters. Also $D$ is the length of the data written with letters of~$M$.

Accordingly, the total number of meta-tiles is \hbox{$7.I$+$12.\ell$+$D$+8}.

Combining that result with the previous countings we get:

\begin{lem}\label{countundec}
For each Turing machine $M$ with $s$ states and $\ell$ letters whose program contains
$I$ instructions exactly, and whose data requires $D$ letters, there is a set 
${\mathcal P}_{M,\mathcal D}$ of \hbox{$240$$+$$7.I$$+$$12.\ell$$+$$D$} prototiles, so that the 
tiling problem is undecidable for the set of all ${\mathcal P}_{M,\mathcal D}$ applied to data 
written with letters of~$M$.
\end{lem}

That completes the proof of Theorem~\ref{undec}.

\section{A few corollaries for connected tiling problems}\label{s_corol}

   For the convenience of the reader, that section reproduces the similar section of
\cite{mmarXiv22}.

   The construction leading to the proof of Theorem~\ref{undec} allows to get
a few results in the same line of problems.

   As indicated in \cite{goodmana,goodmanb}, there is a connection between 
the general tiling problem and the {\bf Heesch number} of a tiling. That
number is defined as the maximum number of {\bf coronas} of a disc which 
can be formed with the tiles of a given set of tiles, see \cite{mann} for 
more information. As indicated in \cite{goodmanb}, and as our construction 
fits in the case of domino tilings, we have the following corollary of 
Theorem~\ref{undec}.

\begin{thm}\label{heesch}
There is no computable function which bounds the Heesch number for the tilings
of the hyperbolic plane.
\end{thm}

   The construction of~\cite{mmarXiv1} gives the following result,
see \cite{mmarXiv3,mmrp07}.

\begin{thm}\label{finite}
The finite tiling problem is undecidable for the hyperbolic plane.
\end{thm}

   Indeed, when the Turing machine halts, the halting state triggers a signal
which encloses the computation area and which compels the tiling to be completed
by blank tiles only, see~\cite{mmrp07}. 

   Combining the construction proving Theorem~\ref{finite} and the partition
theorem which is proved in~\cite{mmbook1}, chapter~4, section~4.5.2 about
the splitting of Fibonacci patchworks, also see~\cite{mmJCA}, the construction 
of this paper allows us to establish the following result, see~\cite{mmarXiv4}.

\begin{thm}\label{periodictiling}
The periodic tiling problem is undecidable for the hyperbolic plane, also
in its domino version.
\end{thm}

   Note that the analog of Theorem~\ref{periodictiling} for the Euclidean plane
was proved by Gurevich and Koriakov, see \cite{gurevich}. 

   In the statement of Theorem~\ref{periodictiling}, {\bf periodic} means that 
there is a shift which leaves the tiling globally invariant. The construction
mimics that of Theorem~\ref{finite} in the fact that if the simulated Turing
machine halts, we also enclose the computing area. But this time, we enlarge
the notion of computing area and of triangles so as to also permits 
black trees to support embedded triangles. In this way, we can
define areas of the kind defined by Fibonacci patchworks and of the size 
dictated by the halting of the machine. We define colours for these surrounding
signals in such a way that they entail a construction of a scaled Fibonacci
tree, see~\cite{mmJCAd}. Next, it is not difficult to construct a tiling of the hyperbolic plane 
in this way, periodically, applying the shifts already mentioned in~\cite{mmbook1}, chapter~4, 
section~4.5.3, also see~\cite{mmJCA}. 
\vskip 7pt
   At last, in another direction, we may apply the arguments of Hanf and Myers,
see~\cite{hanf,myers}, and prove the following result. 

\begin{thm}\label{nonrec}
There is a finite set of tiles such that it generates only non-recursive tilings
of the hyperbolic plane.
\end{thm}

   The proof makes use of the construction of two incomparable recursively
enumerable sets~$A$ and~$B$ of integers. The set of tiles defines the 
computation of these sets by a Turing machine. Moreover, the set of tiles 
tiles the plane if and only if there is a set to separate~$A$ from~$B$.
As such a set cannot be constructed by an algorithm, we obtain the
result stated in Theorem~\ref{nonrec}.

\subsection{Conclusion}

   It seems to me that the construction of Section~\ref{s_aperiod} could be applied to
prove undecidability problems on cellular automata. Of course, the
halting problem for cellular automata is undecidable, but this is a simple
consequence of the undecidability of the same problem for Turing machines.

   In fact, it is interesting to notice that the construction of Section~\ref{s_aperiod} 
which is based on Construction~\ref{c_interwov} can be performed by a cellular automaton.

   The working of the automaton could be devised as follows.

   We consider that the automaton works on three layers. On the first one,
it tries to construct the tiling. The initial configuration of this layer
is a blank plane, except at a tile called central, which is an active
seed. On the second layer, the cellular automaton updates a ball around
the central tile which coincides with that of the first layer. The third
layer is a `working sheet' for intermediate computations performed by the
automaton.

   It is plain that if the Turing machine implemented in the set of tiles
does not halt, the cellular automaton will tile the plane in infinite time.
If the Turing machine halts, the cellular automaton will take notice that
the construction is stopped at some point.
\vskip 7pt
   A last consequence of the construction of Section~\ref{s_aperiod} leads us back to
hyperbolic geometry.

   We used Figure \ref{f_interwov}, in order to better understand 
Construction~\ref{c_interwov}. It is worth noticing that both figures are indeed
Euclidean constructions. However, Construction~\ref{c_interwov} proceeds in a hyperbolic 
tiling.  It seems to me that the fact that this transfer is possible has an important 
meaning. From my humble point of view, it means that a construction which seems to be 
purely Euclidean has indeed a purely combinatoric character. It belongs to absolute 
geometry and it mainly requires Archimedes' axiom. Note that absolute geometry itself 
has no pure model. A realisation is necessarily either Euclidean or hyperbolic. 
To conclude with it, we suggest that probably the extent of absolute geometry is
somehow under-estimated.

    As indicated in the Introduction, the construction of the paper is inspired by the
construction of the paper I wrote in 2007 to prove Theorem~\ref{undec}. However, and it 
was the main goal of the present paper, the number of needed prototiles is significantly reduced.

From Lemma~\ref{countundec}, simulating a Turing machine~$M$ whose program contains $I$ 
instructions for $s$ states and $\ell$ and which letters applied to a data $\mathcal D$ of length 
$D$ with a tiling, \hbox{$240$$+$$7.I$$+$$12.\ell$$+$$D$} prototiles are needed in our simulation.
In contrast, paper \cite{mmarXiv22} required \hbox{556+20.$I$+136.$s$+12.$\ell$+$D$} for the same 
goal while my 2007 paper~\cite{mmundecTCS} required \hbox{$18870$+$880.s$+$1852.\ell$+$192.I$$+$$D$}
prototiles. Note that the importance of the parameters involved in those formulas seriously 
depends on~$M$ and on its data. For the same Turing machine $M$ there are infinitely many possible 
data so that $D$ is the single variable. If we consider tiny universal Turing machines, 
see~\cite{neary_woods} for example, $D$ is enormous in comparison with~$I$. As far for a single 
Turing machine there are infinitely many possible data, it makes sense to focus our attention on 
the program of~$M$. If we apply those formulas to the universal Turing machine with 6 states and 4 
letters from~\cite{neary_woods} we get 449 prototiles with the present paper, while 1884 prototiles 
are required in \cite{mmarXiv22} and 35782 of them for \cite{mmundecTCS}. The present result is at 
least four times better than that of \cite{mmarXiv22} and more than seventy nine times better than 
that of \cite{mmundecTCS}.

If we consider Turing machines with a high number of instructions and if we ignore the size of the 
data, then in the present paper the amount of prototiles is of order 7.$I$, it is 12.$I$ in 
\cite{mmarXiv22} and 192.$I$ in \cite{mmundecTCS}. Accordingly, in magnitude, the present paper 
is a bit more than 1.71 better than \cite{mmarXiv22} and it divides by more than 27 that 
of~\cite{mmundecTCS}. 

Note that if we consider a fixed universal Turing machine $U$ and if we consider the halting 
problem for $U$ applied to all its possible data, that problem is also undecidable. If $U$ is the 
considered tiny universal Turing machine, the single variable is then $D$. In that case, the 
previous formulas are all of the order of~$D$.

\ifnum 1=0 {
    The present paper is probably the last or the penultimate I shall write.  I take that
occasion to thank all my friends in science, namely Andrew Adamatsky, Serge Grigorieff, 
Genaro Juarez Martinez and George Paun.
\subsection*{Acknowledgement}

I am very pleased to acknowledge the interest of several colleagues and
friends to the main result of this paper. Let me especially thank Andr\'e
Barb\'e, Jean-Paul Delahaye, Chaim Goodman-Strauss, Serge Grigorieff, 
Yuri Gurevich, Tero Harju, Oscar Ibarra, Hermann Maurer, Gheorghe P\u aun, 
Grzegorz Rozenberg and Klaus Sutner. I am specially in debt to Professor Chaim 
Goodman-Strauss for most valuable comments to improve the presentation of
the proof. 
} \fi

\def\kvs{\vspace{-1pt}}

\newpage
\ligne{\hfill\bf\huge Appendix\hfill}
\vskip 10pt
We first reproduce the pictures for the construction of the interwoven triangles at a larger scale.
\vskip 0pt
\ligne{\hfill
\vtop{\leftskip 0pt\parindent 0pt\hsize=320pt
\ligne{\hfill\hskip-10pt
\includegraphics[scale=0.4]{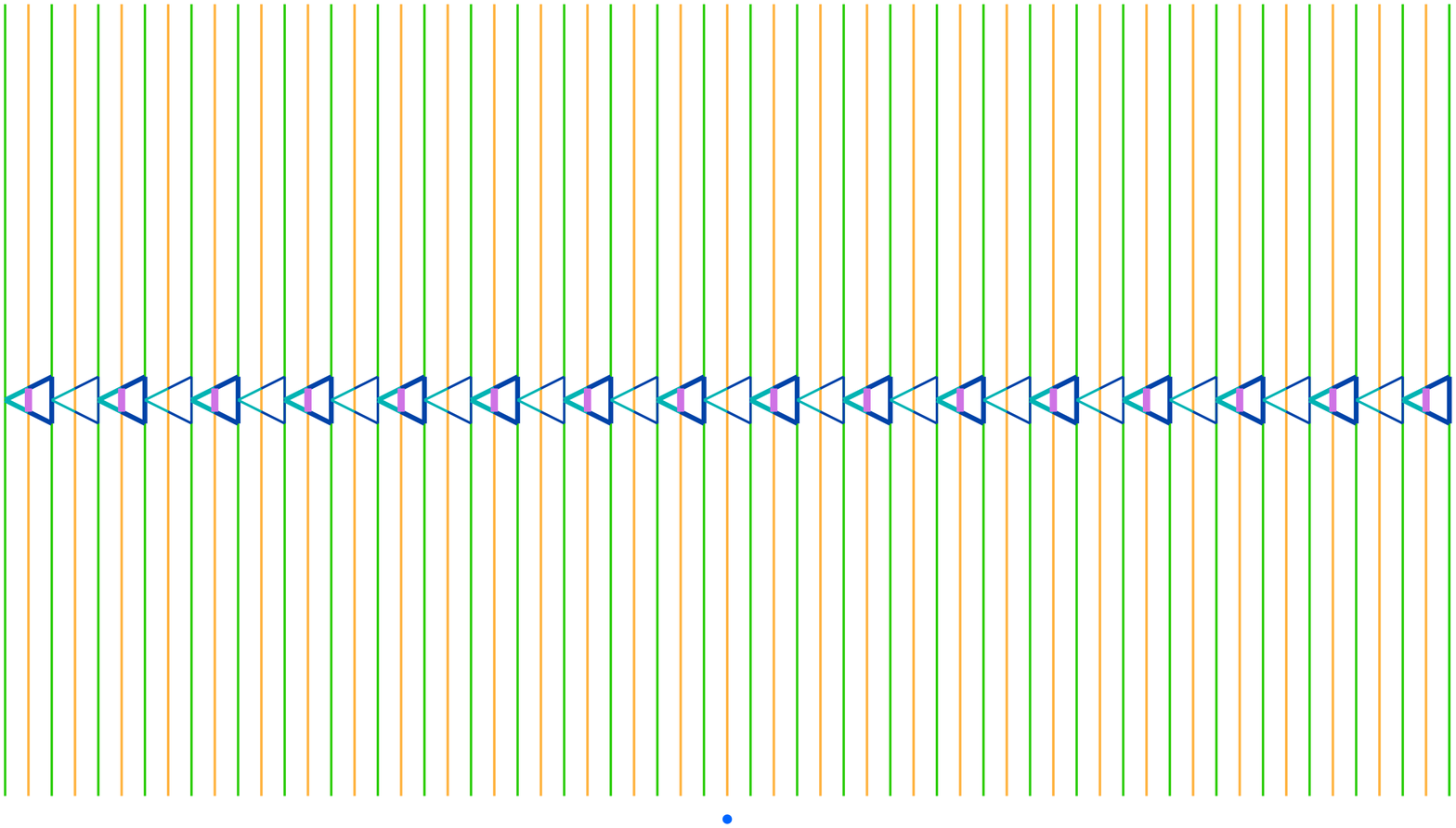}
\hfill}
\begin{fig}\label{t_interwov_0}
\leurre
The isoclines and the generation~$0$ of the interwoven triangles. Note the alternation of triangles and
phantoms.
\end{fig}
}
\hfill}
\vskip 0pt
\ligne{\hfill
\vtop{\leftskip 0pt\parindent 0pt\hsize=320pt
\ligne{\hfill\hskip-10pt
\includegraphics[scale=0.4]{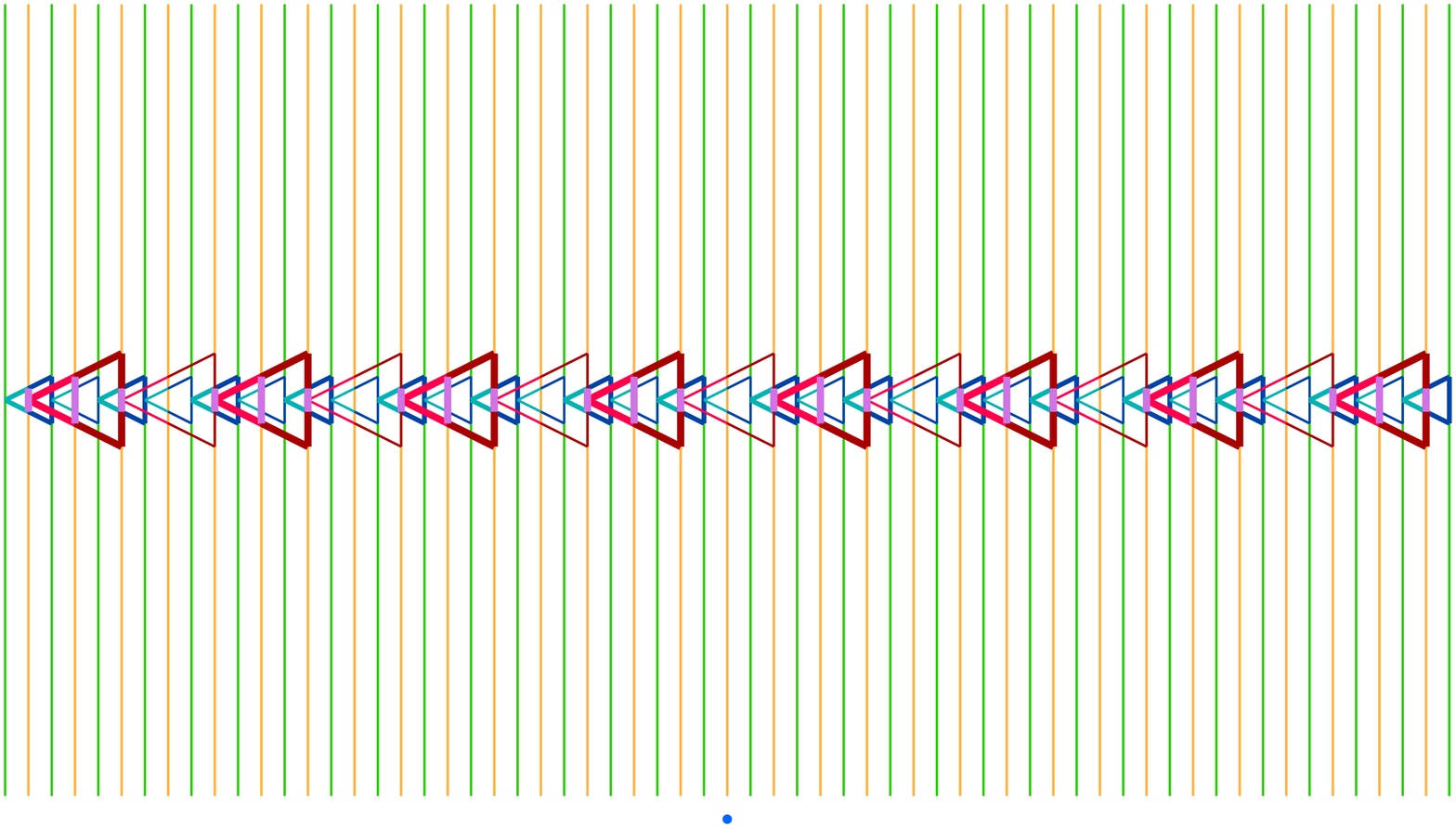}
\hfill}
\begin{fig}\label{f_interwov_1}
\leurre
The isoclines and the generations~$0$ and~1 of the interwoven triangles. Note the alternation of 
triangles and phantoms of the same colour
\end{fig}
}
\hfill}
\newpage
\ligne{\hfill
\vtop{\leftskip 0pt\parindent 0pt\hsize=320pt
\ligne{\hfill\hskip-10pt
\includegraphics[scale=0.4]{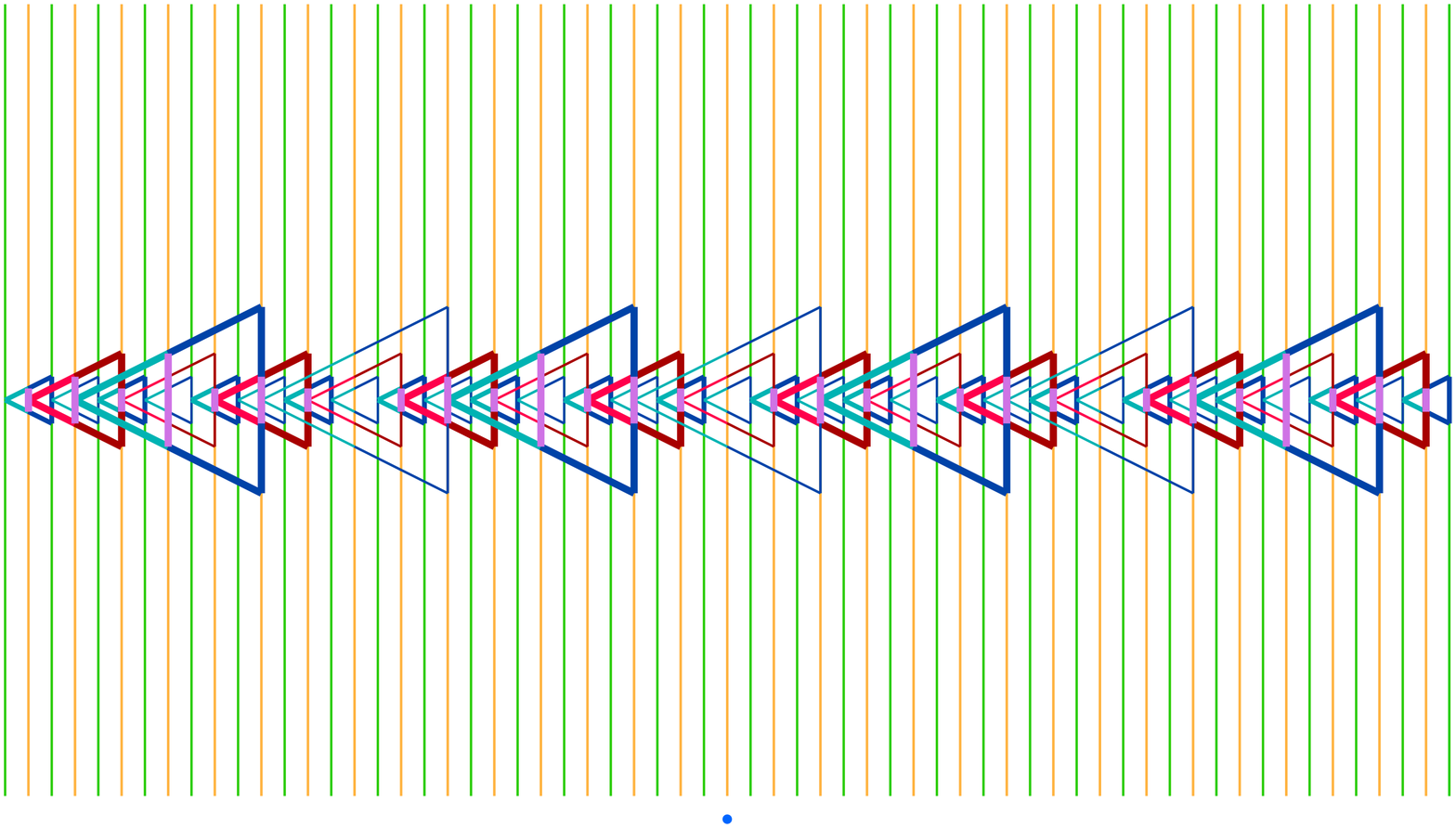}
\hfill}
\begin{fig}\label{f_interwov_2}
\leurre
The isoclines and the generations~$0$, 1 and~2 of the interwoven triangles. Note the alternation of 
triangles and phantoms of the same colour
\end{fig}
}
\hfill}
\vskip 0pt
\ligne{\hfill
\vtop{\leftskip 0pt\parindent 0pt\hsize=320pt
\ligne{\hfill\hskip-10pt
\includegraphics[scale=0.4]{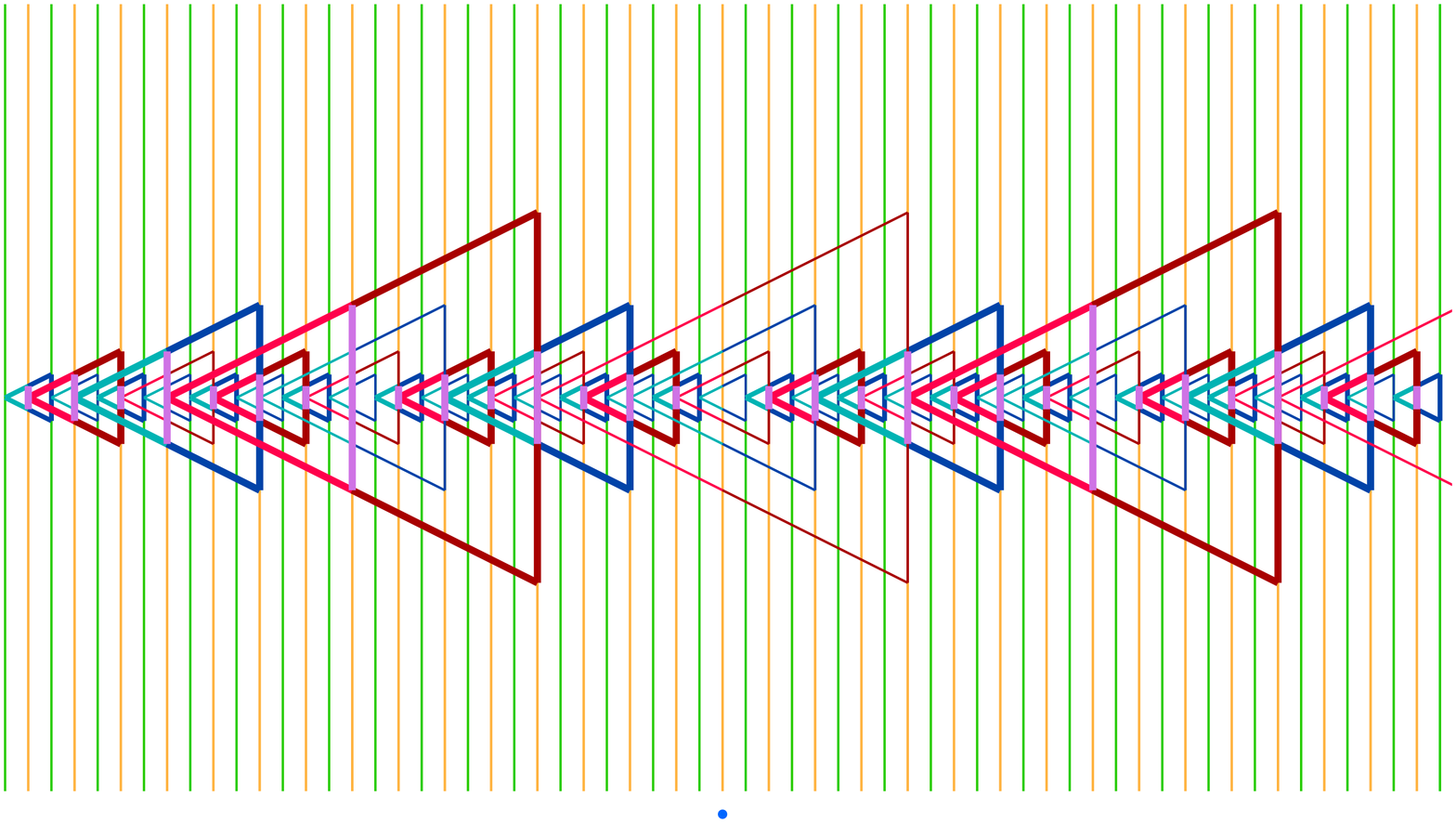}
\hfill}
\begin{fig}\label{f_interwov_3}
\leurre
The isoclines and the generations~$0$, 1, 2 and~3 of the interwoven triangles. Note the alternation of 
triangles and phantoms of the same colour
\end{fig}
}
\hfill}
\newpage
\ligne{\hfill
\vtop{\leftskip 0pt\parindent 0pt\hsize=320pt
\ligne{\hfill\hskip-10pt
\includegraphics[scale=0.4]{tttt_interwov_4.ps}
\hfill}
\begin{fig}\label{f_interwov_4}
\leurre
The isoclines and the generations~$0$, 1, 2, 3 and~4 of the interwoven triangles. Note the alternation of 
triangles and phantoms of the same colour
\end{fig}
}
\hfill}

\newpage
\vskip 20pt
As announced in Subsection~\ref{ss_iso_thr} we give several figures where the 
neighbourhood of the central tile occurs in Figure~\ref{f_proto}.
Figure~\ref{f_pavBB} illustrates how the rules are applicated starting from a fixed 
central and its father according to Figure~\ref{f_proto}. The following figures apply the
same principle.
\vskip 10pt
\ligne{\hfill
\vtop{\leftskip 0pt\parindent 0pt\hsize=300pt
\ligne{\hfill
\includegraphics[scale=0.35]{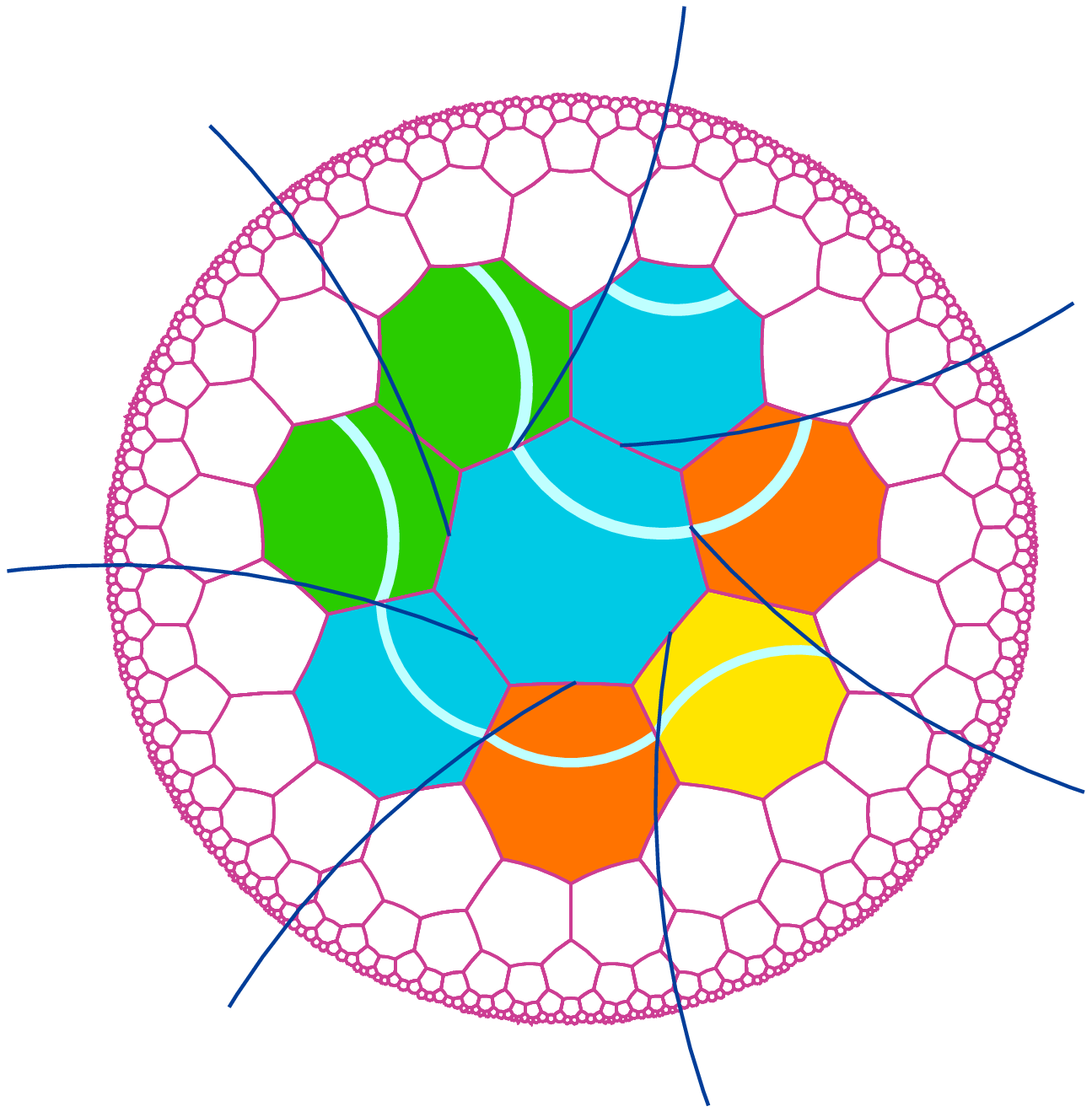}
\hfill}
\begin{fig}\label{f_pavsch}
\leurre
The central tile and its immediate neighbourhood is that of Figure~\ref{f_pavBB}.
\end{fig}
}
\hfill}
\vskip 10pt
\vskip 10pt
\ligne{\hfill
\vtop{\leftskip 0pt\parindent 0pt\hsize=300pt
\ligne{\hfill
\includegraphics[scale=0.55]{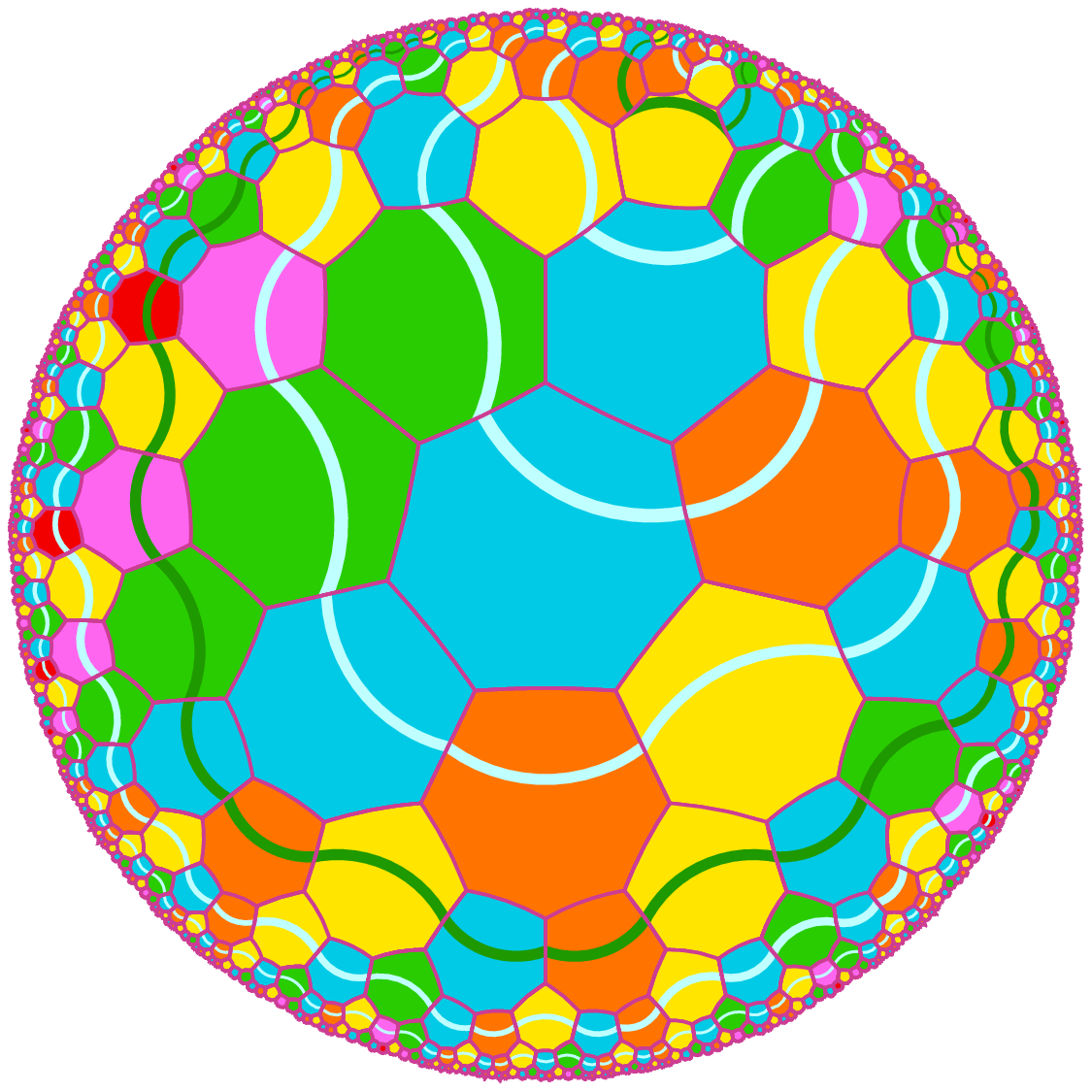}
\hfill}
\begin{fig}\label{f_pavBB}
\leurre
Central tile: a \BB-tile whose father is also a \BB-tile. The rules of~$(R_1)$
are applied.
\end{fig}
}
\hfill}
\newpage
Presently, the \BB-tiles with, as father, an \OO-tile then a \YY-one. There cannot be
a \GG-father: in that case, the \BB-son is replaced by an \MM-one, see 
Figure~\ref{f_pavMG}
\vskip 10pt
\ligne{\hfill
\vtop{\leftskip 0pt\parindent 0pt\hsize=300pt
\ligne{\hfill
\includegraphics[scale=0.55]{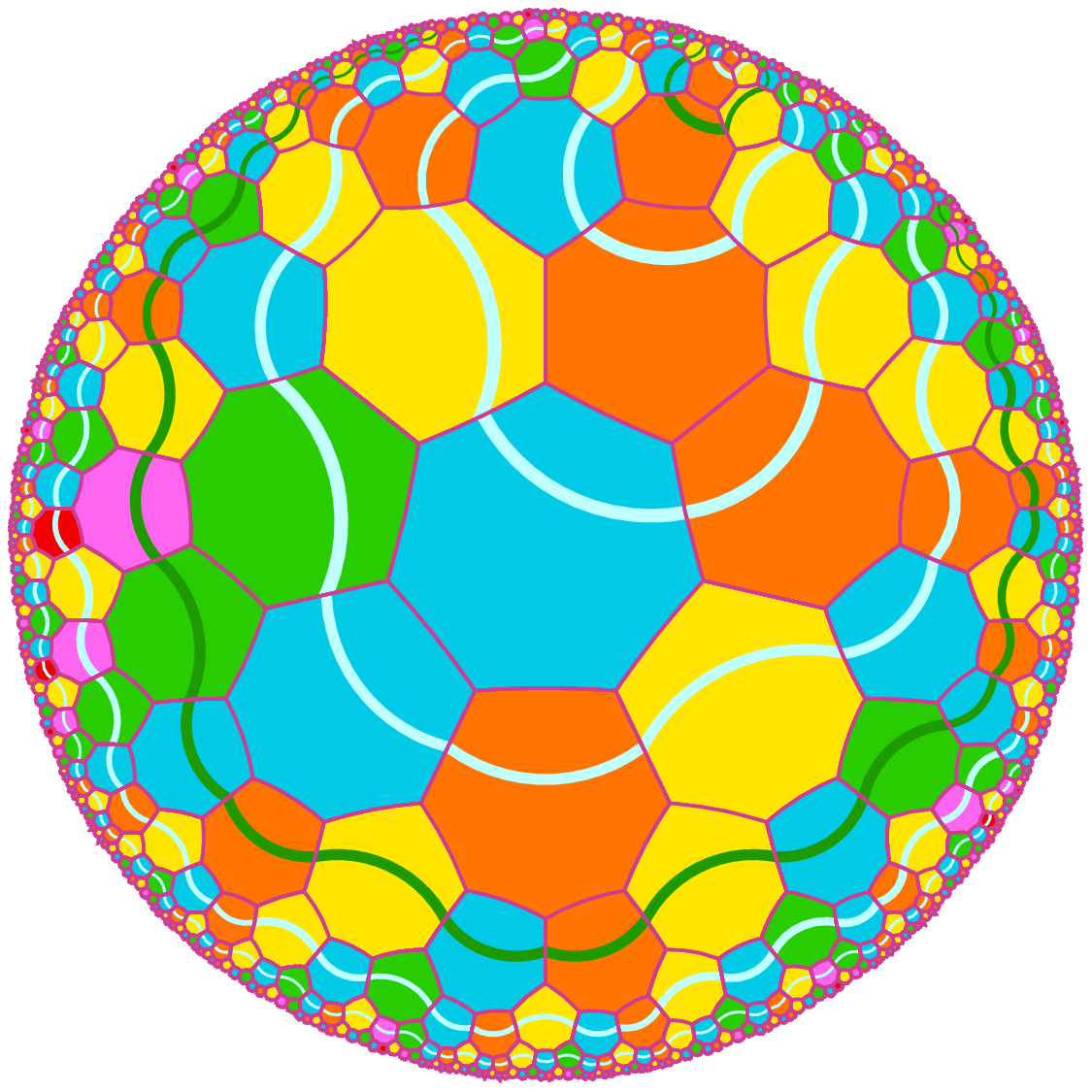}
\hfill}
\begin{fig}\label{f_pavBO}
\leurre
Central tile: a \BB-tile whose father is an \OO-tile. The rules of~$(R_1)$
are applied.
\end{fig}
}
\hfill}
\vskip 10pt
\ligne{\hfill
\vtop{\leftskip 0pt\parindent 0pt\hsize=300pt
\ligne{\hfill
\includegraphics[scale=0.6]{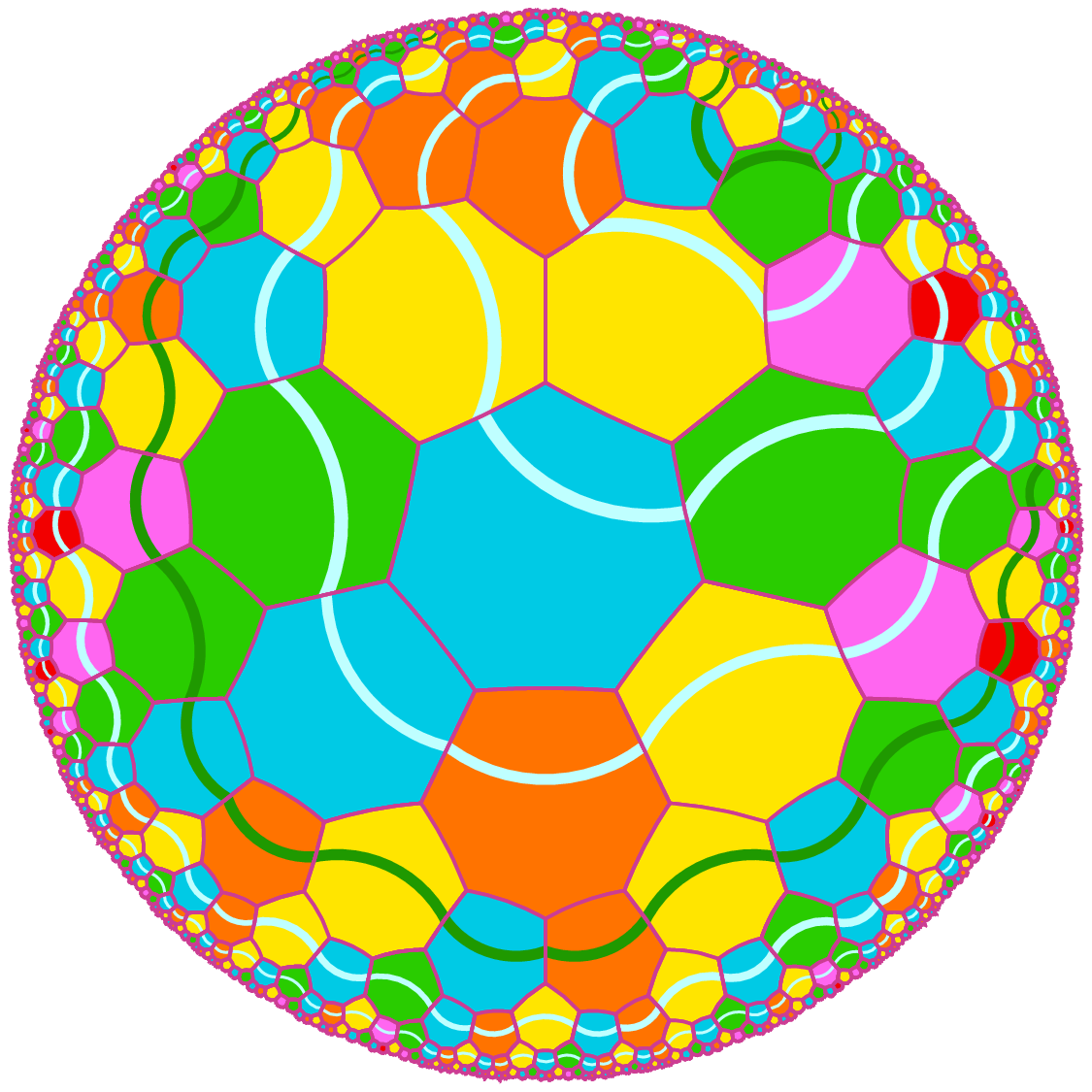}
\hfill}
\begin{fig}\label{f_pavBY}
\leurre
Central tile: a \BB-tile whose father is a \YY-tile. The rules of~$(R_1)$
are applied.
\end{fig}
}
\hfill}
\newpage
Presently, the case of an \MM-tile whose father is necessarily a \GG-tile, according
to rules~$(R_1)$.
\vskip 10pt
\ligne{\hfill
\vtop{\leftskip 0pt\parindent 0pt\hsize=300pt
\ligne{\hfill
\includegraphics[scale=0.6]{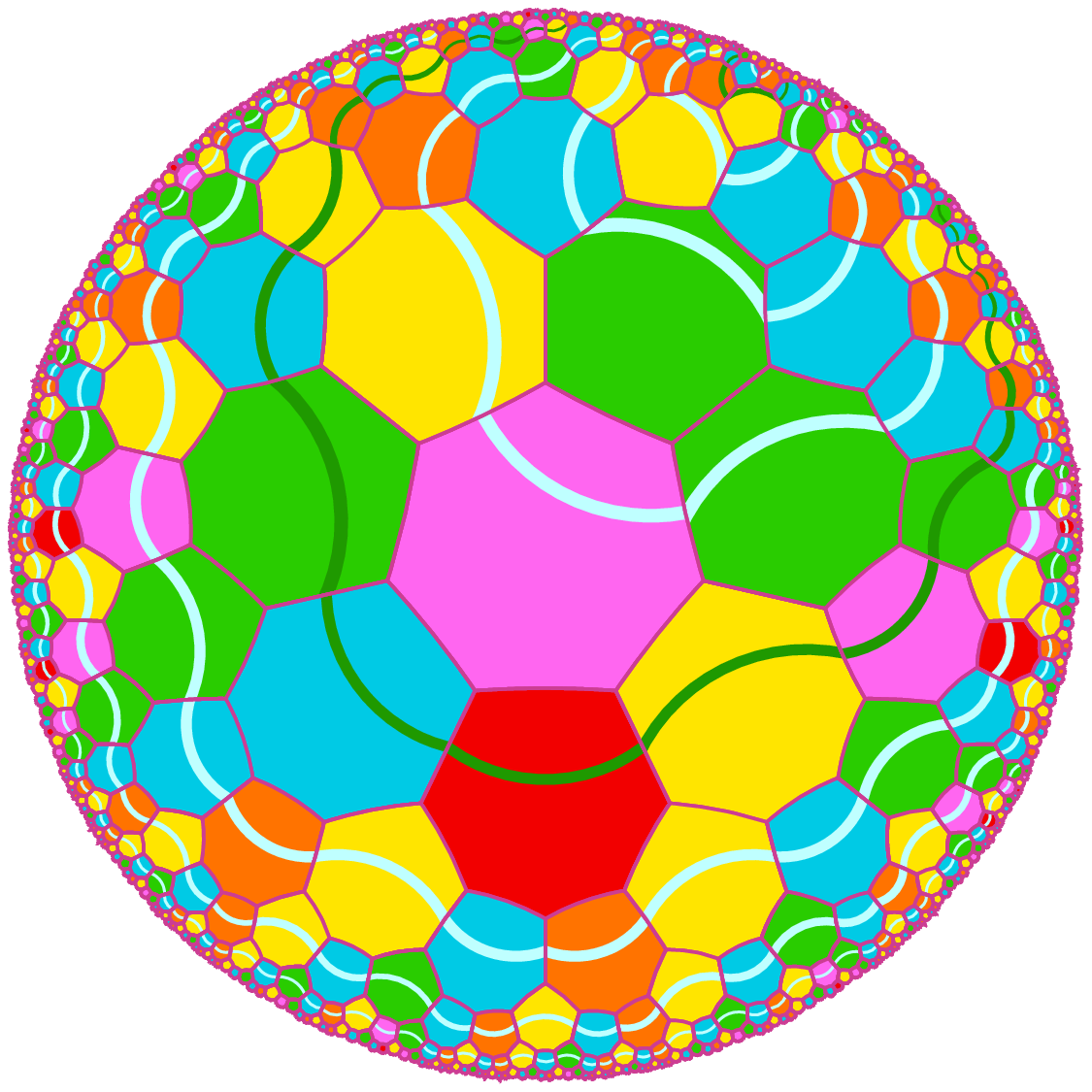}
\hfill}
\begin{fig}\label{f_pavMG}
\leurre
Central tile: an \MM-tile. Its father is necessarily a \GG-tile. The rules of~$(R_1)$
are applied. Note that the neighbourhood is different from those of Figures~\ref{f_pavBB}
and~\ref{f_pavBO}.
\end{fig}
}
\hfill}
\vskip 5pt
Presently, the central tile are \YY-tiles in the four following figures.
\vskip 10pt
\ligne{\hfill
\vtop{\leftskip 0pt\parindent 0pt\hsize=300pt
\ligne{\hfill
\includegraphics[scale=0.6]{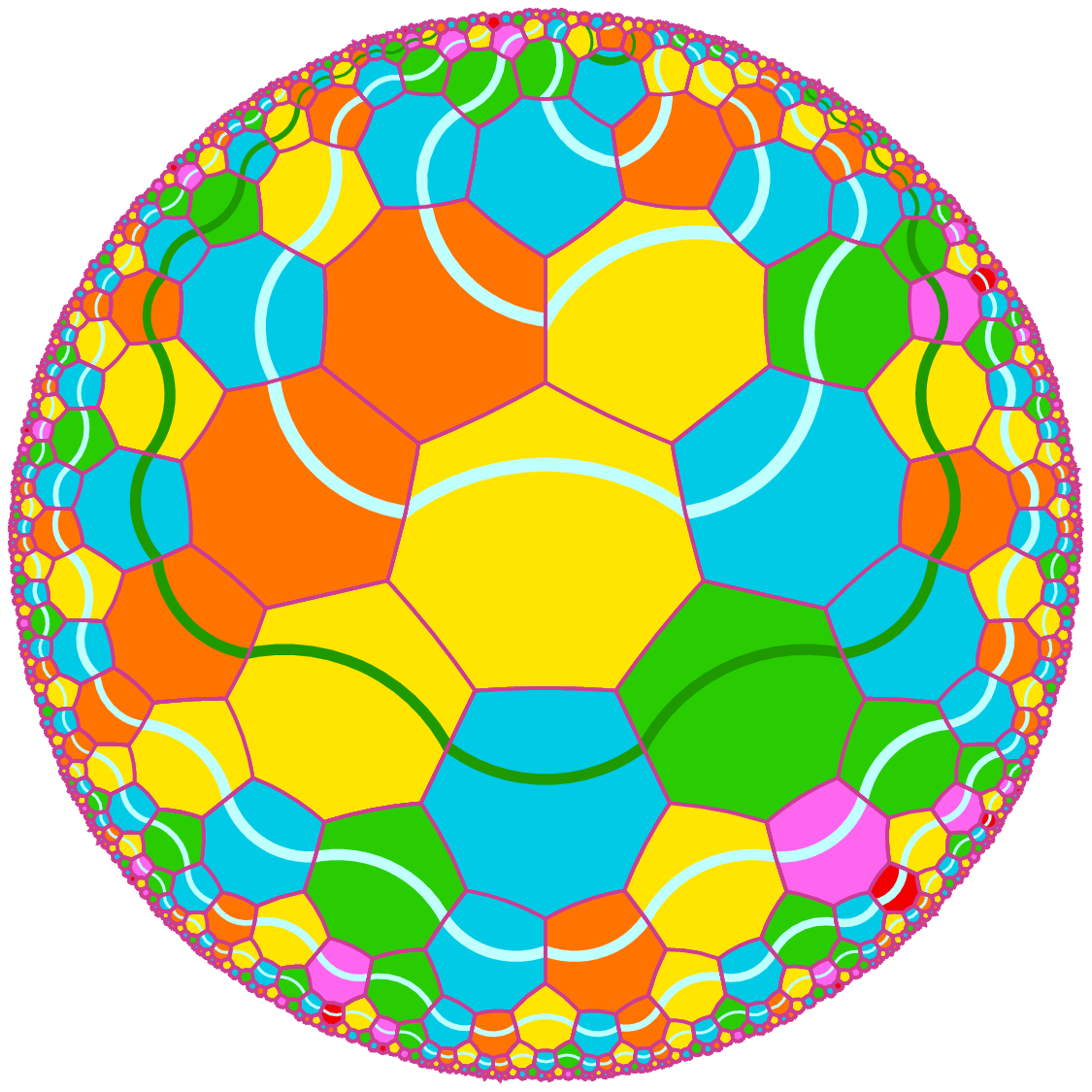}
\hfill}
\begin{fig}\label{f_pavYY}
\leurre
Central tile: a \YY-tile whose father is also a \YY-tile. The rules of~$(R_1)$
are applied.
\end{fig}
}
\hfill}
\newpage
\vskip 10pt
\ligne{\hfill
\vtop{\leftskip 0pt\parindent 0pt\hsize=300pt
\ligne{\hfill
\includegraphics[scale=0.6]{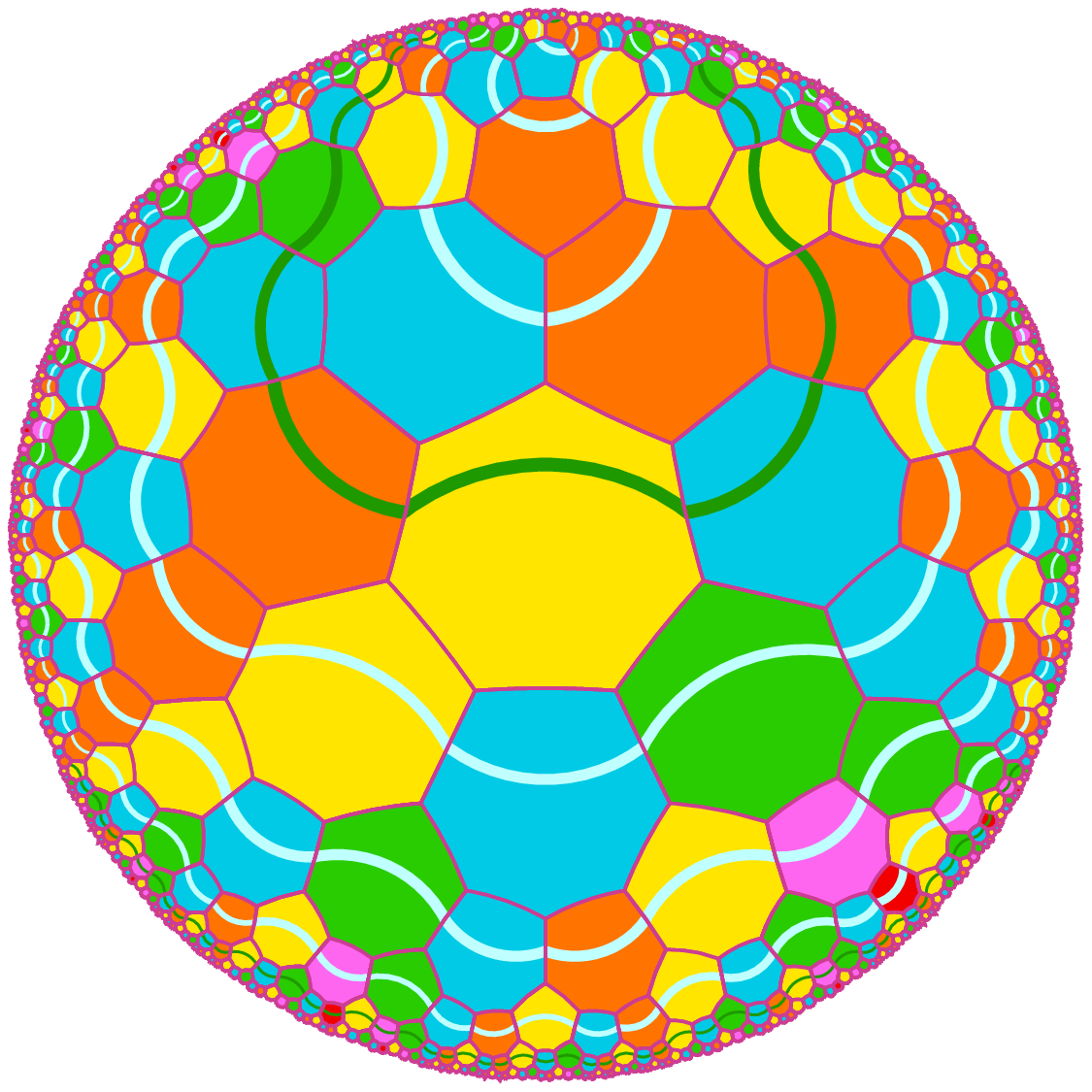}
\hfill}
\begin{fig}\label{f_pavYO}
\leurre
Central tile: a \YY-tile whose father is an \OO-tile. The rules of~$(R_1)$
are applied. The neighbourhood is different from that of Figure~\ref{f_pavYY}.
\end{fig}
}
\hfill}
\vskip 10pt
In Figures~\ref{f_pavYGM} and \ref{f_pavYGB}, the father is a \GG-tile.
\vskip 10pt
\ligne{\hfill
\vtop{\leftskip 0pt\parindent 0pt\hsize=300pt
\ligne{\hfill
\includegraphics[scale=0.6]{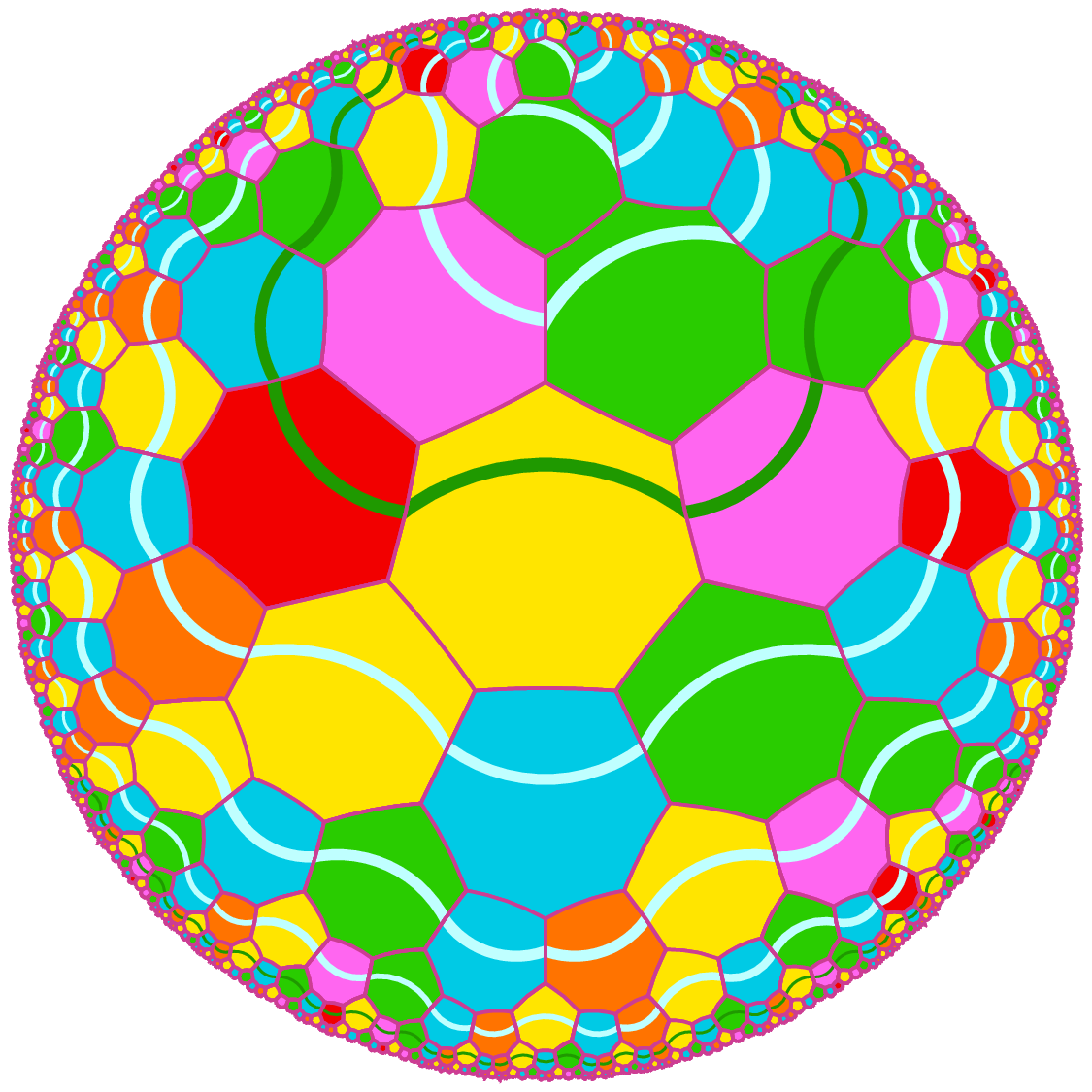}
\hfill}
\begin{fig}\label{f_pavYGM}
\leurre
Central tile: a \YY-tile whose father is a \GG-tile. The rules of~$(R_1)$
are applied. Neighbours~2 and~7 of the central tile are an \MM-tile.
\end{fig}
}
\hfill}
\vskip 10pt
\newpage
\ligne{\hfill
\vtop{\leftskip 0pt\parindent 0pt\hsize=300pt
\ligne{\hfill
\includegraphics[scale=0.6]{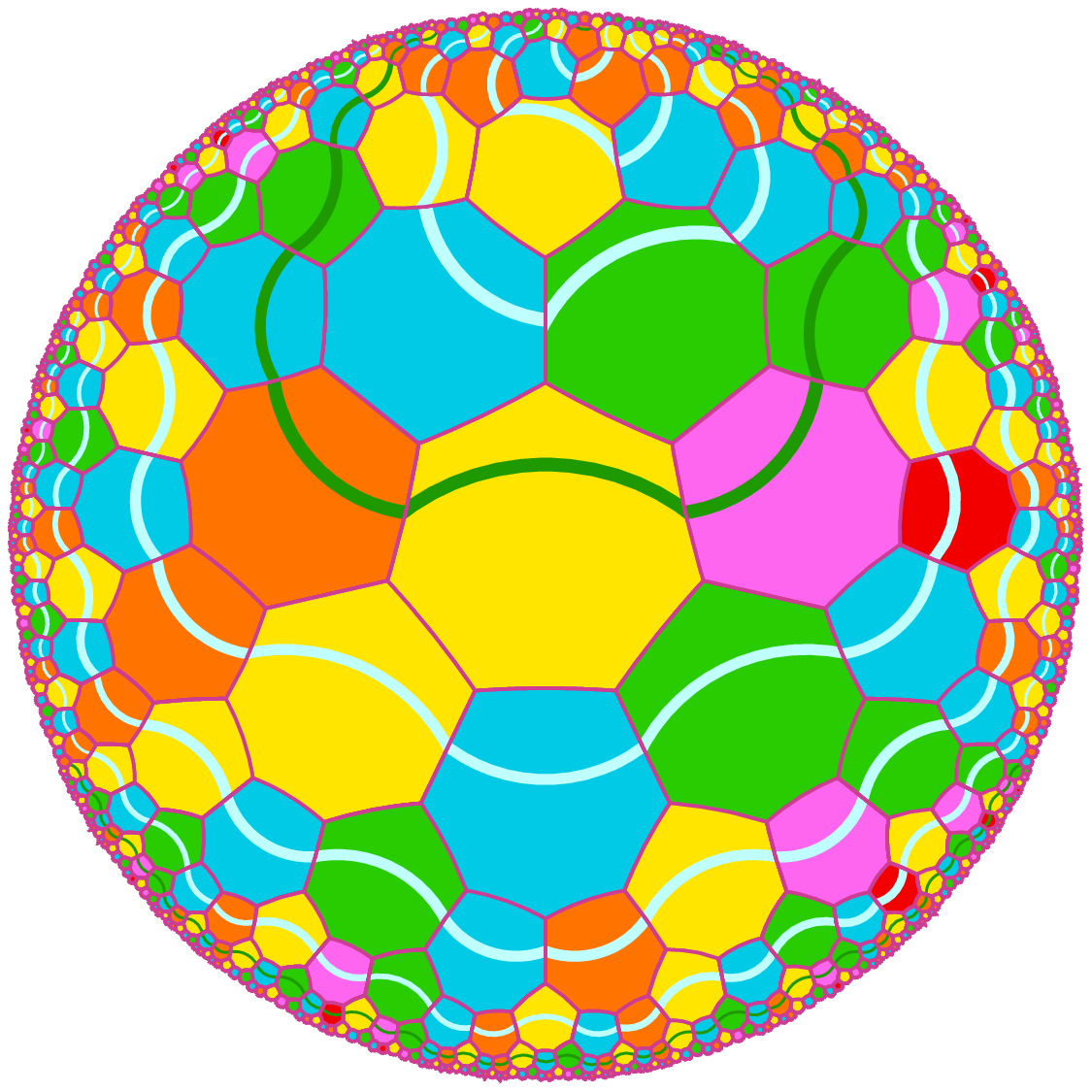}
\hfill}
\begin{fig}\label{f_pavYGB}
\leurre
Central tile: a \YY-tile whose father is a \GG-tile. The rules of~$(R_1)$
are applied. Neighbour~2 of the central tile is a \BB-tile, neighbour~7 is an \MM-one. 
\end{fig}
}
\hfill}
\vskip 5pt
Presently, the central tile is a \GG-tile. The father is either a \YY-tile or
a \GG-one.
\vskip 10pt
\ligne{\hfill
\vtop{\leftskip 0pt\parindent 0pt\hsize=300pt
\ligne{\hfill
\includegraphics[scale=0.6]{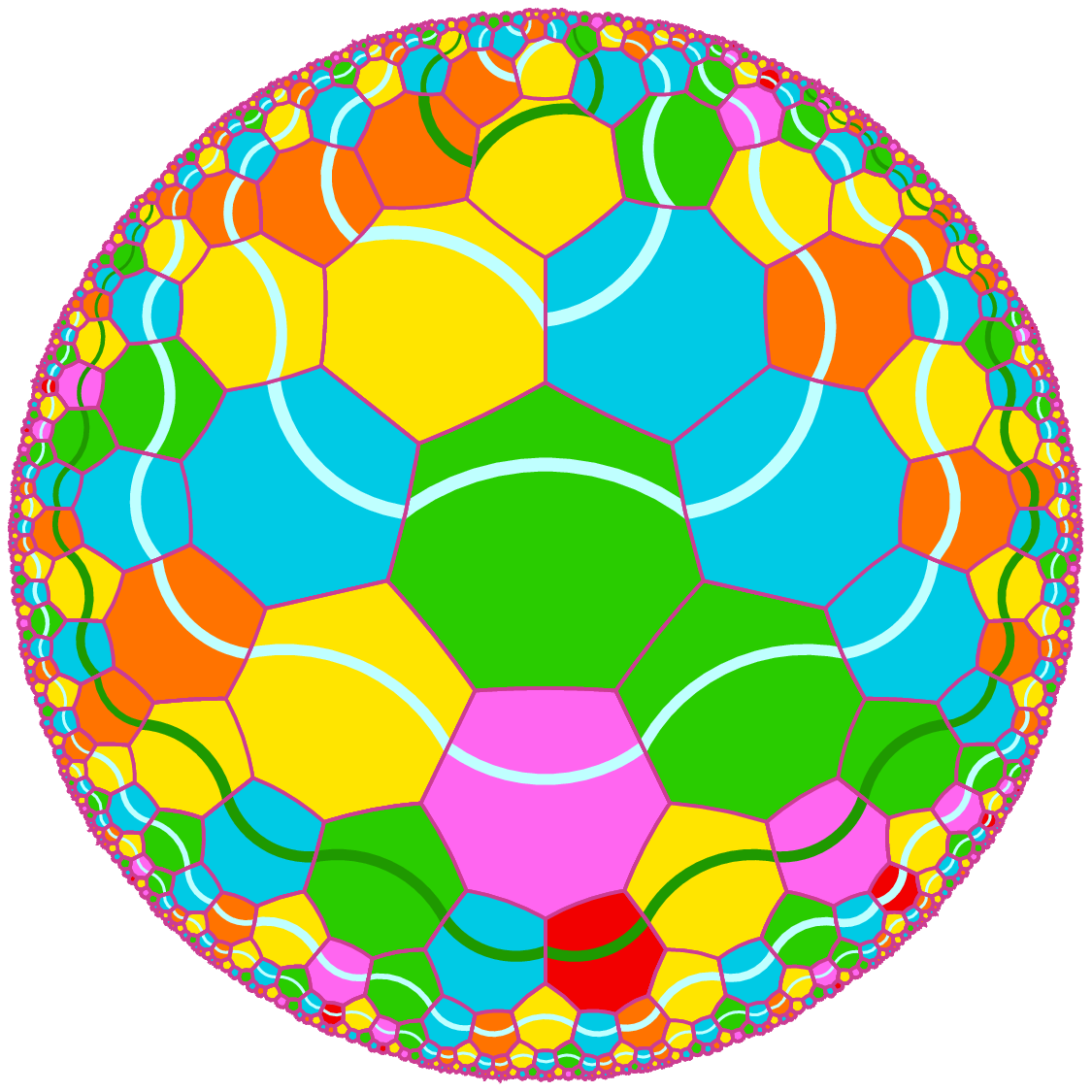}
\hfill}
\begin{fig}\label{f_pavGY}
\leurre
Central tile: a \GG-tile whose father is a \YY-tile. The rules of~$(R_1)$
are applied. Neighbours~4 of the central tile is an \MM-tile. Neighbour~2 is a \BB-one.
\end{fig}
}
\hfill}
\vskip 10pt
\ligne{\hfill
\vtop{\leftskip 0pt\parindent 0pt\hsize=300pt
\ligne{\hfill
\includegraphics[scale=0.6]{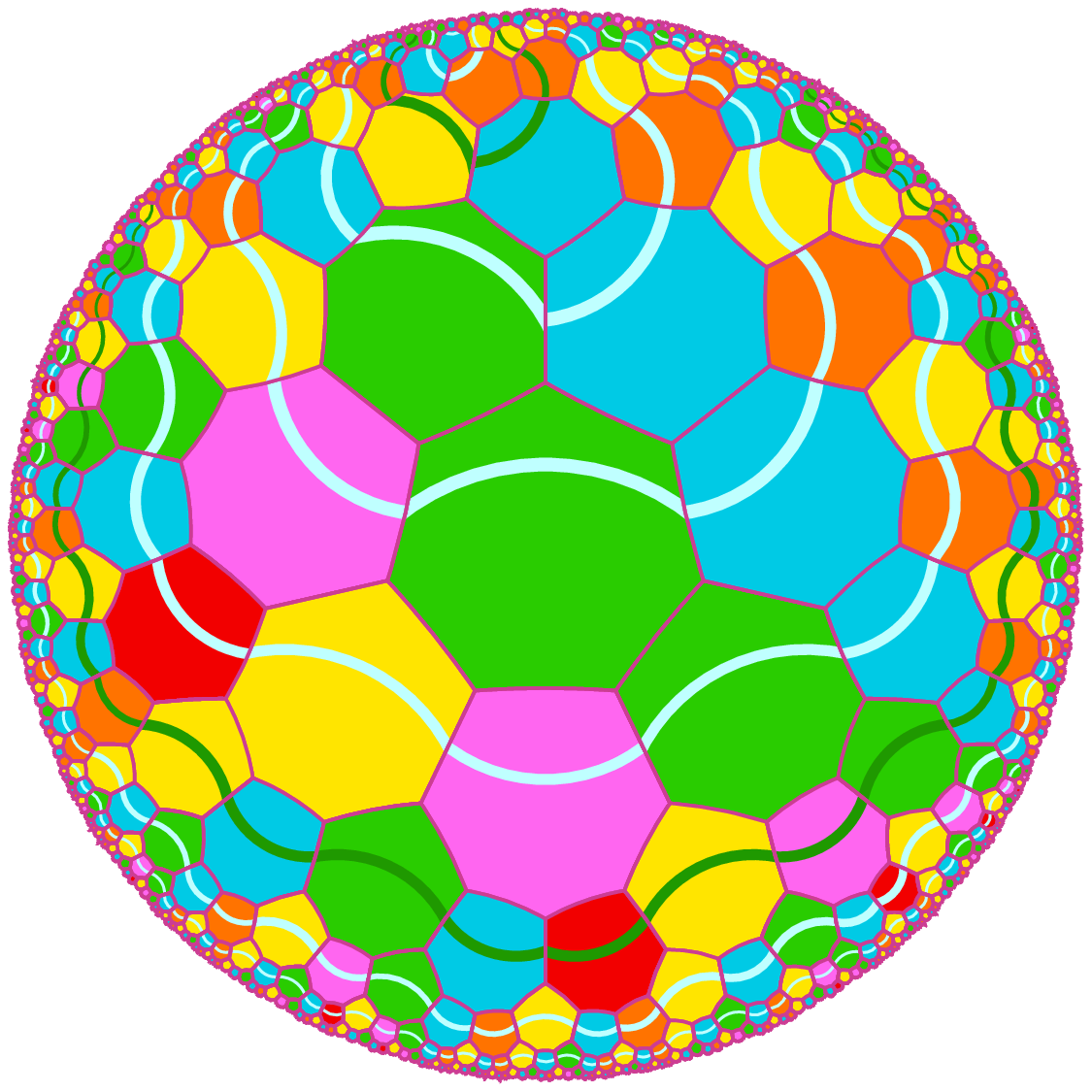}
\hfill}
\begin{fig}\label{f_pavGG}
\leurre
Central tile: a \GG-tile whose father is also a \GG-tile. The rules of~$(R_1)$
are applied. Neighbour~2 of the central tile is a \BB-tile, neighbour~3 is an \MM-one. 
\end{fig}
}
\hfill}
\vskip 10pt
Presently, the central tile is an \OO-tile. The father is either a \BB-tile or
an \OO-one.
\vskip 10pt
\ligne{\hfill
\vtop{\leftskip 0pt\parindent 0pt\hsize=300pt
\ligne{\hfill
\includegraphics[scale=0.6]{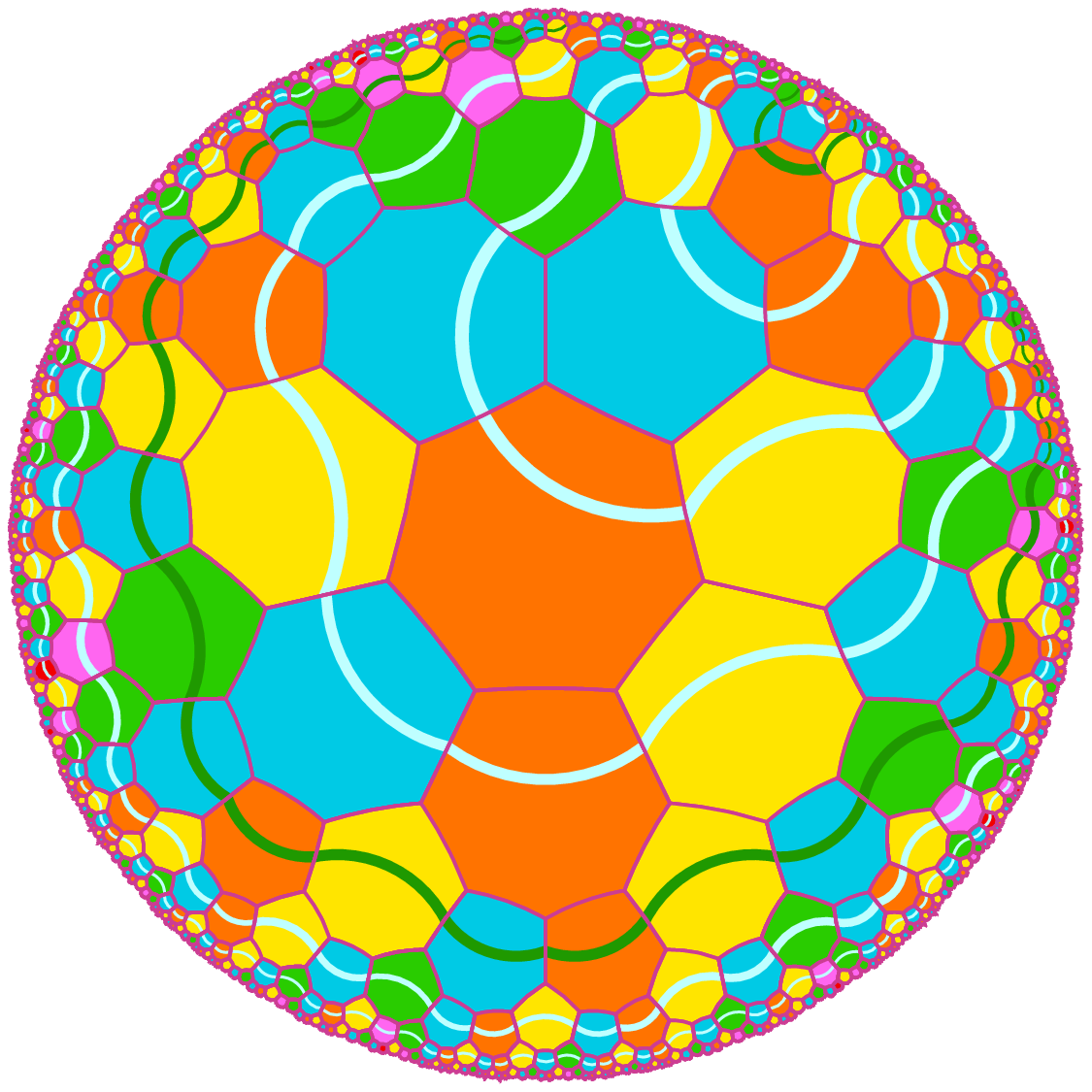}
\hfill}
\begin{fig}\label{f_pavOB}
\leurre
Central tile: an \OO-tile whose father is a \BB-tile. The rules of~$(R_1)$
are applied. 
\end{fig}
}
\hfill}
\newpage
\ligne{\hfill
\vtop{\leftskip 0pt\parindent 0pt\hsize=300pt
\ligne{\hfill
\includegraphics[scale=0.6]{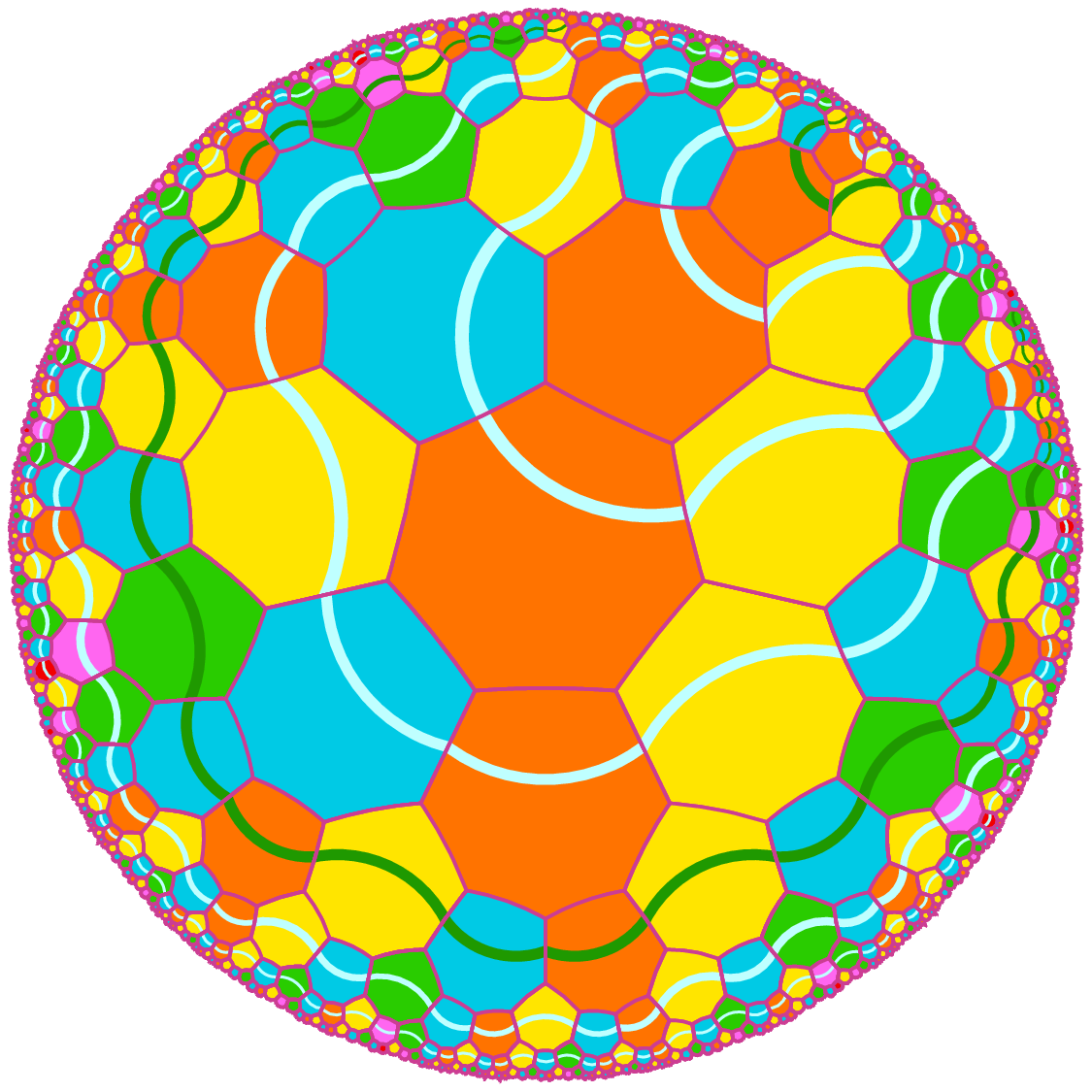}
\hfill}
\begin{fig}\label{f_pavOO}
\leurre
Central tile: an \OO-tile whose father is an \OO-tile. The rules of~$(R_1)$
are applied. Besides the father, the neighbourhood is the same as in Figure~\ref{f_pavOB}.
\end{fig}
}
\hfill}
\vskip 10pt
Presently, the central tile is an \RR-tile. The father is necessarily an \MM-tile.
\vskip 10pt
\ligne{\hfill
\vtop{\leftskip 0pt\parindent 0pt\hsize=300pt
\ligne{\hfill
\includegraphics[scale=0.6]{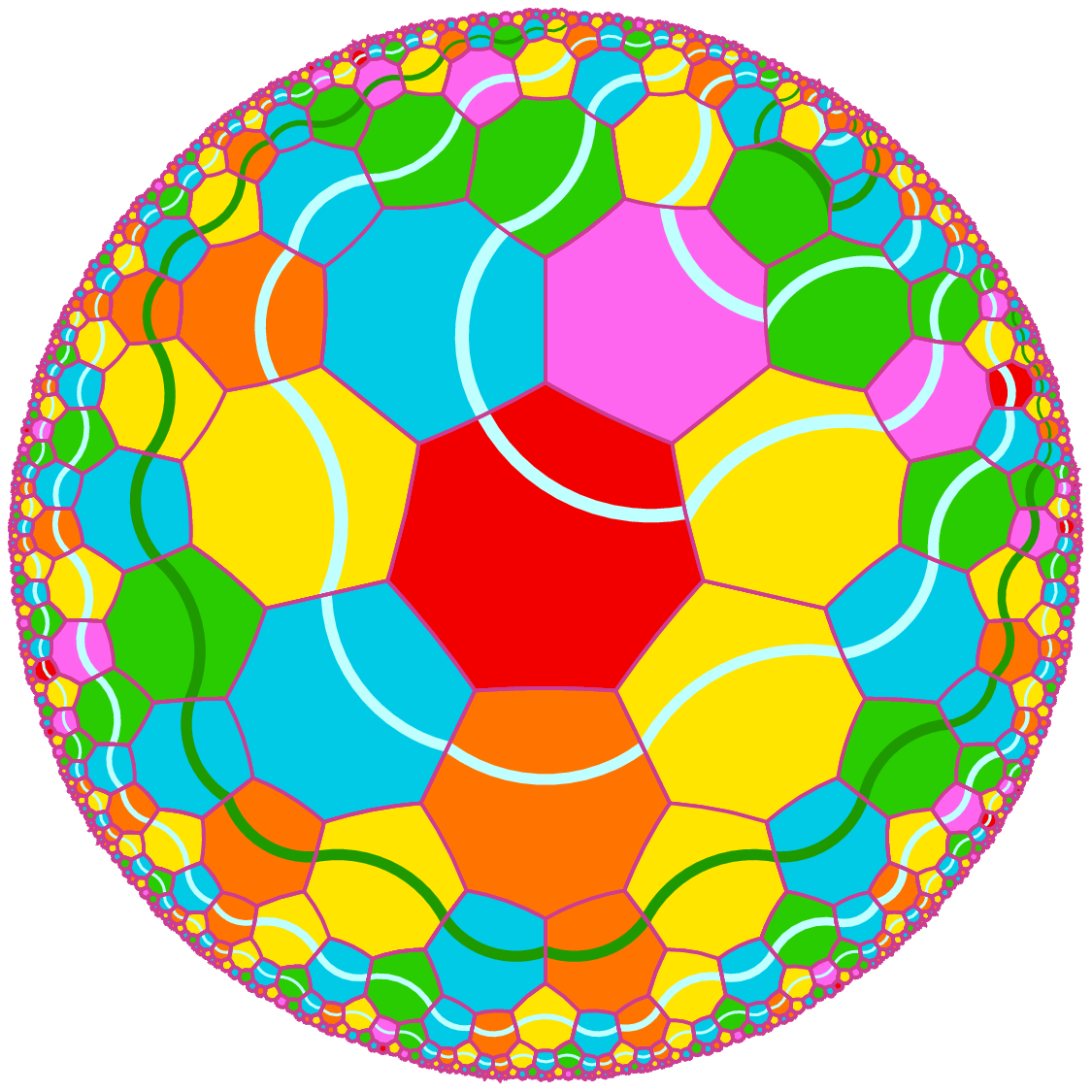}
\hfill}
\begin{fig}\label{f_pavRM}
\leurre
Central tile: an \RR-tile. The father is an \MM-tile. The rules of~$(R_1)$
are applied. The neighbourhood is different from those of Figures~\ref{f_pavOB} 
and~\ref{f_pavOO} despite the fact that the rules are similar to those involving 
\OO-tiles.
\end{fig}
}
\hfill}

\begin{thebibliography}{5}
\font\itviii=cmti10
\font\bfviii=cmbx10

\bibitem{berger}
Berger~R., The undecidability of the domino problem,
{\it Memoirs of the American Mathematical Society}, {\bf 66}, (1966), 1-72.

\kvs
\bibitem{ibkmACRI}
Chelghoum~K., Margenstern~M., Martin~B., Pecci~I.,
Cellular automata in the hyperbolic plane: proposal for a new environment,
{\it Lecture Notes in Computer Sciences},
{\bf 3305}, (2004), 678-687, proceedings of ACRI'2004, Amsterdam, October,
25-27, 2004.

\kvs
\bibitem{goodmana}
Goodman-Strauss, Ch.,
A strongly aperiodic set of tiles in the hyperbolic plane,
{\it Inventiones Mathematicae}, {\bf 159}(1), (2005), 119-132.

\kvs
\bibitem{goodmanb}
Goodman-Strauss, Ch.,
Growth, aperiodicity and undecidability,
{\it invited address at the AMS meeting at Davidson, NC}, March, 3-4, (2007).

\kvs
\bibitem{gurevich}
Gurevich~Yu., Koriakov~I.,
A remark on Berger's paper on the domino problem,
{\it Siberian Mathematical Journal}, {\bf 13}, (1972), 459--463.

\kvs
\bibitem{hanf}
Hanf~W.,
Nonrecursive tilings of the plane. I.
{\it Journal of Symbolic Logic}, {\bf 39}, (1974), 283-285.

\kvs
\bibitem{hopcroft}
Hopcroft, J.E., Motwani, R., 
Ullman, J.D.,
{\it Introduction to Automata Theory, Languages, and Computation},
Addison Wesley, Boston/San Francisco/New York, (2001).

\kvs
\bibitem{mann}
Mann~C., Heesch's tiling problem, {\it American Mathematical Monthly},
{\bf 111}(6), (2004), 509-517.

\kvs
\bibitem{mmJUCSii}
Margenstern~M.,
New Tools for Cellular Automata of the Hyperbolic Plane,
{\it Journal of Universal Computer Science}
{\bf 6{\rm N$^\circ$12}}, 1226--1252, (2000)
%

\kvs
\bibitem{mmDMTCS}
Margenstern~M.,
Cellular Automata and Combinatoric Tilings in Hyperbolic Spaces, a
survey,
{\it Lecture Notes in Computer Sciences},
{\bf 2731}, (2003), 48-72.

\kvs
\bibitem{mmJCA}
Margenstern~M.,
A new way to implement cellular automata on the penta- and heptagrids,
{\it Journal of Cellular Automata} {\bf 1}, N$^\circ1$, (2006), 1-24.

\kvs
\bibitem{mmarXiv1}
Margenstern~M.,
About the domino problem in the hyperbolic plane
from an algorithmic point of view,
{\it iarXiv:cs.CG/$0603093$ v$1$}, (2006), 11p.

\kvs
\bibitem{mmarXiv3}
Margenstern~M.,
The finite tiling problem is undecidable in the hyperbolic plane,
{\it arxiv$:$cs.CG/$0703147v1$}, (2007), March, 8p.

\kvs
\bibitem{mmarXiv4}
Margenstern~M.,
The periodic domino problem is undecidable in the hyperbolic plane,
{\it arxiv$:$cs.CG/$0703153v1$}, (2007), March, 12p.

\kvs
\bibitem{mmrp07}
Margenstern~M.,
The Finite Tiling Problem Is Undecidable in the Hyperbolic Plane,
{\it Workshop on Reachability Problems}, {\bf RP07}, July 2007, Turku, Finland,

\kvs
\bibitem{mmbook1}
Margenstern~M.,
Cellular Automata in Hyperbolic Spaces, Volume 1, Theory,
{\it OCP}, Philadelphia, (2007), 422p.

\kvs
\bibitem{mmBEATCS}
Margenstern~M.,
The Domino Problem of the Hyperbolic Plane is Undecidable,
{\it Bulletin of the EATCS}, {\bf 93}, October, (2007), 220-237.

\kvs
\bibitem{mmundecTCS}
Margenstern~M.,
The domino problem of the hyperbolic plane is undecidable, {\it Theoretical Computer
Science}, {\bf 407}(1-3), (2008), 29-84.

\kvs
\bibitem{mmJCAd}
Margenstern~M.,
 A Uniform and Intrinsic Proof that there are Universal Cellular Automata
in Hyperbolic Spaces,
{\it Journal of Cellular Automata} {\bf 3}, N$^\circ2$, (2008), 157-180.

\kvs
\bibitem{mmarXiv22}
Margenstern~M.,
The domino problem of the hyperbolic plane is undecidable, new proof,
{\it arXiv$:$cs.DM/$2205.07317v1$}, (2022), May, 49pp.

\kvs
\bibitem{myers}
Myers~D.,
Nonrecursive tilings of the plane. II.
{\it Journal of Symbolic Logic}, {\bf 39}, (1974), 286-294.

\kvs
\bibitem{neary_woods}
Neary~T., Woods~D., Four Small Universal Turing Machines,
{\it Fundamenta Informaticae}, {\bf 91}(1), (2009), 123-144.

\kvs
\bibitem{robinson1}
Robinson~R.M.
Undecidability and nonperiodicity for tilings of the plane,
{\it Inventiones Mathematicae}, {\bf 12}, (1971), 177-209.

\kvs
\bibitem{robinson2}
Robinson~R.M.
Undecidable tiling problems in the hyperbolic plane.
{\it Inventiones Mathematicae}, {\bf 44}, (1978), 259-264.

\kvs
\bibitem{turing}
Turing A.M.,
On computable real numbers, with an application to the
Entscheidungsproblem, {\it Proceedings of the London Mathematical
Society}, ser. 2, {\bf 42}, 230-265, (1936).

\kvs
\bibitem{wang}
Wang~H.
Proving theorems by pattern recognition, Bell System Tech. J. vol. {\bf 40}
(1961), 1--41.
\end{thebibliography}
\end{document}